\pdfoutput=1
\documentclass{aa}  

\usepackage{graphicx}
\usepackage{txfonts}
\usepackage[utf8]{inputenc}
\usepackage{threeparttable}
\usepackage{amsmath}
\usepackage{hyperref}
\usepackage{float}
\usepackage[toc,page]{appendix} 
\usepackage[caption = false]{subfig}
\usepackage{newfloat}
\DeclareFloatingEnvironment[name={Supplementary Figure}]{suppfigure}

\begin{document}

   \title{The massive star population of the Virgo Cluster galaxy NGC 4535}

   \subtitle{Discovery of new massive variable candidates with the \textit{Hubble Space Telescope}}
	\author{ Z.T. Spetsieri\inst{1,2} et al.}
   \author{Z.T. Spetsieri\inst{1,2}, A.Z. Bonanos \inst{1}, M. Kourniotis\inst{1,3}, M. Yang\inst{1}, S. Lianou \inst{4}, I. Bellas-Velidis \inst{1}, P. Gavras \inst{1}, \\D. Hatzidimitriou \inst{2,1}, M. Kopsacheili\inst{5}, M.I. Moretti \inst{1,6}, A. Nota\inst{7,8}, E. Pouliasis\inst{1,2}, K.V. Sokolovsky \inst{1,9,10}}

\authorrunning{Spetsieri et al.} 
\titlerunning{The massive star population of NGC 4535}
\institute{ IAASARS, National Observatory of Athens, 15236 Penteli, Greece\\
\email{zspetsieri@noa.gr}
 \and Department of Astrophysics, Astronomy \& Mechanics, Faculty of
Physics, University of Athens, 15783 Athens, Greece
\and Czech Academy of Sciences, Astronomical Institute, Fricova 298, Ondrejov, 251 65, Czech Republic
\and Laboratoire AIM, CEA/IRFU/Service d'Astrophysique, Université Paris Diderot, Bat. 709, F-91191 Gif-sur-Yvette, France
\and IESL/FORTH \& University of Crete, Greece
\and INAF Osservatorio Astronomico di Capodimonte, via Moiariello 16, 80131 Naples, Italy 
\and Space Telescope Science Institute, 3700 San Martin Drive, Baltimore, MD 21218, USA
\and ESA, SRE Operations Division, Spain
\and Sternberg Astronomical Institute, Moscow State University,
Universitetskii pr. 13, 119992 Moscow, Russia
\and Astro Space Center of Lebedev Physical Institute, Profsoyuznaya Str.
84/32, 117997 Moscow, Russia}
%
 
\abstract {We analyzed the massive star population of the Virgo Cluster galaxy NGC 4535 using archival \textit{Hubble Space Telescope} Wide Field Planetary Camera 2 images in filters F555W and F814W, equivalent to Johnson \textit{V} and Kron-Cousins \textit{I}. We performed high precision point spread function fitting photometry of 24353 sources including 3762 candidate blue supergiants, 841 candidate yellow supergiants and 370 candidate red supergiants. We estimated the ratio of blue to red supergiants as a decreasing function of galactocentric radius. Using Modules for Experiments in Stellar Astrophysics isochrones at solar metallicity, we defined the luminosity function and estimated the star formation history of the galaxy over the last 60 Myrs. We conducted a variability search in the V and I filters using three variability indexes: the median absolute deviation, the interquartile range and the inverse von-Neumann ratio. This analysis yielded 120 new variable candidates with absolute magnitudes ranging from M$_{V}$ = $-$4 to $-$11 mag. We used the MESA evolutionary tracks at solar metallicity, to classify the variables based on their absolute magnitude and their position on the color-magnitude diagram. Among the new candidate variable sources are eight candidate variable red supergiants, three candidate variable yellow supergiants and one candidate luminous blue variable, which we suggest for follow-up observations.}



    \maketitle
%

\section{Introduction}

\par Identifying massive stars at a range of metallicities is the first step toward constraining stellar evolutionary models, determining the upper limits for stellar masses, temperatures and luminosities, finding supernova (SN) progenitors and providing observational constraints for these sources in different environments. Due to complex mechanisms, such as mass loss, internal mixing and magnetic field effects \citep[see reviews by][]{Langer2012, Smith2014}, predicting and understanding the evolution of stars with masses $>$ 8 M$_{\odot}$ remains a challenge. Owing to their high luminosities, such stars are observable out to large distances, which allows for the investigation of their properties well beyond the Local Group. 

\par The massive star population could shed light on the link between metallicity and star formation. For example, the ratio of the number of blue to red supergiants (B/R) versus galactocentric distance was found to be related to metallicity, which could modulate mass-loss rates and convection processes \citep[]{brunish1986,LangerBsgs}. Supernova (SN) explosions are also related to processes which enrich the interstellar and intergalactic medium with heavy elements affecting both the metallicity and the star formation in galaxies. Finding the link between the progenitors and the SN types is important for determining the mass of the progenitors and understanding the underlying physical mechanisms of a SN explosion \citep{Smartt2009}. The types of massive stars linked with core-collapse SN explosions are Wolf-Rayet stars (WR) stars, red supergiants (RSGs) along with blue supergiants (BSGs) and luminous blue variables (LBVs) \citep[e.g.][]{Kotak2006,Szczygiel2010,Maund2011,Vanbeveren2013,Georgy2012, Groh2013}. Photometric studies and in particular variability studies of stellar populations in galaxies, could significantly contribute to identifying SN progenitors and SNe with multiple eruption events \citep[e.g.][]{Smith2010SN}.

\par Variability studies of stars in nearby galaxies provide insight into the evolutionary states of poorly understood classes of massive stars \citep[e.g.][]{Szczygiel2010, Kourniotis, Conroy2018}. Understanding the causes of stellar variability could shed light on the open questions of stellar evolution and help constrain fundamental stellar parameters \citep[e.g.][]{Torres2010}. 
The high spatial resolution and sensitivity of the \textit{Hubble Space Telescope} (\textit{HST}) has enabled stellar variability searches out to 20 Mpc. For example, the \textit{HST} Extragalactic Distance Scale Key Project \citep{Freedman2001} and, more recently, \citet{Riess2016} have contributed to an accurate determination of the Hubble constant. IZw18 is a distant (19 Mpc), ultra-low metallicity dwarf galaxy (Z = 0.0004), in which classical Cepheids have been identified, among other luminous variables by \textit{HST} observations \citep{Fiorentino2008}. The next step is to examine the population of luminous stars and their variability properties, in particular targeting LBVs \citep[]{Humphreys1994}, YSGs/YHGs \citep[]{deJager98, Neugent2010}, and RSGs \citep[]{Massey2003, Kiss2006, Massey2017} in a larger number of distant galaxies.

\par The goal of this work is to identify new massive stars in the Virgo Cluster galaxy NGC 4535 and investigate whether they display variability by using archival HST data. This galaxy was part of the \textit{HST} Key Project \citep{Freedman2001}. \citet{Macri99} reported 50 Cepheid variables, which were used to derive the period luminosity relation and distance to the galaxy, at 16.0 $\pm$ 1.9 Mpc or $\mu_0$=31.02 $\pm$ 0.26 mag. However, there are bound to be non-periodic massive variable candidates in the data, as \citet{Macri99} only searched for periodic variables that fit Cepheid template light curves.

\par This work is part of the ESA project "Hubble Catalog of Variables" \citep[HCV;][] {Sokolovsky16,HCV1, YangHCV, SokolovskyHCV}. The HCV project aims to identify the variable sources in the HSC \citep[]{Whitmore16, Moretti2018}, out of 92 million objects with over 300 million measurements detected by the WFPC2, ACS and WFC3 cameras on HST, by using an automated pipeline to detect and validate the variable candidates in different fields. The final catalog including all variable sources will consist of point-like and extended sources. NGC 4535 is one of the galaxies that are part of the control sample used to compare known variables with the variable candidates detected by the HCV algorithm in selected fields. 
\par The paper is structured as follows: the observations and data reduction are presented in Section~\ref{Observations}, the massive star population of NGC 4535 along with the blue to red supergiant ratio and the star formation history of the galaxy are described in Section~\ref{analysis}. The selection of variable candidates and the classification of the new massive variable candidates is given in Section~\ref{selection}. In Section~\ref{summary} we summarize and discuss our results.

\section{Observations and Reduction}
\label{Observations}
\subsection{Observations}
We used archival observations of NGC 4535 taken with the HST WFPC2 as described in \citet{Macri99} as part of the HST Key Project \citep{Freedman2001}. We used 13 epochs of F555W (equivalent to Johnson V) and 9 epochs of F814W (equivalent to Kron-Cousins I). The F555W data consisted in total of eight 3x1200 $\sec$ exposures and five 4x1300 $\sec$ exposures, while the F814W of nine 3x1300 $\sec$ exposures. Data were collected at four dithering positions, with one quarter of the observations (four pairs of F814W images and two pairs of F555W images) taken at each position. We retrieved the pre-reduced images through the Mikulvsky Archive for Space Telescopes \url{http://archive.stsci.edu/hst/search.php} and performed Point Spread Function (PSF) photometry using the latest WFPC2 module of \texttt{DOLPHOT} (updated in 2016), which is a modified version of \texttt{HSTphot} \citep{Dolphin2000}.

\subsection{Reduction} 
 We proceeded with the photometric reduction as described in the manual of the WFPC2 module of DOLPHOT and in the studies of \citet[]{Lianou2010,Lianou2013}. Initially, we aligned all images, masked bad pixels and columns, calculated the value of the sky background, rejected cosmic rays and combined all exposures of the same epoch. Distortion and CTE corrections were taken into account based on the equations of \citet{Holtzman95}. DOLPHOT \texttt{wfpc2mask} uses the data quality image (*.c1m.fits in the STScI archive filename convention) to mask bad pixels and other image defects in the sky image (*.c0m.fits). A bad or saturated pixel is not used in the sky determination, photometry, or aperture corrections. In order to achieve accurate results we conducted photometry separately on each chip of WFPC2: PC, WF2, WF3 and WF4. Solutions were also derived in adjacent pixels to determine if the true center of the star is in a different pixel. For each chip, a two-pass search for stars was made, followed by an iterative solution. Each search located peaks in the image and attempted photometry on each one. When two peaks corresponded to one star, a solution was provided for both stars and the sources were added to the star list. This process serves for de-blending. The first pass was made on the image minus the sky; in the second pass the stars located in the first pass were subtracted. The iterative solution was performed up to 25 times, depending on the number of iterations needed for the solutions to converge. For each iteration, stars whose neighbors’ photometry changed with respect to the previous iteration were remeasured. Each star was measured with the sky and all other stars subtracted, meaning that in principle, the frame only contained the star being measured and stars below the detection threshold.
 
 \par
\begin{figure*}
   \centering
   \includegraphics[width=10cm]{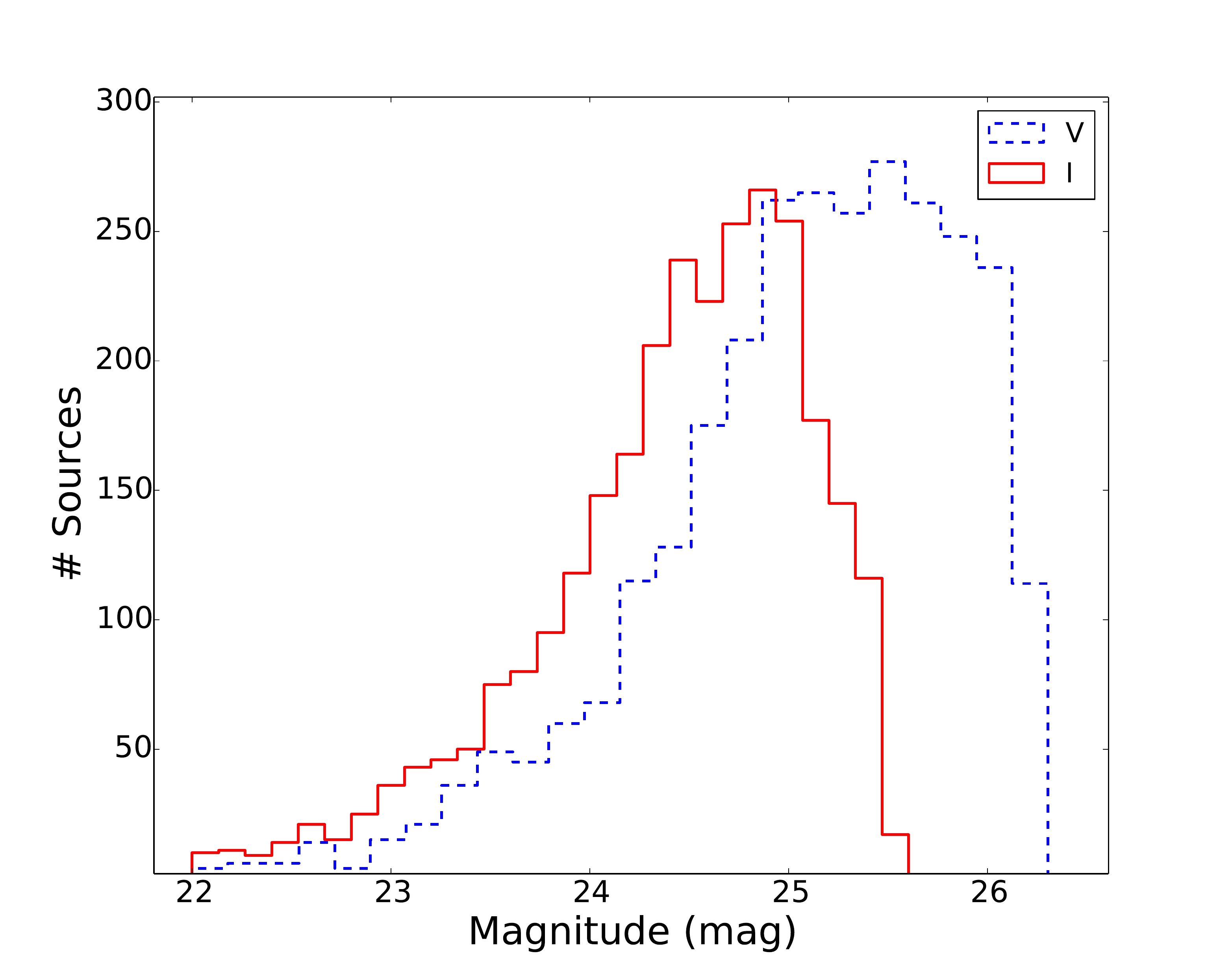}
   \caption{Histogram of the number of sources versus the mean magnitude
for the I filter (red solid line) and the V filter (blue dashed line). The faint limits in V and I magnitude indicate the magnitude up to which our sample is 50 $\%$ complete.}
\label{histsources}
\end{figure*}

 As each of the .c0m frames consisted of four extensions (one for each chip) the individual exposure images were separated with the routine \texttt{splitgroups}. We used the deepest F555W image as a reference frame of each chip. The main \texttt{dolphot} routine ran photometry for all images simultaneously. The program computed both the flight system magnitude for each filter and the standard system magnitudes (Johnson V and Kron-Cousins I) based on the equations of \citet{Holtzman95}. As the raw output of \texttt{DOLPHOT} included all sources of light in one frame, marginal/bad detections were often evident. We used photometric quality cuts to exclude bad detections. The cuts used were: Signal to noise ratio, S/N $\geq$ 5, $-$0.3$\leq$ sharpness $\leq$0.3, $\chi$$^{2}$ $\leq$ 2.5 and Object type=1. The reduced $\chi$$^{2}$, hereafter $\chi$$^{2}$, provides an indication of the best PSF fit with values of $\chi$$^{2}$ < 1.5$-$2.5 being reasonable for isolated stars and de-blended objects. According to the \texttt{DOLPHOT} WFPC2 module manual the $\chi$$^{2}$ parameter is usually set to less than 1.5. The sharpness is zero for a perfectly-fit star, positive for a star that is too sharp (perhaps a cosmic ray), and negative for a star that is too broad (perhaps a blend, cluster, or galaxy). In an uncrowded field, point sources should have sharpness values between $-$0.3 or 0.3. The object types are graded from 1 to 5 with 1 corresponding to round isolated sources and 5 to extended sources. Since we are interested in point sources, our final catalog includes sources of object type 1. The output from \texttt{DOLPHOT} automatically includes the transformation from flight-system F555W, and F814W to the corresponding Johnson-Cousins \citep{Bessell90} magnitudes in V, I following the conversion provided by \citet{Sirianni2005}. A histogram of the magnitudes of the detected sources in both observed filters is shown in Figure~\ref{histsources}.
 
Table~\ref{photcuts} shows the number of the 24353 sources included in each chip after the photometry quality cuts. We created a catalog with the coordinates, mean magnitudes and errors in V and I filters. The first fifty sources of this catalog are shown in Table~\ref{starcat}. A full version is available in a machine-readable form at the CDS.

\par
\begin{figure*}
   \centering
    \includegraphics[scale=0.30]{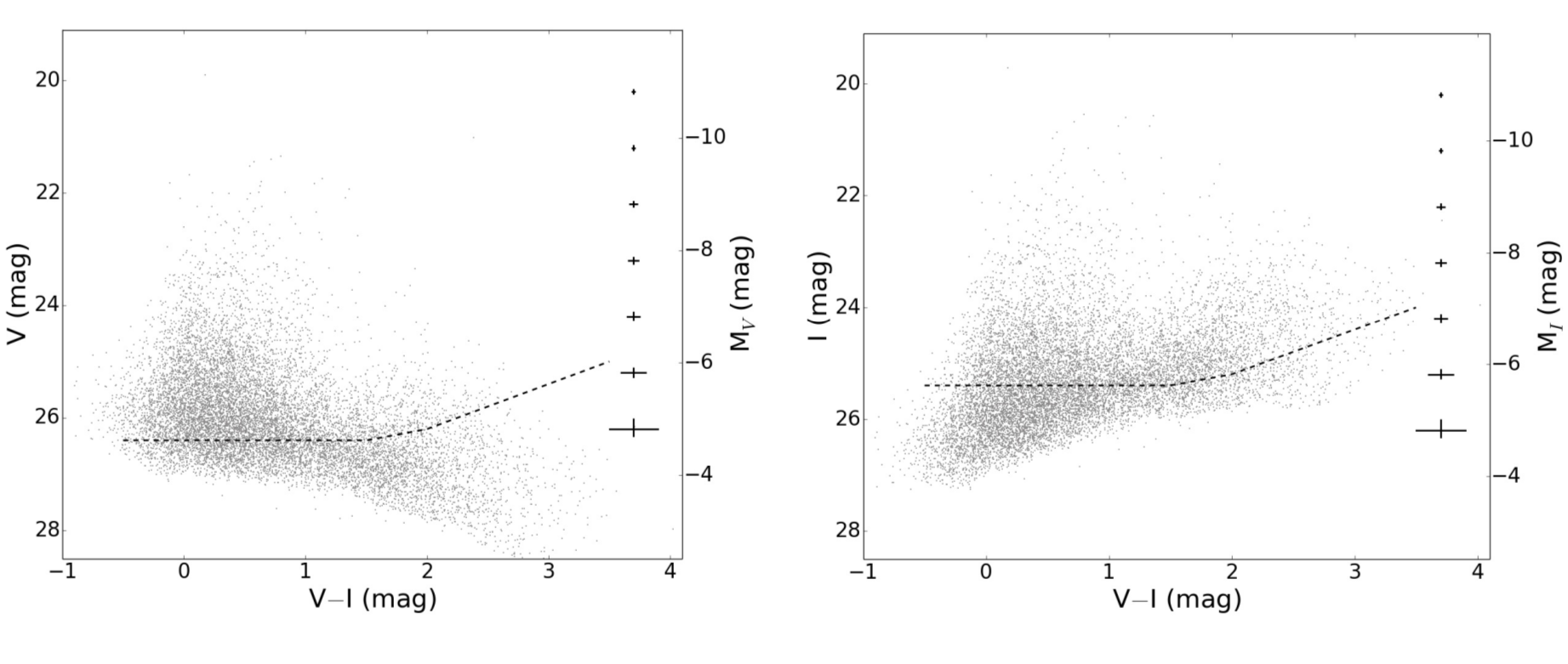}
\caption{CMDs for the stellar sources in NGC 4535 for both V $\&$ I filters. The error bars correspond to the photometric errors as derived from artificial star tests. The gray dotted
line indicates the 50$\%$ completeness level as a function of color and magnitude.}
\label{artstars}
\end{figure*}

\begin{figure*}
   \centering
    \includegraphics[scale=0.35]{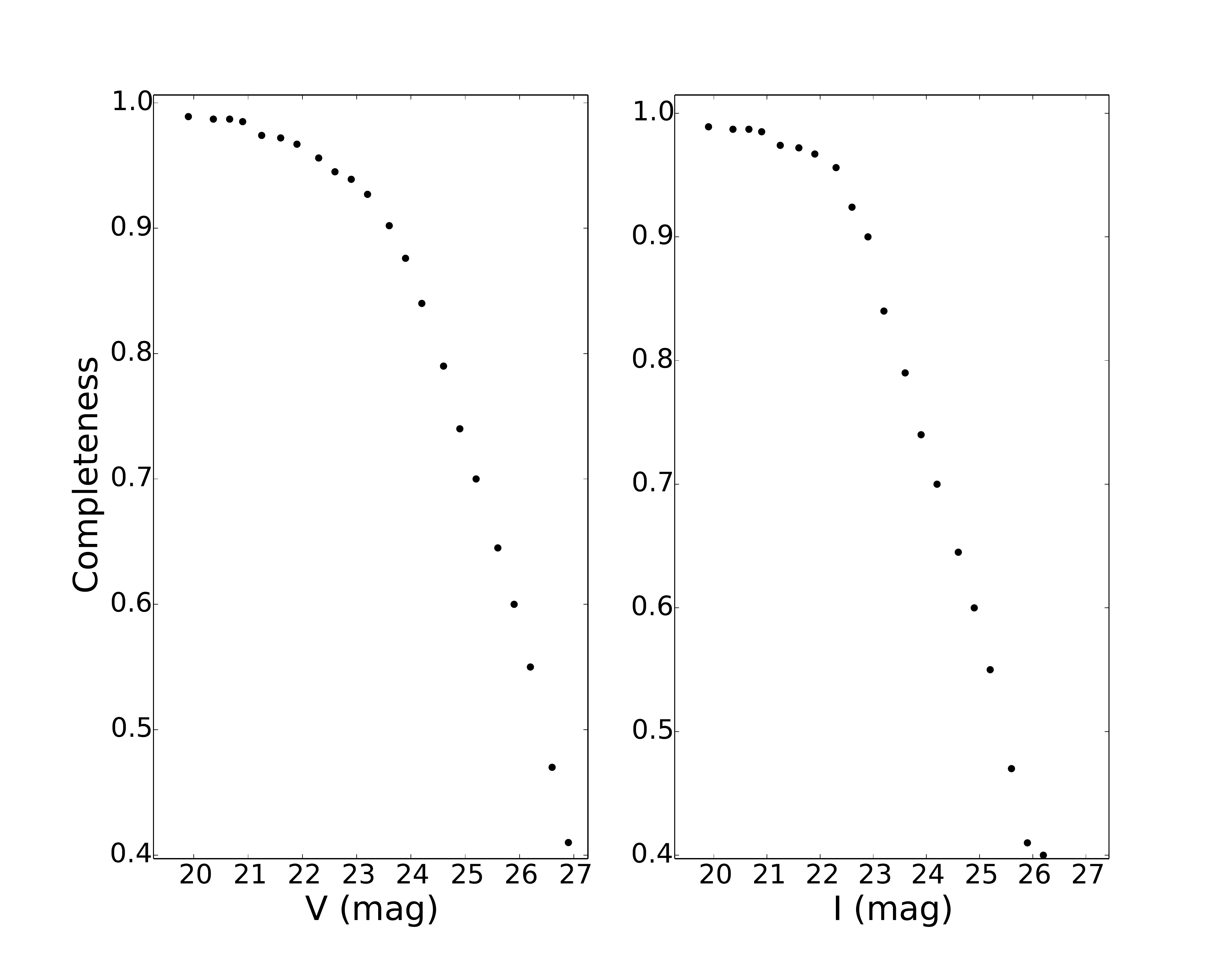}
\caption{Completeness factors as a function of the V and I magnitude.}
\label{completeness}
\end{figure*}

 We conducted artificial star tests to estimate the photometric errors and completeness using the methodology described in \citet{Lianou2013}. We created a catalog of artificial stars that were measured at the same time as the field stars. We ran the tests injecting 5000 stars per chip ranging in magnitude from 19 to 27 mag in both filters. The mean error was estimated, using bins of 0.2 mag, at 0.04 $\pm $ 0.07 mag and 0.03 $\pm$ 0.05 mag in V and I, respectively. The number of stars injected in \texttt{DOLPHOT} for the artificial star tests did not cause overcrowding as the program measured each star one-by-one, without creating a new overcrowded image. The completeness of our catalog was determined by applying to the artificial star photometry the same quality criteria as described above. Figure~\ref{artstars} shows the color-magnitude diagrams (CMDs) for both V and I filters with the errors derived by the artificial star tests. The gray dashed line indicates the completeness factor for both filters from the artificial sources versus the magnitude of the stars that the program recovered (output magnitude) for each filter, while Figure~\ref{completeness} shows the 50$\%$ completeness factor as a function of magnitude and color. The 50$\%$ completeness factor occurs at $\approx$ 26.2 mag and at $\approx$ 25.0 mag for the V and I filters respectively.

\subsection{Astrometry}
Astrometry was performed by using the Hubble Source Catalog (HSC) \citep{Whitmore16} and applying the astrometric correction suggested by \citet{Anderson2000}. After cosmic-ray flagging and image reduction we used \texttt{DOLPHOT} to find and measure positions of stars on each CCD. Because WFPC2 consists of four separate detectors, it was necessary to define a common coordinate system for all four CCDs. The reference-frame produced by multidrizzle and used for photometry, was also used for astrometry. The \texttt{DOLPHOT} package \texttt{wf2fitdistort} corrects the geometric distortion, rotation and magnification that is caused by the camera's focal features and sets a global chip coordinate system based on the central pixel of the reference frame. Through this routine we inserted the X and Y positions of stars on each image and on the reference image. We used stars from the HSC at various positions across each chip to calculate the transformations from the X,Y positions to the RA, Dec. Relative positions are good to $\sim$ 0.05 arcsec, while the accuracy of the absolute positions is limited by the HST coordinates, which are limited by the accuracy of the HST guide star catalog to $\sim$ 0.1$\arcsec$.

\section{Massive star population}
\label{analysis}
\subsection{Identification of the massive star candidates}
\begin{figure*}[h!tb]
\centering
\begin{tabular}{l c  }
\includegraphics[scale=0.6]{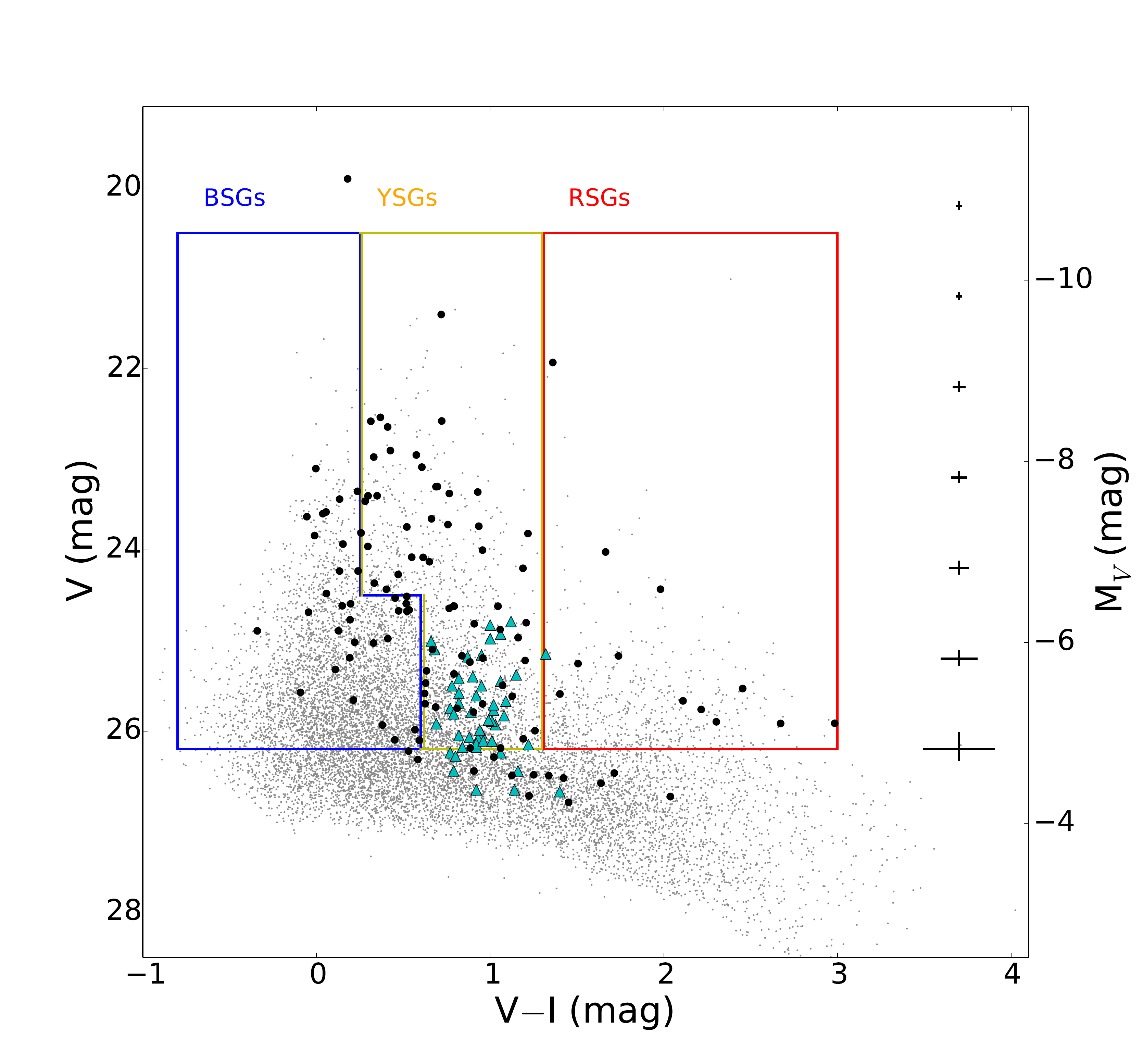} 

\end{tabular}
\caption{V$-$I vs. V color-magnitude diagram of NGC 4535. Both absolute and mean magnitudes of the stars are plotted on the y-axis, based on the distance modulus 31.02 mag. The new variable candidates are shown as black dots, while the known Cepheids as blue triangles.}
\label{cmdbsgs}

\end{figure*}

Figure~\ref{cmdbsgs} shows the V$-$I versus M$_{V}$ CMD, for the stars that passed the quality tests mentioned in Section~\ref{Observations}, based on the distance modulus of NGC 4535 \citep[31.02 mag;][]{Macri99}. The error-bars shown were derived from the artificial star tests. The total foreground interstellar reddening of the galaxy is E(V$-$I)= 0.13 $\pm$ 0.04 mag \citep{Allen2004} and was used to correct the observed colors. We limited our investigation of the massive star population in NGC 4535 to the 4973 sources with magnitudes between 19.5 $\leq$ V $\leq$ 26.2 mag. The faint limit of this magnitude range corresponds to the magnitude in the V band where the sample is 50 $\%$ complete, M$_{V}$ $\approx$ $-$4.9 mag. 

We used the color and magnitude criteria by \citet{GrammerM101} to separate the regions of BSGs and hot OB dwarfs, hereafter, BSGs, YSGs and RSGs. The three populations range from:

\begin{center}
\begin{tabular}{ c c c c c }
 BSGs: & (V$-$I) $\leq$ 0.25 mag & 20.5 $\leq$ V $\leq$ 24.5 mag &&\\ 
       & (V$-$I) $\leq$ 0.60 mag & 24.5 $<$ V $\leq$ 26.2 mag &&\\  
 YSGs: & 0.25 $<$ (V$-$I) $\leq$ 1.3 mag & 20.5 $\leq$ V $\leq$ 24.5 mag &&\\    
       & 0.6 $<$ (V$-$I) $\leq$ 1.3 mag & 24.5 $<$ V $\leq$ 26.2 mag &&\\    
 RSGs: & (V$-$I) $>$ 1.3 mag & 20.5 $\leq$ V $\leq$ 26.2 mag &&\\    

\end{tabular}
\end{center}

\par In total we identified 3762 candidate BSGs, 841 candidate YSGs and 370 candidate RSGs.

\par The degree of foreground contamination for each group of luminous massive star candidates was estimated through the Besan\c{c}on Galactic population synthesis model \citep{Besancon2003} in the direction of NGC 4535. The Besan\c{c}on model predicts a 18$\%$ contamination for the total catalog of bright stars, which corresponds to $\approx$ 900 stars. The majority of foreground stars lie in the YSG region, while in the BSG region no foreground contamination is predicted. Due to photometric uncertainties and reddening, sources classified in the YSG region could also be part of the neighboring RSG population. However, the high galactic latitude (+70.6414$^{o}$), the distance of NGC 4535 (16 Mpc) and the solid angle of WFPC2 (0.0017 degrees $^{2}$), reduce the possibility of a larger number than 18 $\%$ of our catalog to be affected by foreground contamination. We attempted to check the proper motions of our catalog of bright stars by cross-matching our sources with the UCAC5 catalog \citep{UCAC52017} however, no matches were found as the limit in magnitude for UCAC5 is $\approx$ 16 mag. Consequently, $\sim$100 $\%$ of the blue stars are BSG members of NGC 4535, $\sim$ 85 $\%$ of the yellow stars are in the YSG population of the field and $\sim$ 95 $\%$ of the red stars are RSGs.

\subsection{The blue to red supergiant ratio}

\par
\begin{figure*}[h]
\centering
\includegraphics[scale=0.7]{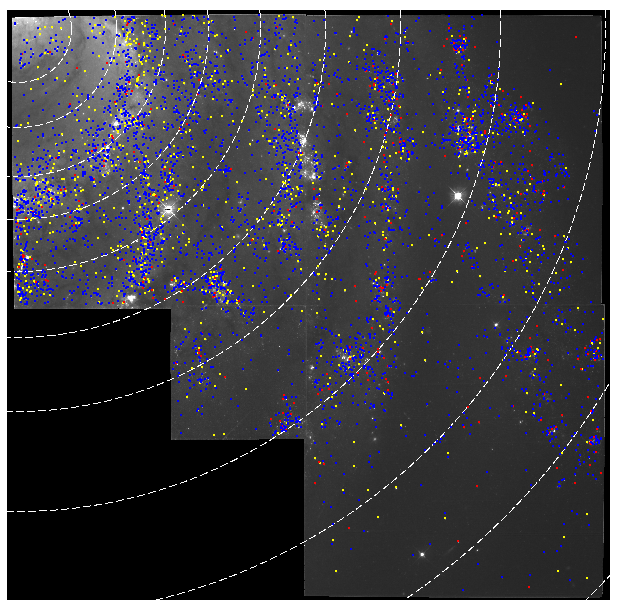} 

\caption{Spatial distribution of the candidate blue and red supergiants in NGC 4535. The candidate blue and red supergiants are shown as blue and red circles. The dashed white annuli indicate the 0.35$^{\prime}$ annuli used for the estimation of the B/R ratio.} 
\label{b2red}
\end{figure*}

\begin{figure*}[h!tb]
\centering
\begin{tabular}{l c c c c }
\includegraphics[trim=0cm 0cm 0.cm 0cm, width=0.45\textwidth]{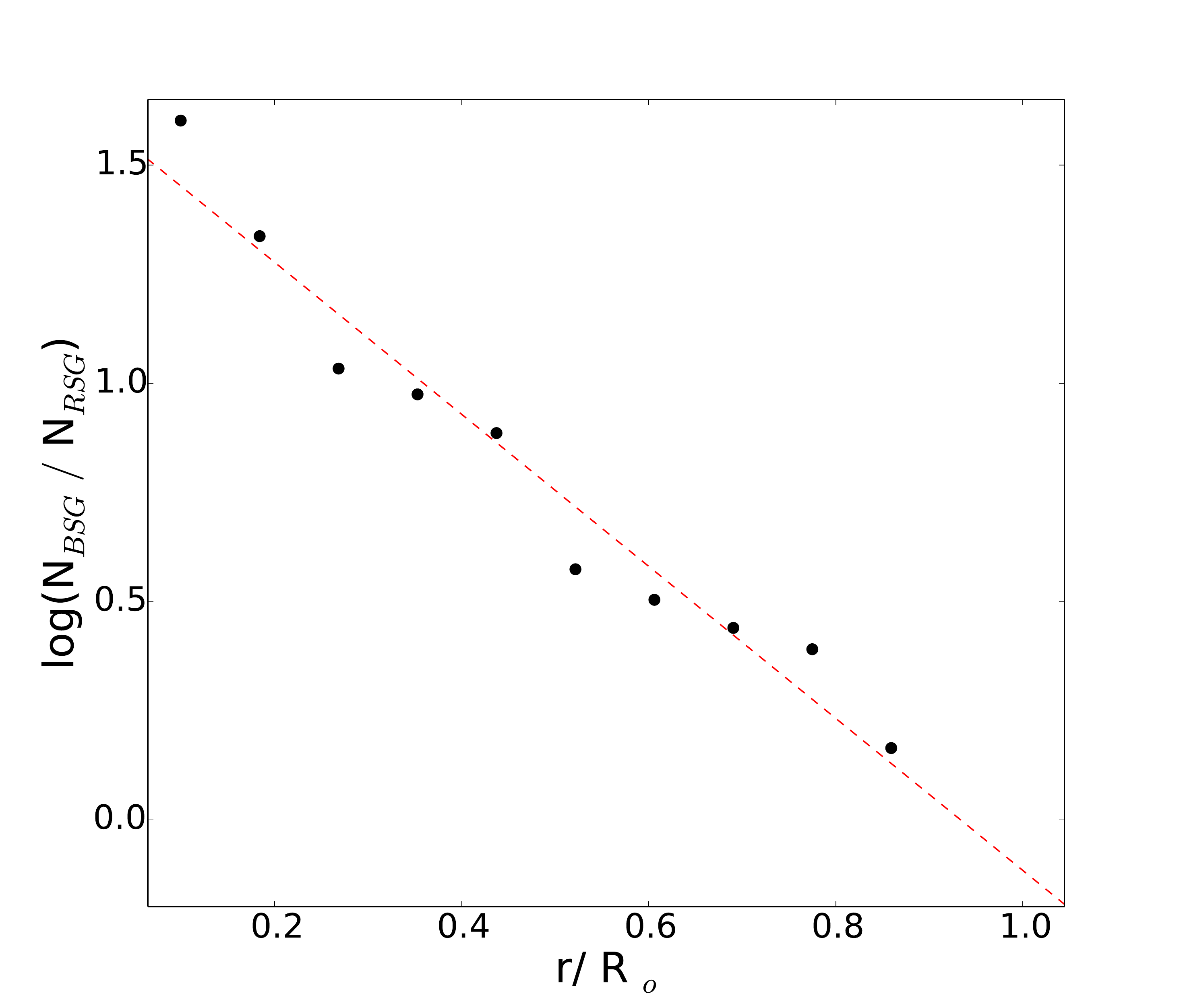} &
\includegraphics[trim=0cm 0cm 0.cm 0cm, width=0.45\textwidth]{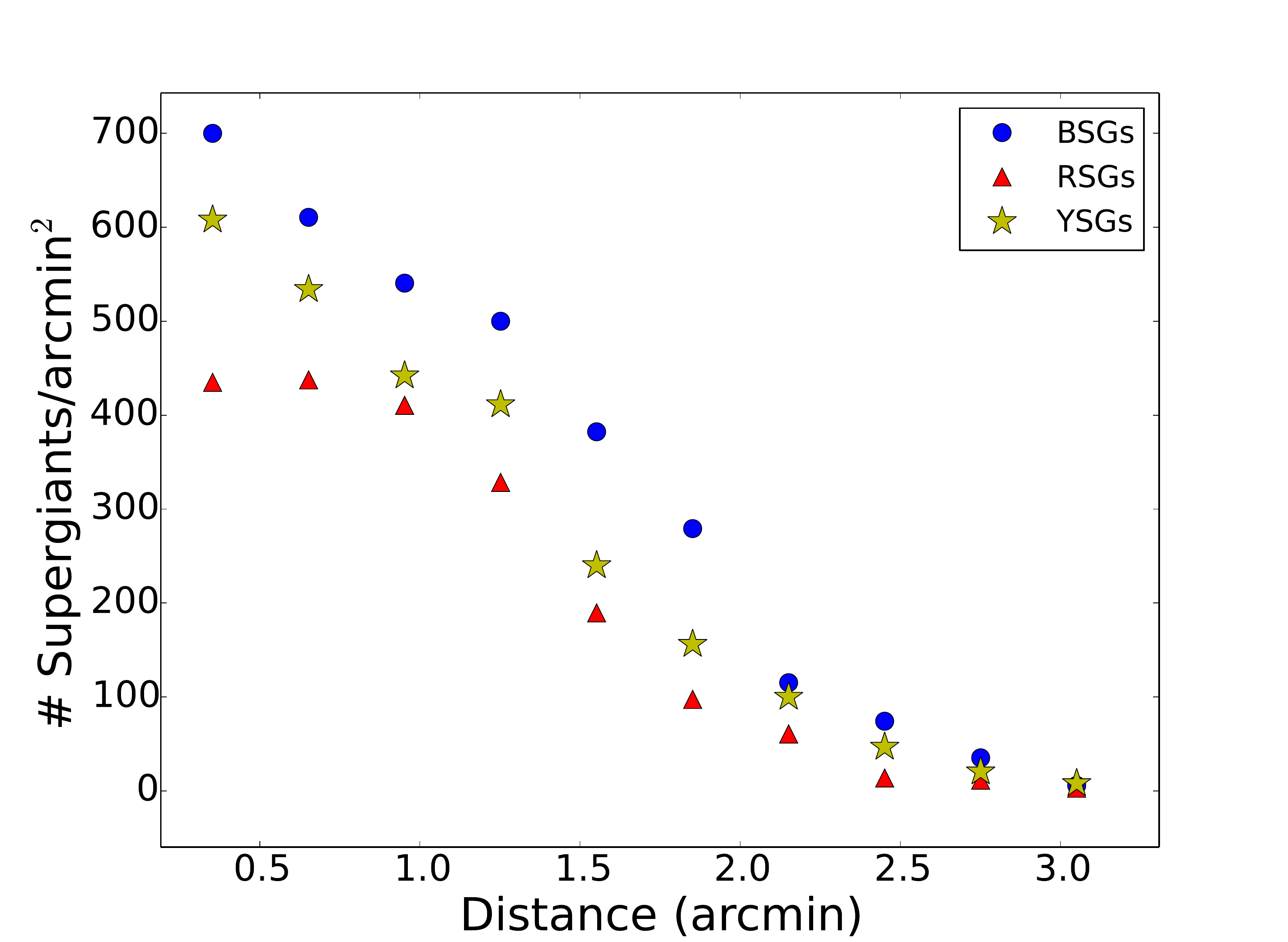}& 
\\
\end{tabular}
\caption{Left panel: The blue to red supergiant ratio for NGC 4535. The dashed red line is a linear fit indicating the monotonic radial decline of the blue to red supergiants. Right panel: The number of candidate BSGs, RSGs, YSGs/arcmin$^{2}$ vs. the distance from the center. The candidate BSGs are shown as blue dots the RSGs as red triangles and the YSGs as yellow stars.}
\label{b2redratio}
\end{figure*}

 We distributed the luminous stars in BSG, YSG, and RSG groups, adopting an inclination angle of 43$^{o}$ \citep{Egusa}, to calculate the de-projected distances for the RSGs and BSGs of the field, from the center of NGC 4535. We divided the field of view into 10 annuli of $~0.35$$^{\prime}$ width, normalized the number of BSGs and RSGs to the same area taking into account the foreground contamination towards the edges of our field and propagated the related errors to calculate the B/R ratio. The spatial distribution of the BSGs and RSGs within the 10 annuli is shown in Figure~\ref{b2red}. The left panel in Figure~\ref{b2redratio} shows the B/R ratio versus galactocentric radius. Each point in the plot corresponds to an annulus in Figure~\ref{b2red}. A decline of the B/R ratio with increasing radius, is reported by \citet[]{HumphreysM33, Eggenberger2002} for M33.
The studies of \citet{Maeder81} and \citet{Massey98} on RSGs and WR stars state that at higher metallicities RSGs have a brief He burning phase accompanied by high mass-loss rates, which is shown to drive the star back to the blue part of the HR diagram. Hence, the B/R ratio is highly dependent on the radial changes of the metallicity of the field. The determination of the correlation between metallicity, distance and the B/R ratio also requires knowledge of the radial star formation history of the galaxy. For NGC 4535 no accurate value of metallicity or any information on the element abundances is available in the literature, which does not allow for further interpretation of the B/R ratio trend of the field. There is evidence of HII regions \citep{Ulvestadetal} that indicates massive star formation in NGC 4535, with a star formation rate of 3.29 M$_{\odot}$/yr \citep{Boselli2015}, determined by correcting the H$\alpha$ luminosity using the 24$\mu$m emission. 

\par The decrease of the BSG population with increasing radial distance is shown in the right panel in Figure~\ref{b2redratio}. The plot shows a radial decline for the population of all three groups of massive stars, which is related both to the decreasing metallicity gradient and the number of stars at the outer regions of a field and to the limits used for separating the three populations. For increasing radial distance the number of BSGs is decreasing while the RSGs become more prevalent.

We compare our results with the work of \citet{GrammerM101}, who have examined photometrically the massive star population in M101. In Table~\ref{comparison} we present a comparison of our results to the massive star population in M101. In the study of \citet{GrammerM101} the number of bright stars is larger as the observations covered the whole galaxy while the observations used in our study cover a small area of NGC 4535.  

As shown in Table~\ref{comparison} $\sim$70$\%$ of the bright stars in M101 are BSGs, $\sim$9$\%$ are YSGs and $\sim$6$\%$ are RSGs. The remaining 15$\%$ of the sources are foreground stars. In NGC 4535, $\sim$76$\%$ lie in the BSG region, $\sim$16 $\%$ in the YSG $\sim$8$\%$ in the RSG region. Although the number of bright stars is larger in M101 as expected, the ratio of the BSGs to RSGs and the ratio of YSGs to RSGs is comparable in both galaxies. This could indicate that both NGC 4535 and M101 have similar metallicities and recent star formation histories as the B/R ratio decreases monotonically with radius in both fields. With decreasing radius the number of BSGs decreases while the number of RSGs is increasing.
We are hesitant to interpret the observed radial decline in the B/R ratio for NGC 4535, due to the lack of the oxygen abundance gradient in order to convert the distances to metallicity as done by \citet{GrammerM101}.

\subsection{Luminosity function and star formation rate}
\par
\begin{figure*}[h!tb]
\centering
\begin{tabular}{l c c c c }
\includegraphics[trim=0cm 0cm 0.cm 0cm, width=0.5\textwidth]{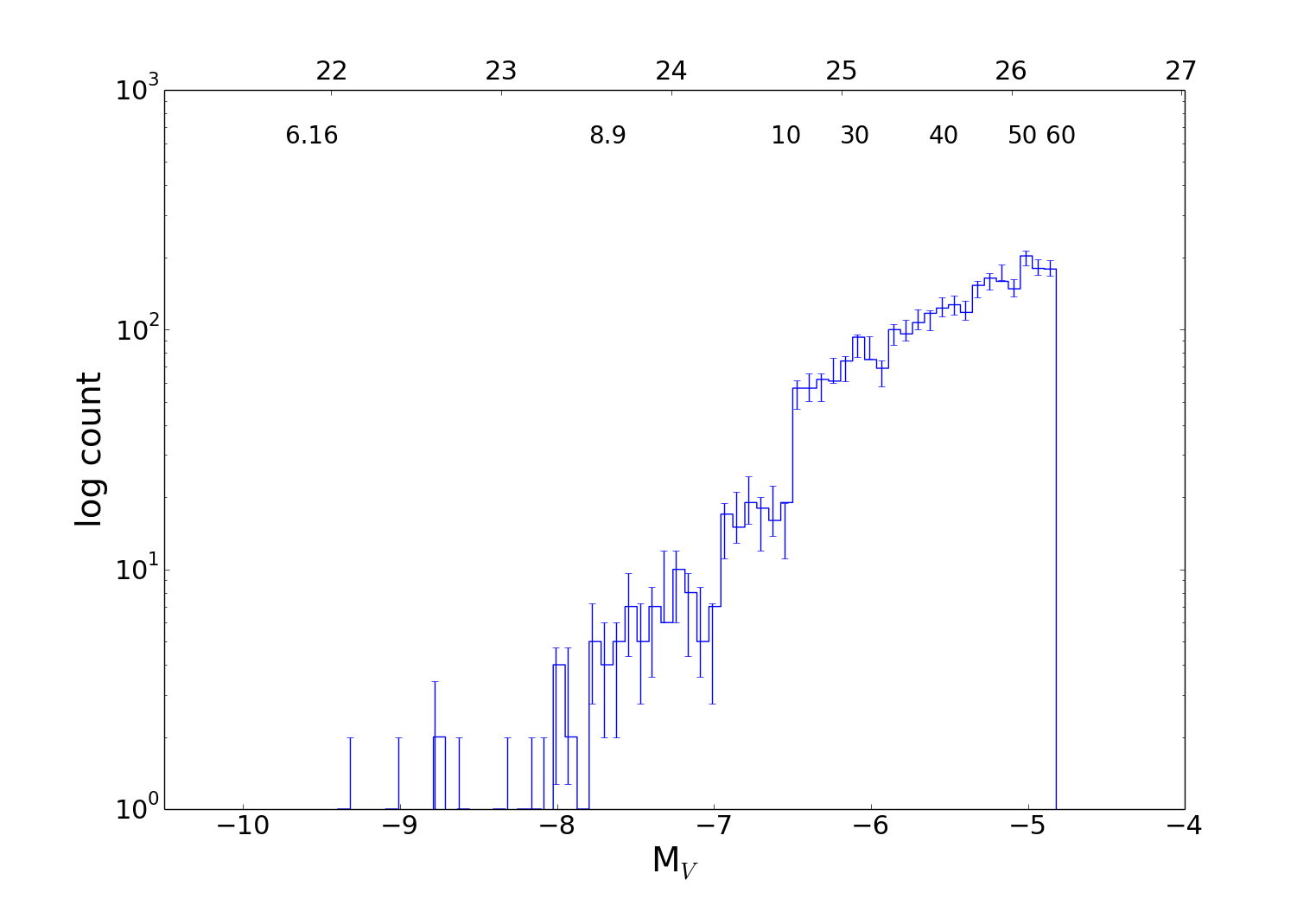} &
\includegraphics[trim=0cm 0cm 0.cm 0cm, width=0.5\textwidth]{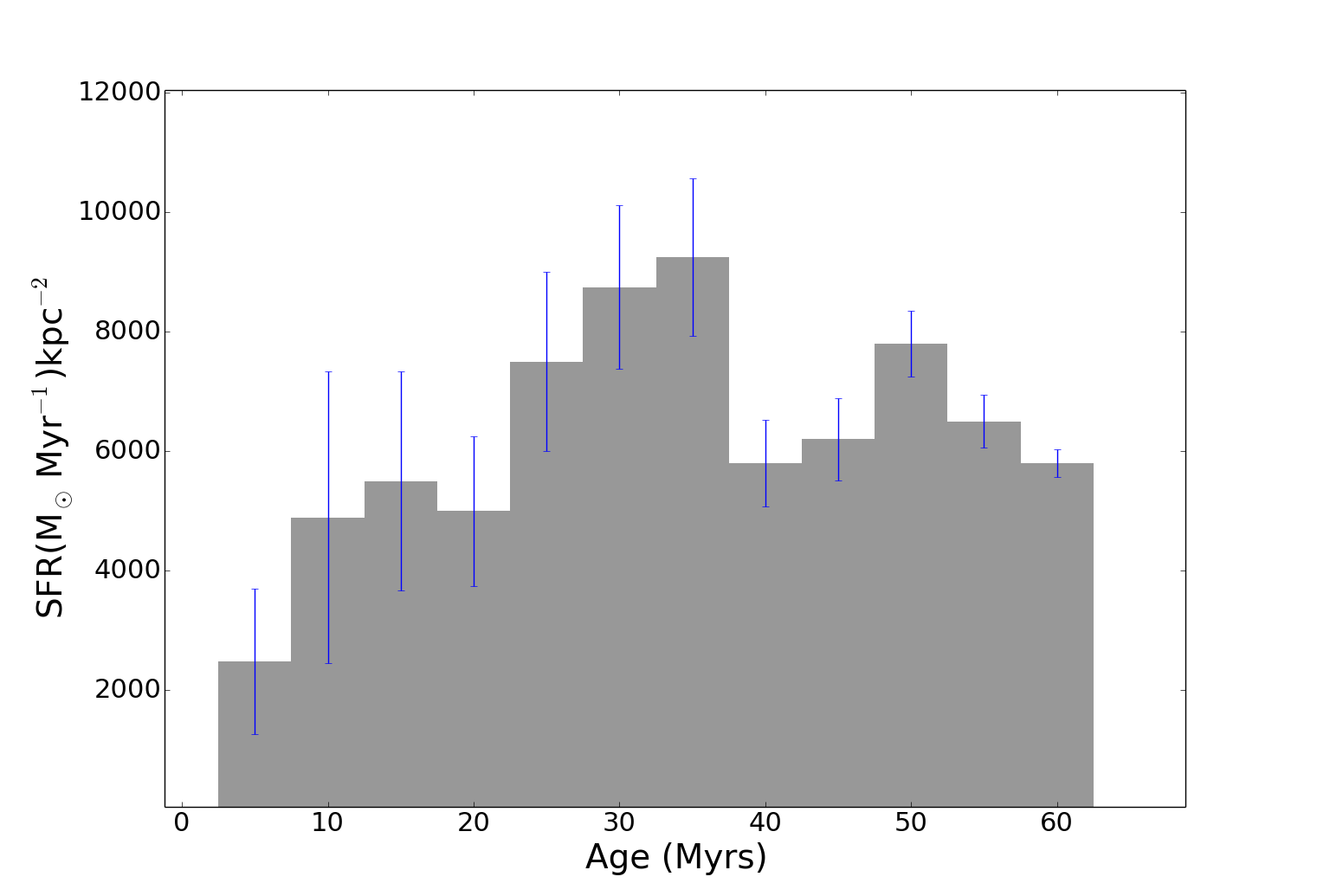}& 
\\
\end{tabular}
\caption{Left panel: The V luminosity function of the blue HeB stars. The bin width is 0.15 mag. The errors shown are derived by the artificial star tests. Plotted above the histogram is the age in Myrs for the blue HeB stars as given by the MESA models. Right Panel: The SFR in M$_{\odot}$ Myr$^{-1}$ kpc$^{-2}$ for NGC 4535 within our field over the last 60 Myr, based on the blue HeB stars. The blue HeB luminosity function has been normalized to account for the IMF and the changing lifetime in this phase with mass. We used the Salpeter IMF slope and the MESA evolutionary tracks for Z= 0.0014.}
\label{lumfunc}
\end{figure*}

 We used the Modules for Experiments in Stellar Astrophysics (MESA) Isochrones and Stellar Tracks \citep[MIST;]{MESA2016} at solar metallicity, to estimate the masses and temperatures, ages and evolutionary state of the massive stars. MESA models use solar-scaled abundances covering a wide range of ages, masses, phases, and metallicities computed within a single framework. The models were generated assuming rotation at 40$\%$ of the critical speed and taking into account mass loss M$_{max}$ 10$^{-3}$ M$_\odot$ yr$^{-1}$. We adopted the method described in \citet{Palmer1997} to derive the V luminosity function for the blue Helium-burning stars (blue HeB stars) of our sample based on the MESA isochrones. We divided our sample in magnitude bins of 0.15 mag from M$_{V}$= $-$4.9 up to $-$11 mag where our sample is 50 $\%$ complete. The MESA models provide the stellar evolutionary stage on each isochrone. The sample of blue HeB stars includes all sources more massive than 8 M$_{\odot}$, within the BSG region and at the Center Helium burning evolutionary stage. The ages of the sources range from 6-60 Myrs. Figure~\ref{lumfunc} (left panel) shows the luminosity function in the V band of the blue HeB stars. The bin width is 0.15 mag, and the data have been corrected for incompleteness with the errors given by the artificial star tests. The numbers plotted above the luminosity function are the ages of the isochrones in Myrs used based on the MESA models. In order to associate the counts from the luminosity function with the star formation rate of the field we used the equation given in \citet{Palmer1997}

\begin{equation}
C (M_{V}, V$-$I)=  \int_{logm_{1}}^{logm_{2}} \int_{t_{1}(log m)} ^{t2logm} \phi (logm,t)\times R(t) dt dlogm
\end{equation}

\noindent adjusted for the stars, where $\phi$ is the initial mass function (IMF) normalized to unity, m is mass, R is the star formation rate in units of M$_{\odot}$ yr$^{-1}$, and t is time. We used the Salpeter slope for the IMF and the normalized IMF is

\begin{equation}
\phi(logm)d logm=0.394 (\frac{m}{M_\odot})^{-1.35} dlogm
 \end{equation}

\noindent After estimating the star formation rate we divided with the area covered by WFPC2 to find the SFR/kpc$^{2}$. The SFR/kpc$^{2}$ is shown in Figure~\ref{lumfunc} (right panel). For the region within the field of view the SFR appears suppressed before 10 Myrs and it rises after $\approx$ 30 Myrs.     

\section{Variability}
\label{selection}
We examined the stellar population included in our field for variability. We considered only the stars with more than 5 measurements in both V and I filters. From the initial catalog of 24353 sources, 8832 met this criterion. These sources were checked for variability based on three different variability indexes: the median absolute deviation (MAD), the interquartile range (IQR) and the inverse von Neumann ratio (1/$\eta$) \citep{Sokolovsky16}. MAD is an outlier-resistant measure of light curve scatter \citep{Zhang2016}. However, it is equally insensitive to real variations that occur only occasionally. The IQR is another measure of scatter providing a different trade-off (compared to MAD) between the outlier resistance and sensitivity to changes in brightness that occur only occasionally. It determines the difference between the median values computed for the upper and lower halves of the data set and is considered a robust variability indicator \citep{Sokolovsky16}. Unlike MAD, it does not assume that the distribution of measured magnitudes is symmetric and is more sensitive to regular excursions in the same direction while still being relatively insensitive to individual outlier measurements. Although both MAD and IQR are similar scatter-based indexes, we chose to use both in order to recover as many variable candidates as possible. The von Neumann ratio $\eta$, is defined as the ratio of the mean square successive difference to the distribution variance. This index describes how smooth the light curve is. The disadvantage of 1/$\eta$ compared to the IQR and MAD is that it is not sensitive to fast variability on timescales shorter than the observing cadence as these kind of variations will not produce a smooth light curve. The value 1/$\eta$ is used so that higher values of the index correspond to a smoother light curve. According to \citet{Sokolovsky16}, the scatter-based variability indicators MAD and IQR as well as the correlation-based 1/$\eta$ perform well in selecting variable candidates from data sets affected by outliers.

\par
\begin{figure*}[h!tb]
\centering
\begin{tabular}{l c c c c}
\includegraphics[trim=0cm 0cm 0.cm 0cm, width=0.45\textwidth]{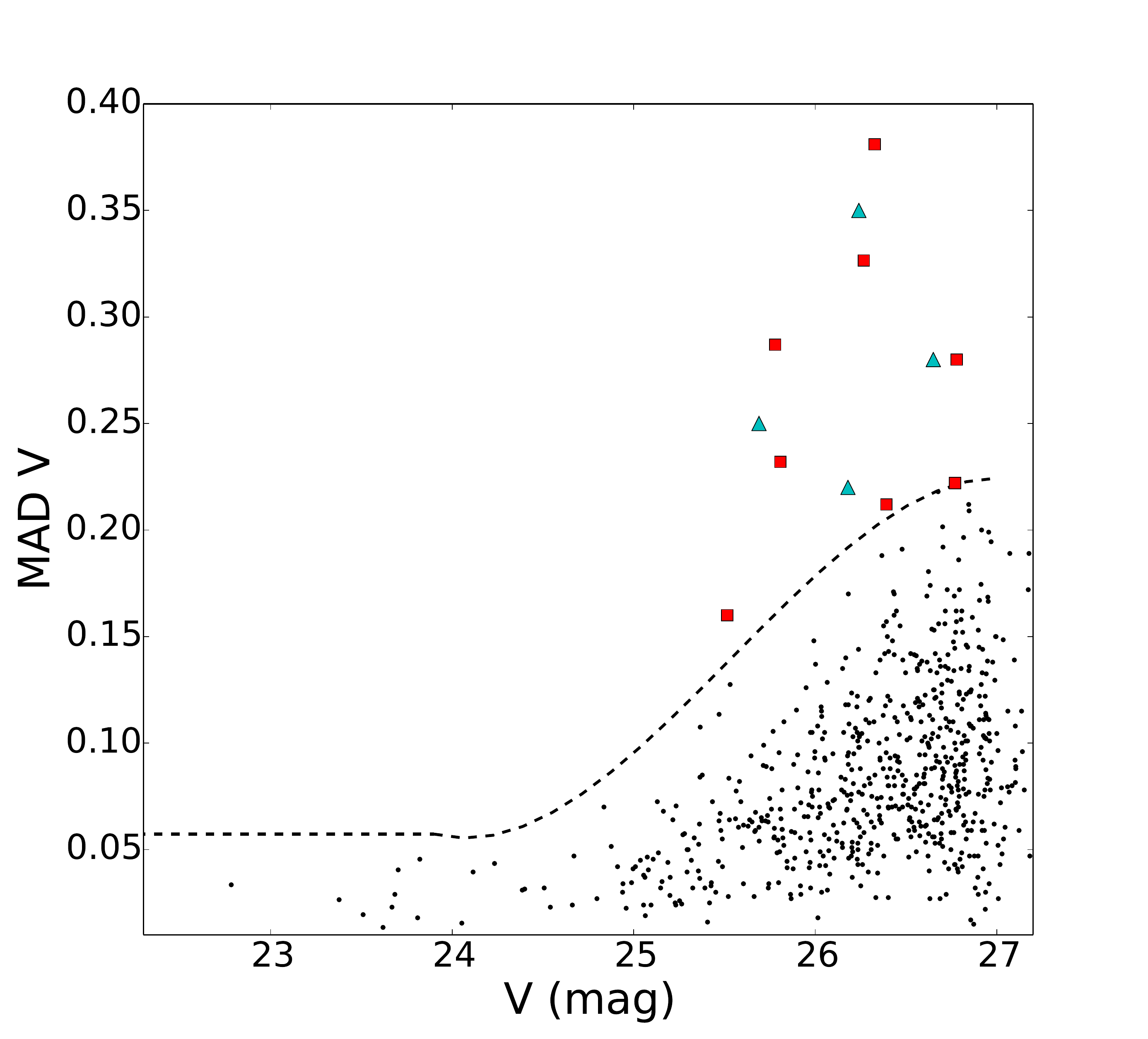} &
\includegraphics[trim=0cm 0cm 0.cm 0cm, width=0.45\textwidth]{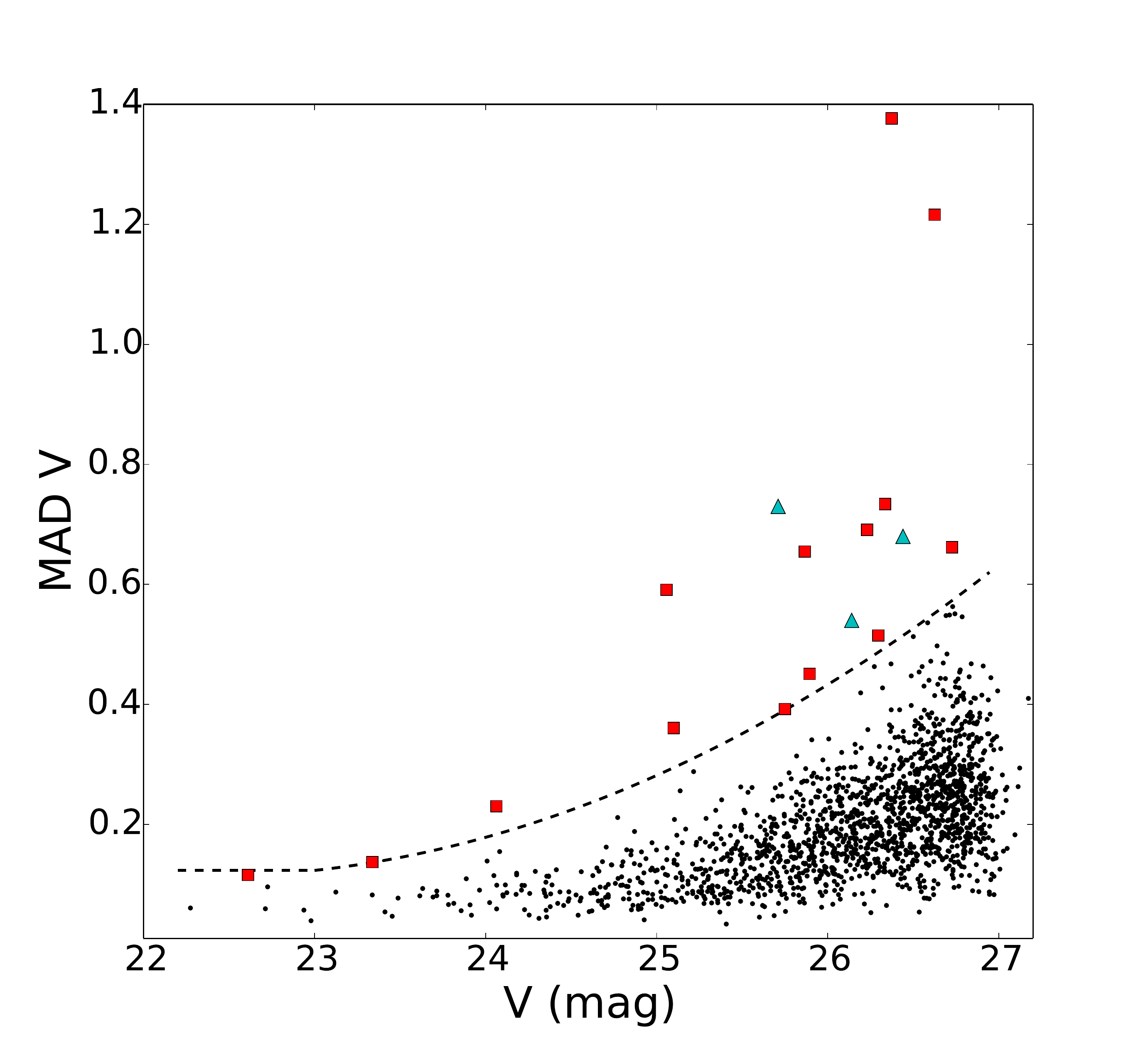}& \\
\includegraphics[trim=0cm 0cm 0.cm 0cm, width=0.45\textwidth]{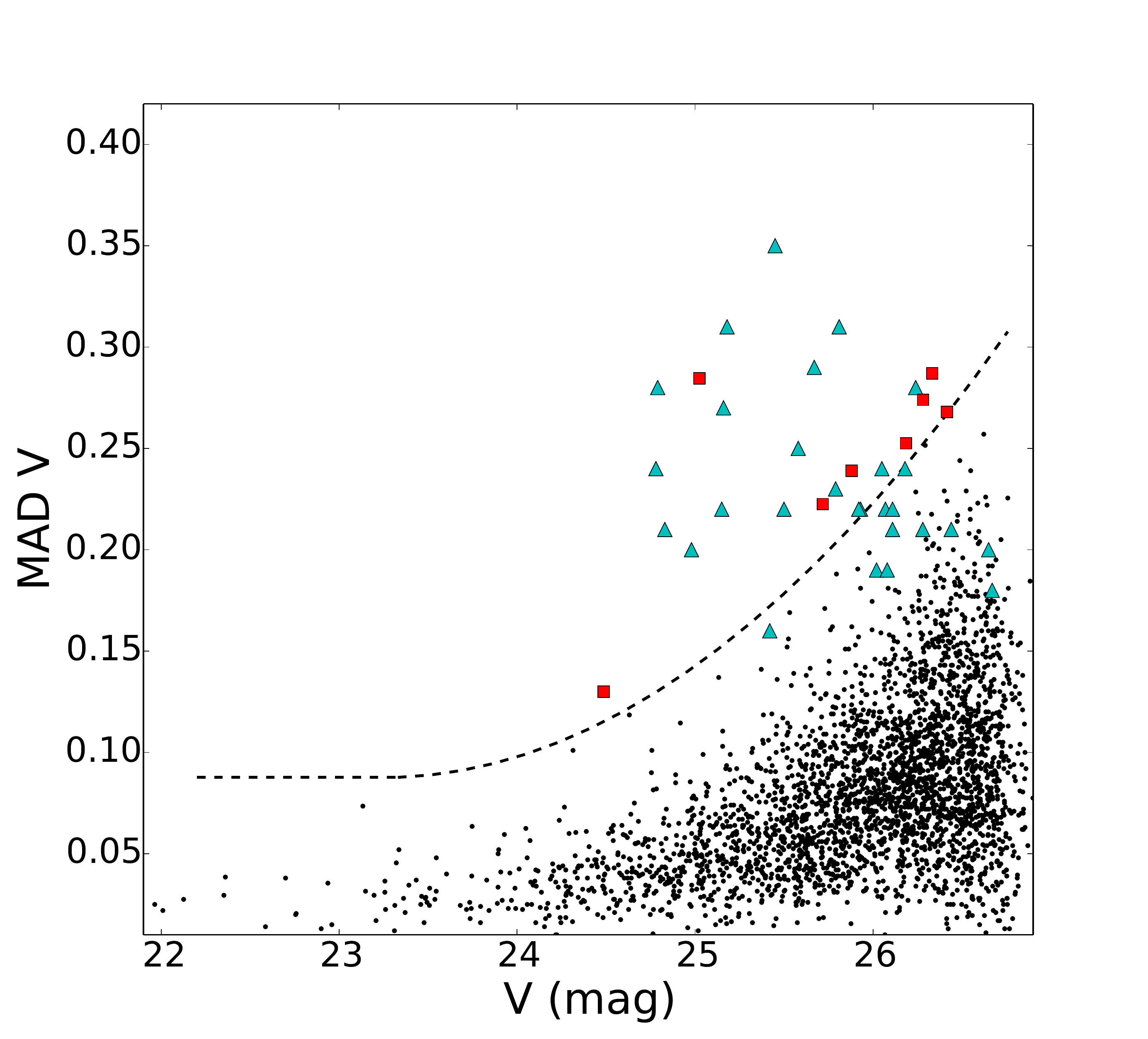} &
\includegraphics[trim=0cm 0cm 0.cm 0cm, width=0.45\textwidth]{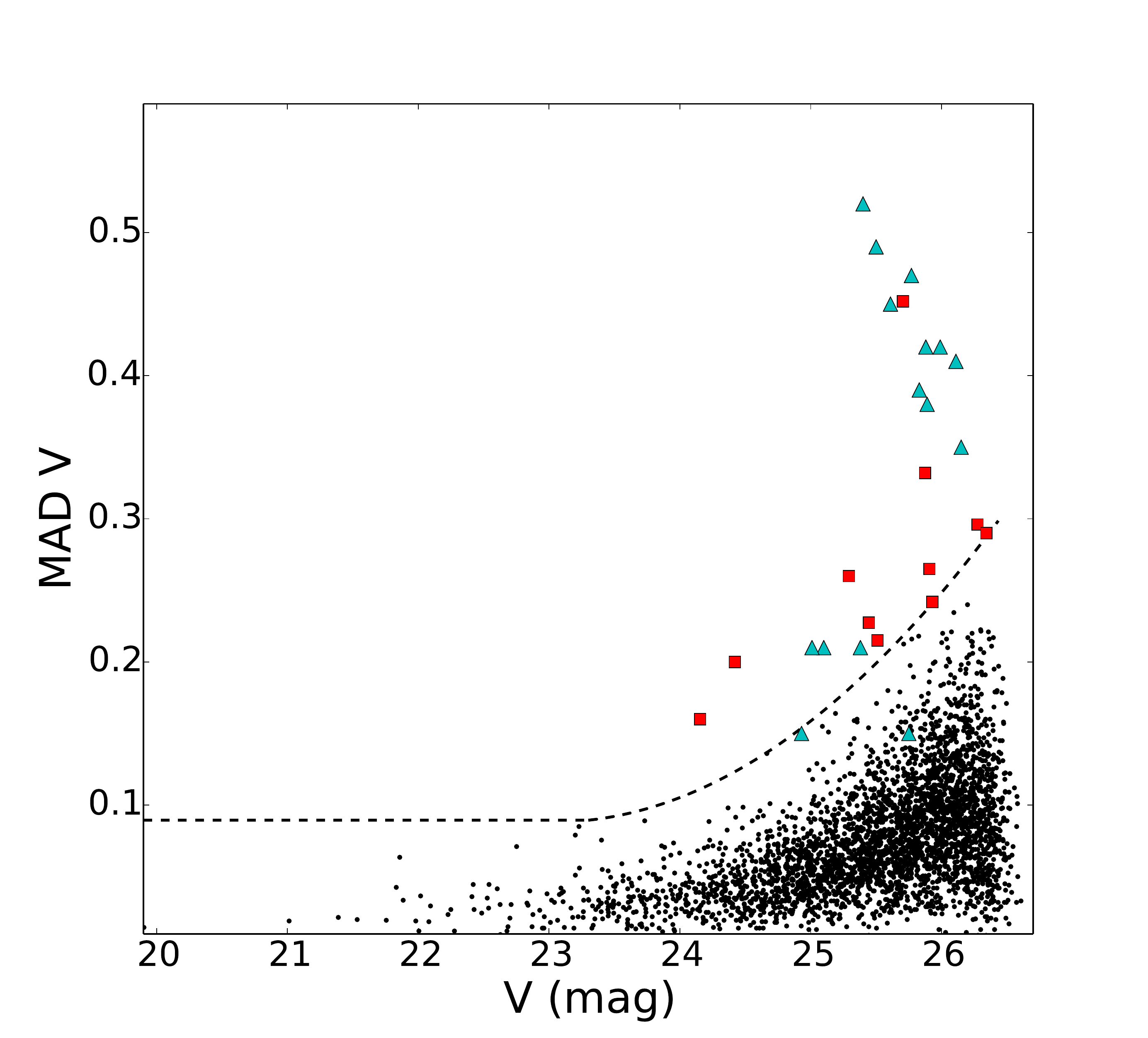}&\\

\\
\end{tabular}
\caption{Mean absolute deviation versus mean magnitude. The dashed line shows the stars over the 4$\sigma$ threshold. The variable candidates are shown in red squares and the known variables are shown in blue triangles. The four panels per column correspond to the chips PC, WF2, WF3 and WF4 respectively.}
\label{indexes}
\end{figure*}

 We calculated variability indexes separately for each chip, i.e. for 681 sources in the PC chip, 1685 stars in WF2, 3064 in WF3 and 3402 in WF4. For each chip we sorted the stars by increasing magnitude and we divided the sources into magnitude bins with at least 5$\%$ of the total number of sources per bin to maintain the statistical significance. For each bin, after calculating the median value of the index and the standard deviation ($\sigma$), we set a 4$\sigma$ threshold and considered as variable candidates all sources above that threshold in either filter.
Figure~\ref{indexes} shows the index versus magnitude plots for the MAD in the V filter for all 4 chips. The dashed black curve indicates the 4$\sigma$ threshold. Candidate variable sources are shown as red squares while the known Cepheids are shown as blue triangles. 

\par
\begin{figure*}[h!tb]
\centering
\begin{tabular}{l c c c c}
\includegraphics[trim=0cm 0cm 0.cm 0cm, width=0.50\textwidth]{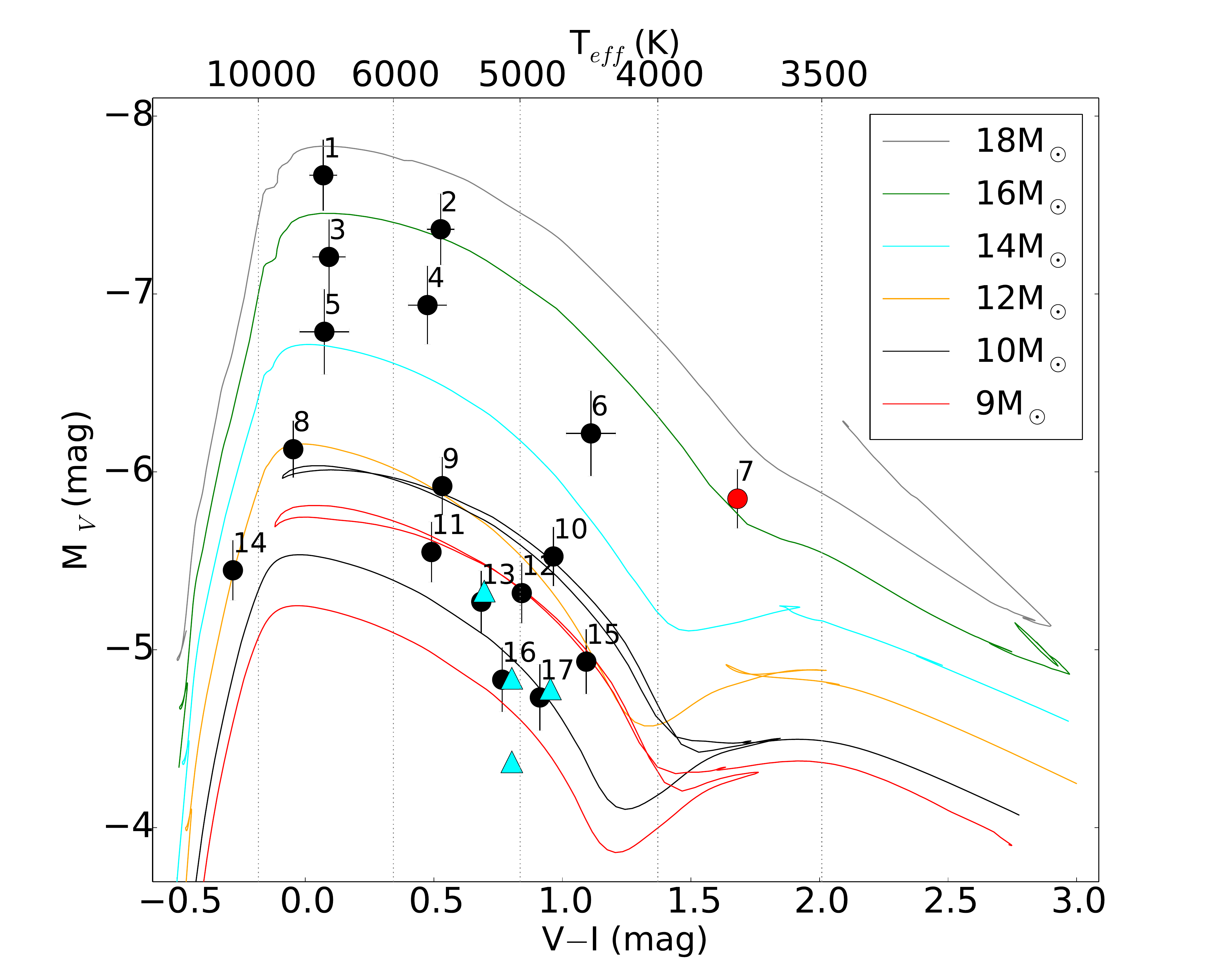} &
\includegraphics[trim=0cm 0cm 0.cm 0cm, width=0.50\textwidth]{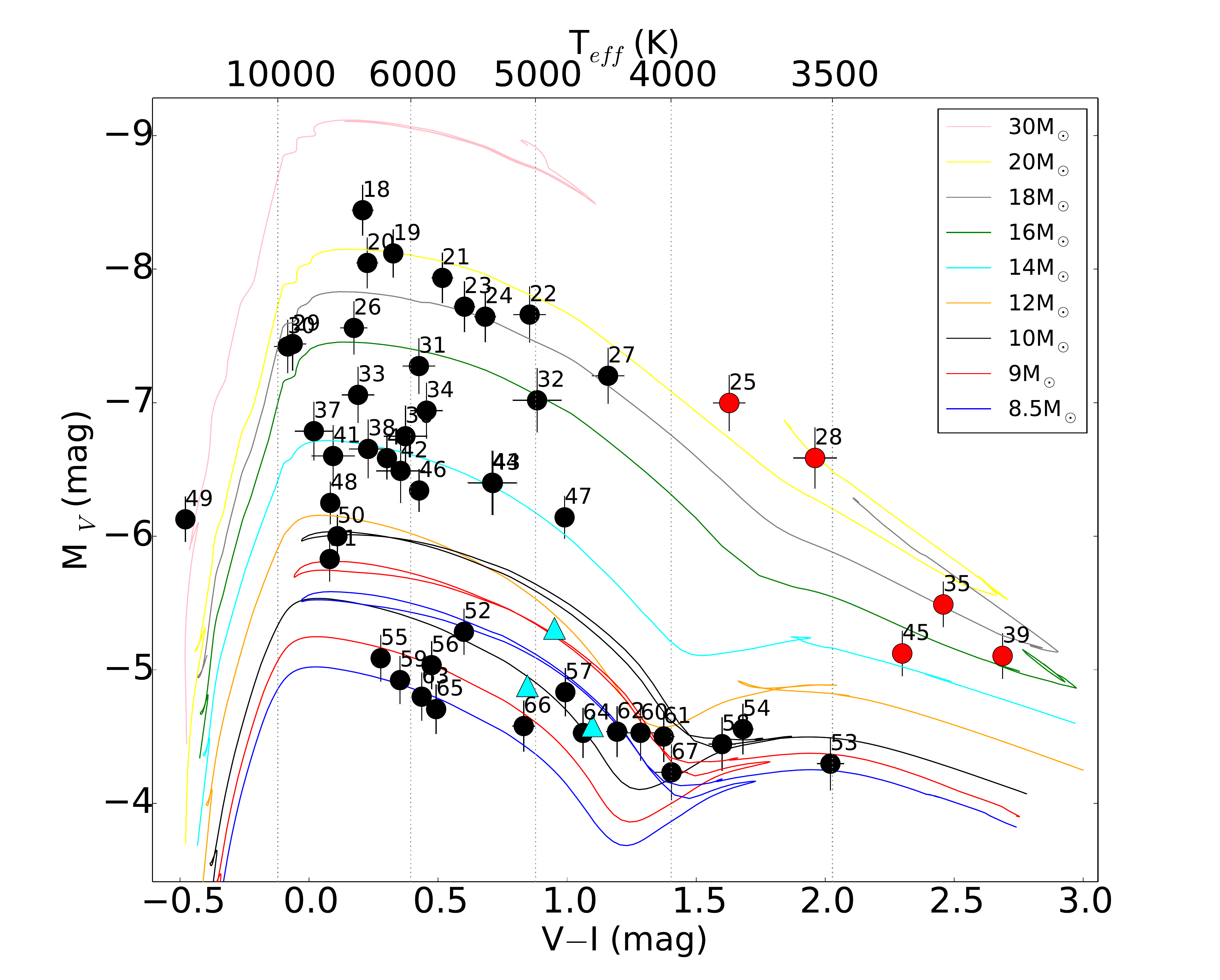}& \\
\includegraphics[trim=0cm 0cm 0.cm 0cm, width=0.50\textwidth]{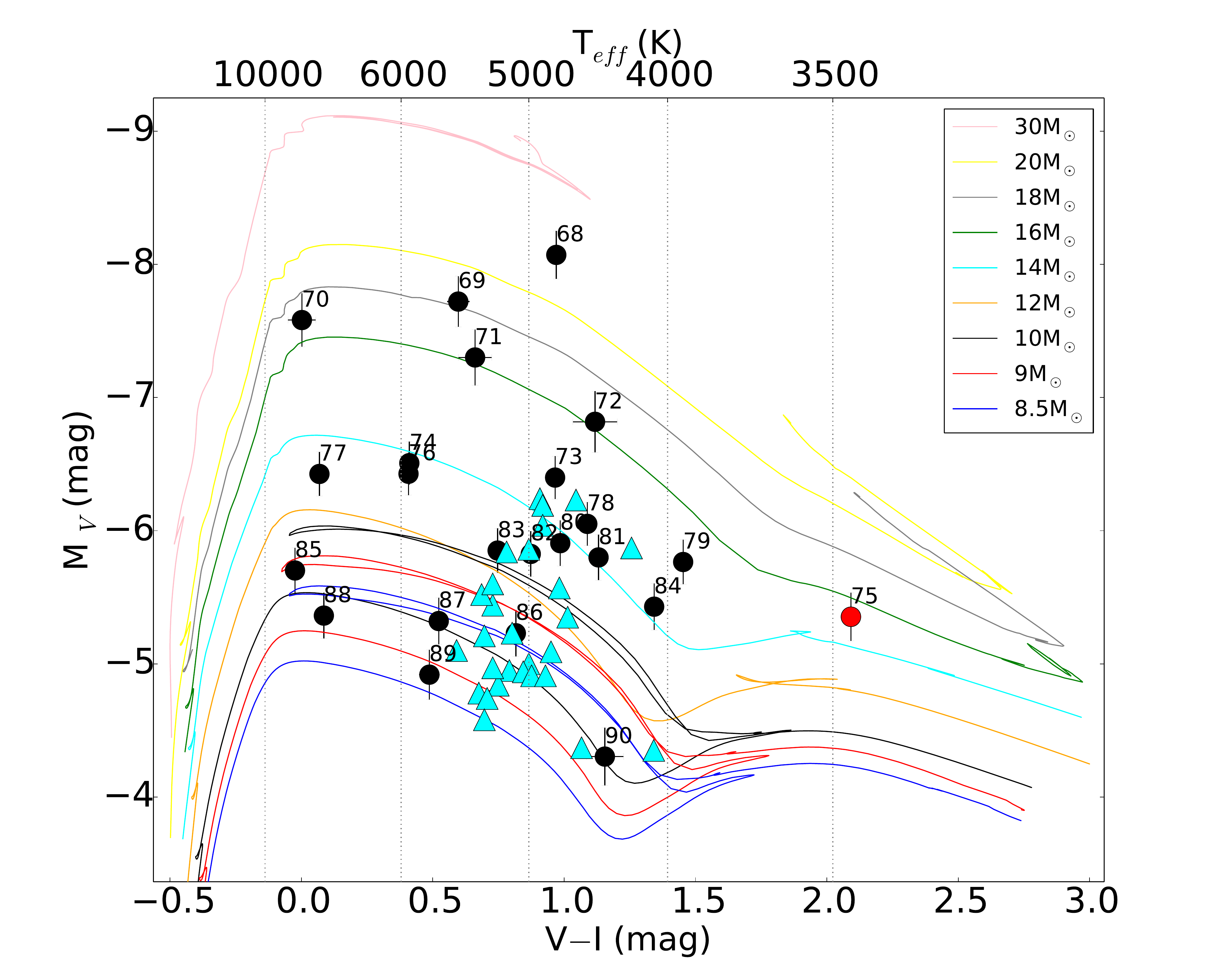} &
\includegraphics[trim=0cm 0cm 0.cm 0cm, width=0.50\textwidth]{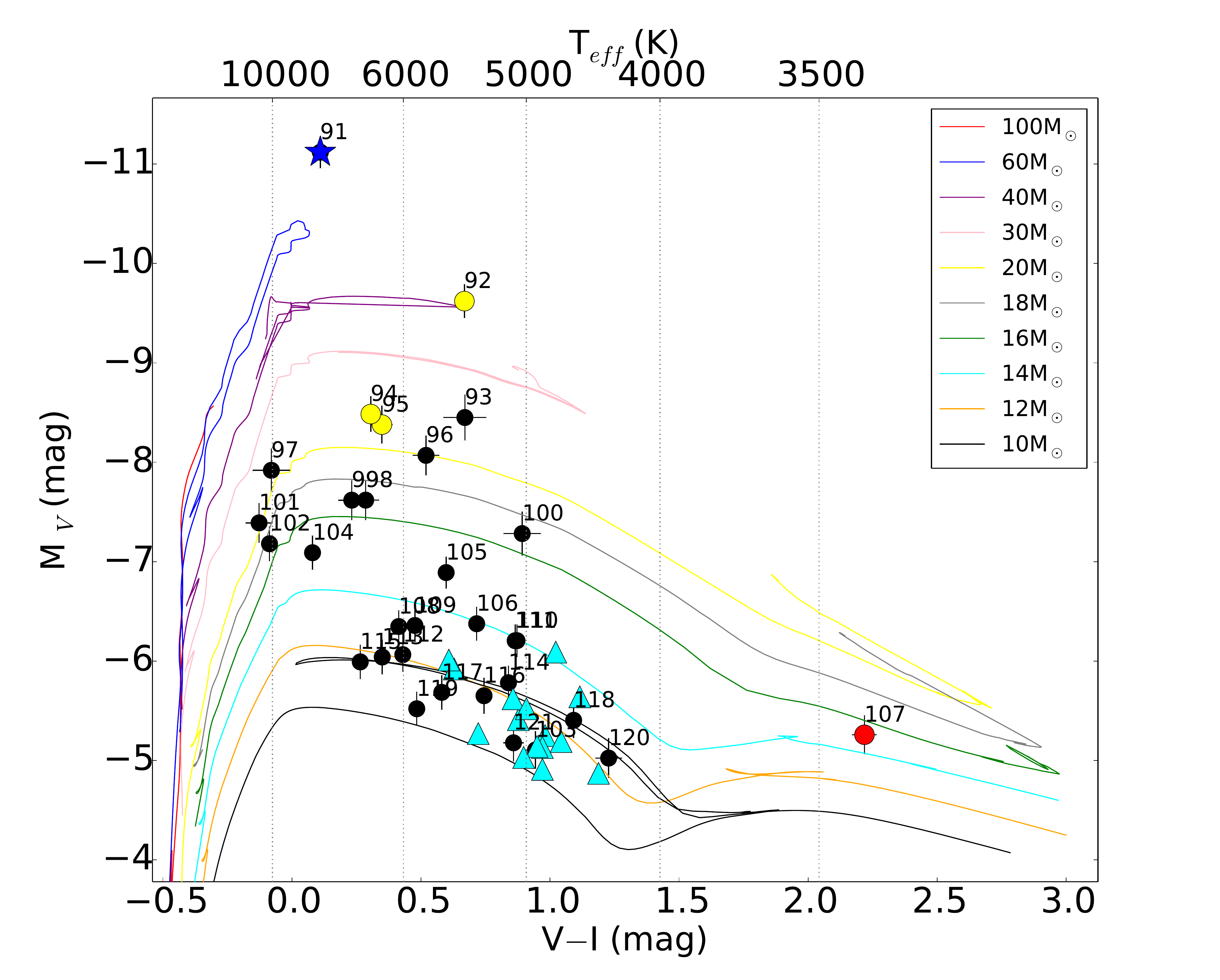}\\

\end{tabular}

\caption{Color-magnitude diagrams per chip with MESA evolutionary models over-plotted. The two upper panels correspond to variables in chips PC and WF2 while the lower panels to those in chips WF3 and WF4.}
\label{evtracks}
\end{figure*}

 The total number of new variable candidates is 120, 15 of which have M$_{V}$ $>$ $-$4.9 mag by taking into account all stars that adhere to the above criteria in V or I. Tables~\ref{pcvars}, \ref{wf2vars}, \ref{wf3vars} and \ref{wf4svars} list the IDs, coordinates, mean magnitude, magnitude errors and variability indexes of the variables. We compared the mean magnitudes of the published Cepheids in \citet{Macri99} and from \texttt{DOLPHOT} to check the quality of our photometry. The mean difference in magnitude is 0.015 $\pm$ 0.008 mag, which indicates consistency between both studies. The tables also include the amplitude of variability (also shown in the light curves) and notes about their classification based on their color (CMD position) and mass as estimated by the models (see Figure~\ref{evtracks}). We use the digit 1 to show that the source was selected by the corresponding index and the digit 0 to show that it was not. 
In many cases a source is only flagged as a candidate variable in one filter and only on the basis of one or two of the variability indexes used. This is partly due to the fact that different types of variability are more efficiently detected by different indexes. Another factor that can lead to different detection in different filters is that the amplitude of variability is often wavelength dependent. The most bright and luminous variable candidates with small variations were selected by the 1/$\eta$ index. All published Cepheids were detected by our photometry, however, not all of them were selected as variable candidates by the three variability indexes. The fact that not all 50 of the published Cepheids are selected as variables by the scatter-based indexes (MAD and IQR) is due to the 4$\sigma$ threshold adopted. The recovered sources that do not exceed the 4$\sigma$ are the ones characterized as lower grade Cepheids in the work of \citet{Macri99}. 
\par From our variable sample 52$\%$ (63) of the variable candidates are selected by any of the MAD or IQR or 1/$\eta$ indexes in both filters. The variable sources included in the HSC and recovered by this analysis, namely V13, V19, V42, V55, V79, V80, V82, V89, V102, V105 and V116 will be later incorporated in the HCV.

\par The index where the largest number of variable candidates was identified is the IQR (66 variable candidates). Thirty nine out of 120 variable candidates ($\sim$47 $\%$) were identified by both MAD and IQR in both or either filters while $\approx$1 $\%$ of the variable sources were common for all three indexes. The fact that MAD and IQR are both scatter-based indexes and selected a different number of variable candidates justifies our decision to use both methods for the identification of variable candidates.

\par As shown in Figure~\ref{cmdbsgs}, the M$_{V}$ of all sources is in the range $-$11 to $-$4.0 mag. The constant stars are shown in gray color, the known Cepheids as blue triangles and the new candidate variables as black dots. The known Cepheids are located within 0.6$\leq$ V$-$I $\leq$1.3 mag with magnitudes between 23.5$-$26.5 mag, which could set the limits of the instability strip for this study. Their masses implied by the models range from 8 $<$ M $<$ 14 M$_{\odot}$. The new candidate variables have magnitudes ranging between $\sim$ 19$-$26.2 mag.     
\par We used the MESA evolutionary tracks to estimate the masses and temperatures of the new variables. Figure~\ref{evtracks} shows the CMDs with evolutionary tracks for the new variables in each chip. The masses corresponding to the tracks span from $\sim$ 8$-$100 M$_\odot$. The temperatures were derived from the models assuming that all sources are at the same distance, age and evolutionary stage (post-AGB stars and supergiants) and each value of T$_{eff}$ corresponds to a specific color index. The error bars indicate the errors corresponding to the reddening E(V$-$I) and the distance modulus. 

\par In Figure~\ref{evtracks} the known Cepheids lie within 0.6 $<$ V$-$I $<$1.3 mag between 8$<$ M $<$ 14 M$_{\odot}$. Identifying the colors and estimating the mass range of the published variables enables the photometric classification of the newly discovered candidate massive variable stars. Most of the new candidate variables lie in crowded regions of the field and along the spiral arms of the galaxy. The models predict that the mass range for the majority of the new variable sources is $\approx$ 9$-$20 M$_{\odot}$ and 0 $<$ V$-$ I $<$ 1.5 mag. We present light curves for all the new variable candidates identified in Figures~\ref{lcpc1}-\ref{lcwf43}. The Figures show that the variable candidates vary with the same trend in both filters and display variations of similar amplitude. 

\par The subsections below provide information about the new variable candidate RSGs, YSGs, the candidate LBV and the candidate Cepheids identified photometrically in NGC 4535.

 \subsection{Variable candidate red supergiants}

Our analysis yielded 8 variable candidate RSGs in the four chips; V7 in the PC chip, V25, V28, V35, V39, V45 in WF2, V75 in WF3 and V107 in WF4. The magnitude range of the candidates is $-$7.0 $\leq$ M$_{V}$ $\leq$ $-$5.0 mag and the V$-$I index is in the range 1.6$-$2.7 mag. According to the MESA models, the masses and temperatures are between 14$-$20 M$_\odot$ and 3000 $<$ T$_{eff}$ $<$ 4000 K, respectively, which correspond to late-K to M types stars. Main-sequence stars with initial masses over 40 M$_\odot$, due to rapidly driven stellar winds that cause mass loss do not expand or drop their temperature to become RSGs \citep{Massey2017}. 
\par Most RSGs show some degree of photometric variability, but without a well-defined period or amplitude. They are usually classified as irregular or semi-regular variables. Their variations are typically slow and of small amplitude in the optical, however, amplitudes up to four magnitudes have been reported \citep{Ming}. In our case, it is difficult to determine whether the candidate RSGs are periodic or not as the observations only cover a short time compared to the typical variability timescale of RSGs (250$-$1000 days). Among the RSG candidates stars V25 and V28 lie at higher masses compared to the others. The values of mass implied by the models, T$_{eff}$ and $\log$L are in good agreement with the ranges calculated for the spectroscopically confirmed RSGs in the SMC, LMC and the Milky Way \citep{Levesque2006}. The studies of \citet{Levesque2006}, \citet{Szczygiel2010} and Yang et al. (2018) provide examples of variations in RSG light curves over a long timescale. Typically, the light variations between 100 days range within 0.6$-$0.8 mag \citep[]{Kiss2006, Ming}. In our work, the amplitude of the light variation is between 0.2$-$0.6 $\pm$ 0.03 mag over $\approx$ 100 days.
AGB stars could be easily confused with RSGs as both types are red and luminous, given the uncertainty in the distance. However, RSGs are more luminous with higher temperatures and lower variability amplitudes. \citet{Wood1983} suggested a criterion for distinguishing RSGs from AGB stars (M$_{bol}$ $\geq$ $-$7.1 mag), so most of our candidates are most likely RSGs. Due to the absence of spectroscopic data we cannot eliminate the possibility of certain stars being super-AGB stars. Nevertheless, the masses inferred by the models are over 16 M$_\odot$, while the mass-range implied for AGB stars is 1 M$_{\odot}$< M <8 M$_{\odot}$ \citep[]{Becker1980,Groenewegen2018}. 

\subsection{Variable candidate yellow supergiants/hypergiants} 

In our sample, stars V92, V94 and V95 are classified as candidate YSGs/YHGs. The stars show M$_V$ brighter than $-$8.5 mag, while the models imply masses and temperatures in the range 20 $<$ M $<$ 40 M$_\odot$ and 5000 $<$ T$_{eff}$ $<$ 8500 K. All stars suggested as candidate YSGs/YHGs vary in both V and I with the same trend. The light curves display variability amplitudes between 0.3$-$0.5 mag, in both filters, in agreement with previous works on these stellar types \citep{Humphreys1979,Humphreys2016,Kourniotis2016}. YSGs are rare stars with spectral types A-G found in a transition state evolving from the main sequence towards the RSG state, or following the opposite path \citep{Humphreys2016}. They are among the most visually luminous stars, with absolute magnitude M$_{V}$ around $-$9 mag. YHGs \citep{deJager98} are evolved massive stars, (post-RSGs) of masses 20$-$40 M$_\odot$ and temperatures that range between 4000$-$7000 K. 
The region between YHGs and LBVs characterized by dynamical instability is known as the yellow void, as only a small number of stars are found in that region due to the brevity of their transition phase \citep{dejager1997}. YHGs are found in the lower-luminosity part of the yellow void while LBVs are in the left of the yellow void, and may have higher luminosities. As spectroscopic classification is essential for the accurate discrimination between YHGs and YSGs, we can not infer on the true state of our yellow stars in the present study.

\subsection{Candidate Luminous Blue Variable}

\par Among the candidate massive stars identified in NGC 4535, star V91 in WF4 deserves special mention, as it is the brightest source in the whole variable sample with M$_{V}$= $-$11 mag. The star is located at V$-$I $\sim$ 0.18 mag and is found above the BSG region shown in Figure~\ref{cmdbsgs}. The position of the source based on its mass, is not consistent with the predictions of the models. The 60$-$100 M$_{\odot}$ tracks imply that the main sequence and post-main sequence evolution of sources as massive is limited by the presence of the Humphreys-Davidson instability limit \citep{Humphreys1994}. The H-D limit is known as the empirical limit of the luminosity or metal-modified Eddington luminosity of cool supergiants, providing information of the general mass-loss behavior of massive stars. The H-D limit signifies the beginning of stellar death for high-mass stars, as the hydrogen-rich envelope of the star, when the lumniosity reaches logL/L$_\odot$ = 5.8 $\pm$ 0.1 is lost before the star evolves blueward in the H-R diagram. This luminosity limit is mainly caused by mass-loss, either through weak stellar winds, or by episodic Luminous Blue Variable (LBV) type eruptions. The information from the CMD in addition to the low-amplitude variability $\sim$ 0.1$-$0.3 mag in the light curve (see Figure~\ref{lcwf41}), leads us to classify this source as a candidate LBV in quiescence \citep[]{Shporer2006LBV, Pastorello2010}. The alternative scenario of a foreground dwarf is unlikely, due to the variability of the source \citep{Ciardi2011}.
\par The production mechanism, the evolutionary path and the duration of the LBV phase still remain unknown, however, the late stages of the evolution of LBVs have been linked with various types of SNe. There is evidence that LBVs at high initial masses, could explode directly as core-collapse type IIn SNe \citep[]{Smith2007,Gal-Yam2007} or dramatically increase their luminosity during outburst and be characterized as SN impostors \citep{Goodrich1989,Filipenko1995,Humphreys2016}. However, more photometric and spectroscopic studies are required to discover the actual causes of photometric and spectroscopic variations occurring during the LBV phase. If confirmed as an LBV, V91 would be the second most distant LBV yet discovered, following PHL 293B at 21.4 Mpc \citep{Izotov2009}. 

\subsection{New candidate Cepheids in NGC 4535}
Other interesting high-amplitude candidate variable massive stars in NGC 4535 are V13, V16, V17 in the PC chip, and V64 in WF2. The sources are located in the instability strip near the known Cepheids of the field \citep[]{Sandage1971, Macri99} as shown in Figure~\ref{cmdbsgs} with : $-$5 $<$ M$_{V}$ $<$ $-$6 mag and 0.8$<$ V$-$I $<$ 1.2 mag. According to the models they are found between the 9$-$10 M$_{\odot}$ track and their temperature is estimated 4000 $<$ T$_{eff}$ $<$ 5000 K. The light curve of all four sources (Figures~\ref{lcpc2} $\&$ \ref{lcwf25}) shows a light variation similar to that displayed by pulsating variable candidates. We roughly estimated the period of the candidate pulsating variable sources based on their light-curve, using the on-line tool of the Nasa Exoplanet Archive \footnote{\url{http://exoplanetarchive.ipac.caltech.edu}}. The period of each source is shown in Tables~\ref{pcvars} and \ref{wf2vars}. These sources are classified as candidate Cepheid variables based on their position in the instability strip near the known Cepheids. However, a larger number of observations is required in order to robustly estimate the type and the period of a candidate Cepheid.

\section{Summary}
\label{summary}

\par We performed PSF photometry on archival HST WFPC2 images and created a catalog of 24353 luminous stars in the Virgo Cluster galaxy NGC 4535. We studied the massive star population of NGC 4535 by creating a sub-catalog of 4973 sources with M$_{V}$ $<$ $-$ 4.9 mag. Using color criteria on the CMD we separated the massive star candidates into three subsets: BSGs, YSGs, and RSGs. We calculated the foreground contamination in the direction of NGC 4535 using the Besan\c{c}on Galactic population synthesis model \citep{Besancon2003} and found that, $\sim$100$\%$ of the 3762 luminous blue supergiants, 85 $\%$ of the 841 yellow supergiants, and 95$\%$ of the 370 red supergiants are members of NGC 4535. We calculated the de-projected distances of the massive stars and estimated the blue to red supergiant ratio at various radial distances. The B/R ratio decreases monotonically with increasing distance. We examined the recent star formation history of the field over the last 60 Myrs within the WFPC2 field of view. We derived the luminosity function based on the blue HeB stars and estimated the SFR using the Salpeter slope for the IMF. Although the star formation events in the field before 10 Myrs seem suppressed, the SFR rises after $\approx$ 30 Myrs. 
\par We conducted a variability search for the initial catalog of 24353 sources, using three different methods: the MAD, IQR and the inverse von Neumann ratio. The variability search yielded 120 new variable sources in addition to the 50 known Cepheid variables of the field \citep{Macri99}. We used the MESA models and evolutionary tracks \citep{MESA2016} to model our results and classify the new variable candidates. Among the luminous massive variable sources are 8 variable candidate RSGs, 3 variable candidate YSGs and 1 variable candidate LBV. However, the high number of BSGs confirms the high star formation rate of the galaxy. 

\par This work demonstrates the science potential of archival data from the Hubble Legacy Archive (HLA) of other HST Key Project galaxies \citep{Freedman2001}, as it provides an analysis of the massive star population and the variability of massive stars beyond the Local Group.
The characterization of massive stars at large distances unveils important information about the surrounding environment and could help identify the progenitors of future SN explosions \citep{Smartt2009}, and their variability properties \citep[e.g.][]{Supernova2} posing a challenge for the new generation of telescopes.

\begin{acknowledgements}
We thank the anonymous referee for the careful reading of the manuscript and the insightful comments and suggestions. We acknowledge financial support by the European Space Agency (ESA) under the ‘Hubble Catalog of Variables’ program, contract no. 4000112940. This research has made use of the VizieR catalogue access tool, CDS, Strasbourg, France and NASA's Astrophysics Data System (ADS). The Virtual Observatory tools (VO) tools TOPCAT \citep{Taylor2005} and Aladin were used for image downloading and table manipulation. Versions 1 and 2 of the Hubble Source Catalog (HSC) were also used. The data used in this study were downloaded from the Mikulski Archive for Space Telescopes (MAST). All figures in this work were produced using Matplotlib a Python library for publication graphics.
\end{acknowledgements}

\bibliographystyle{aa}
\bibliography{aanda.bib}

\begin{table*} [ht!]
\caption{Number of stars after photometric quality cuts.}
\begin{tabular}{l c c c c } 
\hline
 {WFPC2 chip} & {\#stars after cuts} \\
\hline\hline

 PC1 & 2223 \\ 

 WF2 & 5290\\

 WF3 & 7507 \\
 
 WF4 & 9333 \\
 \hline
 Sum &24353 \\

\hline
\end{tabular}
\label{photcuts}
\end{table*}

\begin{table*}{}                               
\caption {Catalog of 24353 stars. This table is available in its entirety in an electronic version. The fifty first entries are shown as a guidance for the content of the tables.}
\begin{tabular}{l c c c c c c } 
\hline
{RA} &{DEC} &{V}& {$\rm \sigma$ $_{V}$} & {I} & {$\rm \sigma$ $_{I}$}\\
(J2000) & (J2000) & (mag) & (mag) & (mag) &(mag)\\
\hline
  12:34:19.091 & +8:09:36.63 & 22.757 & 0.005 & 21.328 & 0.005\\
  12:34:20.354 & +8:10:09.04 & 23.334 & 0.005 & 23.117 & 0.005\\
  12:34:19.789 & +8:10:11.66 & 23.485 & 0.004 & 22.876 & 0.004\\
  12:34:20.519 & +8:10:05.67 & 23.597 & 0.005 & 23.508 & 0.005\\
  12:34:19.348 & +8:09:38.36 & 23.639 & 0.005 & 22.993 & 0.005\\
  12:34:19.837 & +8:09:56.44 & 23.650 & 0.004 & 21.791 & 0.004\\
  12:34:20.099 & +8:10:02.93 & 23.661 & 0.007 & 23.476 & 0.007\\
  12:34:19.281 & +8:09:41.86 & 23.784 & 0.005 & 23.731 & 0.005\\
  12:34:19.001 & +8:10:09.12 & 23.793 & 0.006 & 23.555 & 0.006\\
  12:34:20.061 & +8:10:06.24 & 24.030 & 0.006 & 23.598 & 0.006\\
  12:34:19.455 & +8:09:37.46 & 24.072 & 0.007 & 23.470 & 0.007\\
  12:34:19.065 & +8:09:44.53 & 24.204 & 0.009 & 24.000 & 0.009\\
  12:34:19.342 & +8:09:37.39 & 24.352 & 0.009 & 24.203 & 0.009\\
  12:34:19.040 & +8:10:08.80 & 24.373 & 0.008 & 24.165 & 0.008\\
  12:34:18.972 & +8:09:51.85 & 24.493 & 0.009 & 23.812 & 0.009\\
  12:34:18.959 & +8:10:01.91 & 24.525 & 0.008 & 23.949 & 0.008\\
  12:34:20.120 & +8:10:10.13 & 24.627 & 0.009 & 24.362 & 0.009\\
  12:34:19.547 & +8:09:37.61 & 24.632 & 0.010 & 24.685 & 0.010\\
  12:34:19.746 & +8:10:07.93 & 24.763 & 0.012 & 24.876 & 0.012\\
  12:34:19.055 & +8:09:37.79 & 24.783 & 0.013 & 24.640 & 0.013\\
  12:34:19.603 & +8:09:51.30 & 24.797 & 0.009 & 23.600 & 0.009\\
  12:34:19.492 & +8:09:37.98 & 24.872 & 0.012 & 24.736 & 0.012\\
  12:34:20.423 & +8:09:52.41 & 24.883 & 0.010 & 24.768 & 0.010\\
  12:34:20.063 & +8:09:55.08 & 24.888 & 0.010 & 24.840 & 0.010\\
  12:34:19.183 & +8:09:40.99 & 24.888 & 0.012 & 24.784 & 0.012\\
  12:34:20.590 & +8:09:56.25 & 24.909 & 0.012 & 23.841 & 0.012\\
  12:34:20.109 & +8:09:58.07 & 24.951 & 0.012 & 24.886 & 0.012\\
  12:34:19.665 & +8:10:14.62 & 24.963 & 0.015 & 25.042 & 0.015\\
  12:34:20.509 & +8:09:51.97 & 24.972 & 0.012 & 24.104 & 0.012\\
  12:34:19.464 & +8:09:38.19 & 25.003 & 0.013 & 25.011 & 0.013\\
  12:34:19.954 & +8:10:00.05 & 25.022 & 0.011 & 24.291 & 0.011\\
  12:34:19.189 & +8:09:34.66 & 25.022 & 0.016 & 24.689 & 0.016\\
  12:34:20.256 & +8:10:08.99 & 25.026 & 0.013 & 25.074 & 0.013\\
  12:34:18.636 & +8:09:55.16 & 25.041 & 0.011 & 24.821 & 0.011\\
  12:34:20.559 & +8:10:02.07 & 25.042 & 0.012 & 25.323 & 0.012\\
  12:34:20.585 & +8:10:04.26 & 25.076 & 0.014 & 25.070 & 0.014\\
  12:34:18.647 & +8:10:01.52 & 25.094 & 0.013 & 24.432 & 0.013\\
  12:34:18.857 & +8:10:08.94 & 25.101 & 0.015 & 24.950 & 0.015\\
  12:34:19.335 & +8:09:39.90 & 25.105 & 0.013 & 22.818 & 0.013\\
  12:34:20.111 & +8:09:56.89 & 25.113 & 0.012 & 24.701 & 0.012\\
  12:34:19.365 & +8:09:38.41 & 25.122 & 0.014 & 24.852 & 0.014\\
\hline&

\end{tabular}
\begin{tablenotes}
\item \textbf{Note.} -- Units of right ascension are hours, minutes, and seconds, and units of declination are degrees, arcminutes, and arcseconds. 
\end{tablenotes}
\label{starcat}
\end{table*}

\begin{table*}{} 
\caption {The massive star population in NGC 4535 and M101.}
 \begin{tabular}{l c c c c c} 
 \hline
 {Galaxy} &{$\#$ stars} &{$\#$BSGs}& {$\#$YSGs}& {$\#$RSGs} & \\
 \hline\hline\hline
 NGC 4535& 4973 & 3762 (76 $\%$) & 841 (16 $\%$) &370 (8 $\%$) \\
 M101 &35705 &25603 (72 $\%$) & 3105 (8 $\%$)  &2294 (6 $\%$) \\ 
 
\hline&
\end{tabular}
\begin{tablenotes}
\item \textbf{Note.}-- The second column corresponds to stars with M$_{V}$ $<$ $-$4.9 mag.
\end{tablenotes}
\label{comparison}
\end{table*}

\begin{sidewaystable*} [ht!]
\caption{Properties of newly discovered variable candidates in the PC chip ordered by average magnitudes.}
\scalebox{0.8}{
\begin{tabular}{l c c c c c c c c c c c c c c c c c}
\hline

{ID} &{X} &{Y} & {RA} & {Dec} & {V}& {$\rm \sigma$ $_{V}$} & {I}&{$\rm \sigma$ $_{I}$}&  {MAD$_{V}$} & {MAD$_{I}$} & {IQR$_{V}$} & {IQR $_{I}$} & {1/$\eta$ $_{V,I}$} & {$\Delta$V} & {$\Delta$I} &{HSC MatchID} &{Notes}  \\ 

& & & (J2000) & (J2000) & (mag) & (mag) & (mag)& (mag) & & &  &  & & (mag) & (mag) & \\
\hline

\hline
 V1 & 1450.9 & 1928.3 & 12:34:20.37 & +08:10:09.8 & 23.352 & 0.005 & 23.116 & 0.009 & 0 & 0 & 0 & 0 & 1 & 0.2 & 0.1 &  & -\\
  V2 & 1997.3 & 1514.6 & 12:34:19.36 & +08:09:39.1 & 23.655 & 0.005 & 22.993 & 0.009 & 0 & 0 & 0 & 0 & 1 & 0.2 & 0.1 &  & -\\
  V3 & 1811.3 & 2113.3 & 12:34:19.02 & +08:10:09.9 & 23.811 & 0.006 & 23.555 & 0.012 & 0 & 0 & 0 & 0 & 1 & 0.2 & 0.2 &  & -\\
  V4 & 1976.9 & 1482.8 & 12:34:19.46 & +08:09:38.2 & 24.082 & 0.007 & 23.468 & 0.011 & 1 & 0 & 1 & 0 & 1 & 0.2 & 0.2 & 428308.0 & -\\
  V5 & 2045.3 & 1632.4 & 12:34:18.99 & +08:09:42.7 & 24.232 & 0.009 & 23.992 & 0.016 & 1 & 1 & 1 & 1 & 0 & 0.4 & 0.3 & 275787.0 & -\\
  V6 & 1811.8 & 1711.9 & 12:34:19.61 & +08:09:52.1 & 24.803 & 0.009 & 23.596 & 0.01 & 1 & 1 & 1 & 1 & 1 & 0.4 & 0.3 & 275818.0 & -\\
  V7 & 1995.8 & 1500.6 & 12:34:19.37 & +08:09:37.8 & 25.17 & 0.016 & 23.431 & 0.011 & 0 & 0 & 0 & 0 & 1 & 0.2 & 0.2 & 272878.0 & candidate RSG\\
  V8 & 1583.5 & 1621.8 & 12:34:20.42 & +08:09:52.4 & 24.892 & 0.01 & 24.765 & 0.026 & 0 & 0 & 0 & 0 & 1 & 0.1 & 0.2 & 273276.0 & -\\
  V9 & 1974.2 & 2023.9 & 12:34:18.66 & +08:10:02.2 & 25.099 & 0.013 & 24.432 & 0.022 & 0 & 0 & 0 & 0 & 1 & 0.2 & 0.2 &  & -\\
  V10 & 1477.9 & 1790.1 & 12:34:20.30 & +08:10:03.1 & 25.495 & 0.016 & 24.424 & 0.02 & 1 & 0 & 1 & 0 & 0 & 0.5 & 0.5 &  & -\\
  V11 & 1604.4 & 2128.1 & 12:34:19.60 & +08:10:15.2 & 25.47 & 0.019 & 24.842 & 0.029 & 0 & 0 & 1 & 0 & 0 & 0.8 & 1.2 & 275825.0 & -\\
  V12 & 2019.8 & 1535.7 & 12:34:19.26 & +08:09:39.6 & 25.701 & 0.019 & 24.745 & 0.025 & 1 & 0 & 1 & 1 & 0 & 1.5 & 1.5 & 273279.0 & -\\
  V13 & 1519.1 & 1715.7 & 12:34:20.25 & +08:09:59.04 & 25.749 & 0.024 & 24.94 & 0.028 & 1 & 1 & 1 & 1 & 0 & 1.2 & 1.2 & 276120.0 & candidate Cepheid (P$\approx$35 days)\\
  V14 & 2007.1 & 1526.3 & 12:34:19.40 & +08:09:38.1 & 25.572 & 0.019 & 25.665 & 0.061 & 0 & 1 & 0 & 0 & 0 & 0.4 & 1.2 &  & -\\
  V15 & 2001.3 & 1537.8 & 12:34:19.30 & +08:09:39.7 & 26.086 & 0.032 & 24.896 & 0.032 & 0 & 0 & 0 & 1 & 0 & 0.2 & 0.5 &  & -\\
  V16 & 1706.3 & 2107.9 & 12:34:19.34 & +08:10:12.2 & 26.187 & 0.031 & 25.302 & 0.041 & 1 & 0 & 1 & 0 & 0 & 0.8 & 1.2 &  & candidate Cepheid (P$\approx$21 days)\\
  V17 & 2070.2 & 1615.5 & 12:34:18.98 & +08:09:42.0 & 26.287 & 0.036 & 25.266 & 0.045 & 1 & 1 & 1 & 1 & 0 & 1.2 & 1.2 &  & candidate Cepheid (P $\approx$33 days)\\
\hline
\end{tabular}}
\begin{tablenotes}
\item \textbf{Note.} -- Units of right ascension are hours, minutes, and seconds, and units of declination are degrees, arcminutes, and arcseconds. 
\\Numbers 1 or 0 in columns MAD, IQR and inverse von Neumann ratio show whether the star was characterized as a variable in the respective index.
\end{tablenotes}
\label{pcvars}
\end{sidewaystable*}

\begin{sidewaystable*} [ht!]
\caption{Properties of newly discovered variable candidates in the WF2 chip ordered by average magnitudes.}
\scalebox{0.8}{
\begin{tabular}{l c c c c c c c c c c c c c c c c c}
\hline
{ID} & {X}& {Y} &{RA} & {Dec} & {V}& {$\rm \sigma$ $_{V}$} & {I}&{$\rm \sigma$ $_{I}$}&  {MAD$_{V}$} & {MAD$_{I}$} & {IQR$_{V}$} & {IQR $_{I}$} & {1/$\eta$ $_{V,I}$} & {$\Delta$V} & {$\Delta$I} & {HSC MatchID}& {Notes} \\ 

& & & (J2000) & (J2000) & (mag) & (mag) & (mag)& (mag) & & &  &  & & (mag) & (mag) & \\
\hline
V18 & 3604.9 & 1425.3 & 12:34:14.70 & +08:08:59.60 & 22.5785 & 0.004 & 22.265 & 0.007 & 0 & 0 & 0 & 0 & 1 & 0.15 & 0.1 & 270842.0 & -\\
  V19 & 2245.8 & 1735.5 & 12:34:18.28 & +08:09:42.30 & 22.902 & 0.003 & 22.476 & 0.005 & 0 & 0 & 0 & 0 & 1 & 0.12 & 0.15 & 268461.0 & -\\
  V20 & 3157.0 & 1955.4 & 12:34:15.25 & +08:09:32.70 & 22.973 & 0.004 & 22.643 & 0.006 & 0 & 0 & 0 & 0 & 1 & 0.16 & 0.05 & 269524.0 & -\\
  V21 & 2204.7 & 1530.6 & 12:34:18.71 & +08:09:35.20 & 23.085 & 0.004 & 22.478 & 0.006 & 1 & 0 & 0 & 0 & 1 & 0.26 & 0.1 & 272464.0 & -\\
  V22 & 2735.6 & 1861.0 & 12:34:16.62 & +08:09:37.90 & 23.3601 & 0.006 & 22.431 & 0.007 & 0 & 1 & 0 & 0 & 1 & 0.24 & 0.12 & 269744.0 & -\\
  V23 & 2522.8 & 1684.6 & 12:34:17.51 & +08:09:34.30 & 23.3 & 0.004 & 22.612 & 0.006 & 1 & 0 & 0 & 0 & 1 & 0.2 & 0.2 & 267111.0 & -\\
  V24 & 2518.4 & 2016.2 & 12:34:17.04 & +08:09:49.60 & 23.376 & 0.004 & 22.611 & 0.006 & 1 & 1 & 0 & 0 & 1 & 0.1 & 0.1 & 274347.0 & -\\
  V25 & 2906.0 & 925.3 & 12:34:17.53 & +08:08:52.50 & 24.022 & 0.006 & 22.357 & 0.005 & 0 & 0 & 1 & 1 & 1 & 0.08 & 0.1 & 428155.0 & candidate M-type RSG\\
  V26 & 2420.6 & 1829.5 & 12:34:17.61 & +08:09:43.53 & 23.46 & 0.005 & 23.179 & 0.008 & 0 & 0 & 0 & 0 & 1 & 0.2 & 0.12 & 269759.0 & -\\
  V27 & 2290.5 & 1990.4 & 12:34:17.76 & +08:09:53.69 & 23.818 & 0.006 & 22.601 & 0.006 & 0 & 0 & 0 & 0 & 1 & 0.15 & 0.1 & 267882.0 & -\\
  V28 & 2543.5 & 1744.4 & 12:34:17.38 & +08:09:37.03 & 24.433 & 0.008 & 22.453 & 0.005 & 0 & 1 & 0 & 1 & 1 & 0.16 & 0.05 & 267855.0 & candidate M-type RSG\\
  V29 & 2443.7 & 1808.9 & 12:34:17.57 & +08:09:42.52 & 23.58 & 0.005 & 23.525 & 0.01 & 0 & 0 & 0 & 0 & 1 & 0.1 & 0.1 & 268449.0 & -\\
  V30 & 2179.7 & 1740.6 & 12:34:18.48 & +08:09:45.57 & 23.599 & 0.005 & 23.562 & 0.011 & 0 & 0 & 1 & 0 & 0 & 0.16 & 0.1 & 270725.0 & -\\
  V31 & 2719.7 & 2003.3 & 12:34:16.45 & +08:09:44.69 & 23.745 & 0.006 & 23.225 & 0.008 & 1 & 0 & 0 & 0 & 0 & 0.16 & 0.16 & 267849.0 & -\\
  V32 & 2292.3 & 1833.6 & 12:34:17.96 & +08:09:45.90 & 24.001 & 0.009 & 23.045 & 0.008 & 0 & 0 & 1 & 0 & 0 & 0.3 & 0.1 & 428222.0 & -\\
  V33 & 3209.8 & 2013.9 & 12:34:14.97 & +08:09:34.03 & 23.961 & 0.006 & 23.665 & 0.011 & 1 & 0 & 0 & 0 & 0 & 0.3 & 0.3 & 270994.0 & -\\
  V34 & 3259.4 & 1689.6 & 12:34:15.32 & +08:09:18.58 & 24.079 & 0.006 & 23.531 & 0.01 & 0 & 0 & 0 & 0 & 1 & 0.1 & 0.2 & 273057.0 & -\\
  V35 & 3411.4 & 1513.1 & 12:34:15.13 & +08:09:07.28 & 25.53 & 0.022 & 23.077 & 0.007 & 1 & 1 & 1 & 1 & 0 & 0.4 & 0.2 & 270849.0 & candidate RSG\\
  V36 & 3622.9 & 1920.8 & 12:34:13.87 & +08:09:20.58 & 24.27 & 0.008 & 23.8 & 0.013 & 0 & 0 & 0 & 1 & 0 & 0.16 & 0.45 & 274464.0 & -\\
  V37 & 3508.0 & 1269.2 & 12:34:15.20 & +08:08:54.26 & 24.233 & 0.007 & 24.1 & 0.015 & 0 & 0 & 0 & 0 & 1 & 0.16 & 0.16 & 271812.0 & -\\
  V38 & 3535.7 & 1323.1 & 12:34:15.05 & +08:08:56.14 & 24.366 & 0.007 & 24.033 & 0.014 & 0 & 0 & 0 & 0 & 1 & 0.16 & 0.1 & 274461.0 & -\\
  V39 & 3000.0 & 1205.6 & 12:34:16.81 & +08:09:03.04 & 25.916 & 0.022 & 23.244 & 0.012 & 0 & 0 & 0 & 1 & 1 & 0.3 & 0.2 & 273375.0 & candidate RSG\\
  V40 & 2532.8 & 1698.0 & 12:34:17.48 & +08:09:35.22 & 24.434 & 0.01 & 24.032 & 0.016 & 0 & 0 & 1 & 1 & 1 & 0.2 & 0.16 &  & -\\
  V41 & 2829.0 & 1465.0 & 12:34:16.93 & +08:09:18.34 & 24.434 & 0.01 & 24.032 & 0.016 & 0 & 0 & 0 & 0 & 1 & 0.1 & 0.2 &  & -\\
  V42 & 3151.9 & 1915.2 & 12:34:15.30 & +08:09:30.08 & 24.53 & 0.009 & 24.078 & 0.018 & 0 & 0 & 1 & 1 & 1 & 0.2 & 0.3 & 267076.0 & -\\
  V43 & 2468.6 & 2071.2 & 12:34:17.11 & +08:09:53.21 & 24.621 & 0.009 & 23.828 & 0.013 & 1 & 0 & 0 & 0 & 0 & 0.3 & 0.1 & 272401.0 & -\\
  V44 & 2837.5 & 1043.8 & 12:34:17.56 & +08:08:58.60 & 24.618 & 0.009 & 23.828 & 0.013 & 1 & 1 & 0 & 0 & 1 & 0.2 & 0.2 & 269713.0 & -\\
  V45 & 3149.5 & 1775.4 & 12:34:15.51 & +08:09:24.87 & 25.897 & 0.026 & 23.595 & 0.01 & 1 & 1 & 1 & 1 & 0 & 1.2 & 0.6 &  & candidate RSG\\
  V46 & 3266.2 & 1913.4 & 12:34:14.96 & +08:09:28.42 & 24.677 & 0.01 & 24.156 & 0.016 & 0 & 0 & 1 & 1 & 1 & 0.2 & 0.2 &  & -\\
  V47 & 3438.5 & 1377.7 & 12:34:15.24 & +08:09:00.73 & 24.878 & 0.011 & 23.821 & 0.013 & 0 & 0 & 0 & 1 & 0 & 0.2 & 0.2 & 266815.0 & -\\
  V48 & 2551.3 & 2004.7 & 12:34:16.95 & +08:09:48.50 & 24.77 & 0.01 & 24.577 & 0.02 & 0 & 0 & 0 & 1 & 0 & 0.1 & 0.3 & 272436.0 & -\\
  V49 & 3643.0 & 1458.6 & 12:34:14.51 & +08:08:59.57 & 24.893 & 0.014 & 25.234 & 0.051 & 0 & 1 & 0 & 0 & 0 & 0.2 & 0.5 &  & -\\
  V50 & 3571.8 & 1389.4 & 12:34:14.82 & +08:08:58.29 & 25.019 & 0.011 & 24.799 & 0.025 & 1 & 0 & 0 & 0 & 0 & 0.3 & 0.1 & 271157.0 & -\\
  V51 & 2568.6 & 2093.7 & 12:34:16.78 & +08:09:51.00 & 25.19 & 0.018 & 24.999 & 0.034 & 1 & 0 & 0 & 1 & 1 & 0.4 & 0.2 & 272439.0 & -\\
  V52 & 3262.5 & 1861.2 & 12:34:15.05 & +08:09:26.27 & 25.735 & 0.023 & 25.049 & 0.034 & 0 & 1 & 0 & 1 & 0 & 0.6 & 0.2 &  & -\\
  V53 & 2175.8 & 1566.7 & 12:34:18.74 & +08:09:38.71 & 26.723 & 0.059 & 24.685 & 0.027 & 0 & 0 & 1 & 0 & 0 & 0.8 & 0.6 &  & -\\
  V54 & 3547.4 & 1616.7 & 12:34:14.58 & +08:09:05.38 & 26.463 & 0.044 & 24.749 & 0.03 & 0 & 0 & 1 & 0 & 0 & 0.5 & 0.4 &  & -\\
  V55 & 2463.1 & 1564.9 & 12:34:17.89 & +08:09:30.89 & 24.933 & 0.025 & 23.554 & 0.056 & 0 & 1 & 0 & 0 & 0 & 0.4 & 0.6 & 274615.0 & -\\
  V56 & 3414.5 & 2043.9 & 12:34:14.31 & +08:09:30.87 & 25.985 & 0.029 & 25.418 & 0.049 & 0 & 1 & 1 & 1 & 0 & 0.4 & 0.8 &  & -\\
  V57 & 2532.1 & 1861.1 & 12:34:17.25 & +08:09:43.07 & 26.187 & 0.029 & 25.127 & 0.035 & 1 & 1 & 0 & 0 & 0 & 0.2 & 0.4 &  & -\\
  V58 & 2365.4 & 1920.2 & 12:34:17.65 & +08:09:47.87 & 26.576 & 0.048 & 24.938 & 0.035 & 0 & 0 & 0 & 1 & 0 & 0.6 & 0.75 &  & -\\
  V59 & 2223.6 & 1668.2 & 12:34:18.46 & +08:09:41.61 & 26.0971 & 0.03 & 25.646 & 0.066 & 0 & 1 & 0 & 1 & 1 & 0.5 & 0.8 &  & -\\
  V60 & 3627.7 & 1445.2 & 12:34:14.59 & +08:08:59.32 & 26.491 & 0.055 & 25.154 & 0.041 & 0 & 0 & 0 & 1 & 0 & 0.25 & 0.5 &  & -\\
  V61 & 2840.3 & 786.8 & 12:34:14.61 & +08:08:59.42 & 26.519 & 0.041 & 25.096 & 0.034 & 0 & 0 & 0 & 1 & 0 & 0.5 & 0.4 &  & -\\
  V62 & 3477.6 & 2001.1 & 12:34:14.18 & +08:09:27.30 & 26.483 & 0.039 & 25.232 & 0.039 & 1 & 0 & 0 & 0 & 0 & 1.0 & 0.5 &  & -\\
  V63 & 2486.9 & 1775.0 & 12:34:17.51 & +08:09:40.18 & 26.221 & 0.031 & 25.691 & 0.063 & 0 & 1 & 0 & 0 & 0 & 0.6 & 0.8 & 275364.0 & -\\
  V64 & 3360.6 & 1038.2 & 12:34:15.97 & +08:08:45.25 & 26.492 & 0.038 & 25.367 & 0.039 & 1 & 1 & 0 & 1 & 0 & 0.8 & 1.0 &  & periodic variable (P $\approx$20 days)\\
  V65 & 3447.1 & 2073.5 & 12:34:14.16 & +08:09:31.77 & 26.314 & 0.035 & 25.73 & 0.064 & 0 & 0 & 1 & 1 & 0 & 0.4 & 1.2 &  & -\\
  V66 & 3225.7 & 1997.6 & 12:34:14.94 & +08:09:33.20 & 26.441 & 0.042 & 25.534 & 0.055 & 1 & 0 & 0 & 0 & 0 & 0.8 & 0.6 &  & -\\
  V67 & 2200.5 & 1936.7 & 12:34:18.12 & +08:09:54.18 & 26.78 & 0.06 & 25.336 & 0.045 & 0 & 1 & 0 & 1 & 0 & 0.6 & 0.8 & 273259.0 & -\\
\hline
\end{tabular}}

\begin{tablenotes}
\item \textbf{Note.} -- Units of right ascension are hours, minutes, and seconds, and units of declination are degrees, arcminutes, and arcseconds. 
\\Numbers 1 or 0 in columns MAD, IQR and inverse von Neumann ratio show whether the star was characterized as a variable in the respective index.
\end{tablenotes}
\label{wf2vars}
\end{sidewaystable*}

\begin{sidewaystable*} [ht!]
\caption{Properties of newly discovered variable candidates in the WF3 chip ordered by average magnitudes.}
\scalebox{0.8}{
\begin{tabular}{l c c c c c c c c c c c c c c c c c c}
\hline
{ID} &{X} &{Y}& {RA} & {Dec} & {V}& {$\rm \sigma$ $_{V}$} & {I}&{$\rm \sigma$ $_{I}$}&  {MAD$_{V}$} & {MAD$_{I}$} & {IQR$_{V}$} & {IQR $_{I}$} & {1/$\eta$ $_{V,I}$} & {$\Delta$V}& {$\Delta$I} &{HSC  MatchID}& {Notes}  \\ 

& & & (J2000) & (J2000) & (mag) & (mag) & (mag)& (mag) & & &  &  & & (mag) & (mag) & \\
\hline
    
V68 & 2288.0 & 2653.1 & 12:34:16.76 & 08:10:23.20 & 21.929 & 0.002 & 20.568 & 0.002 & 0 & 0 & 0 & 0 & 1 & 0.05 & 0.1 & 271855.0 & -\\
  V69 & 2905.8 & 2909.4 & 12:34:14.52 & 08:10:20.80 & 23.299 & 0.004 & 22.603 & 0.006 & 1 & 1 & 1 & 0 & 0 & 0.13 & 0.04 &  & -\\
  V70 & 2956.9 & 2990.5 & 12:34:14.24 & 08:10:23.10 & 23.438 & 0.005 & 23.305 & 0.012 & 1 & 0 & 1 & 0 & 0 & 0.3 & 0.2 &  & -\\
  V71 & 2477.6 & 2564.4 & 12:34:16.33 & 08:10:15.00 & 23.719 & 0.006 & 22.962 & 0.009 & 1 & 0 & 1 & 0 & 0 & 0.3 & 0.2 & 275267.0 & -\\
  V72 & 3347.5 & 2709.0 & 12:34:13.51 & 08:10:01.90 & 24.202 & 0.008 & 23.013 & 0.008 & 0 & 0 & 0 & 0 & 1 & 0.1 & 0.1 & 271199.0 & -\\
  V73 & 2177.1 & 2873.5 & 12:34:16.75 & 08:10:35.50 & 24.621 & 0.011 & 23.576 & 0.014 & 1 & 1 & 1 & 1 & 0 & 0.4 & 0.4 &  & -\\
  V74 & 2574.3 & 2232.4 & 12:34:16.54 & 08:09:58.00 & 24.512 & 0.011 & 23.993 & 0.018 & 0 & 0 & 0 & 1 & 0 & 0.2 & 0.2 &  & -\\
  V75 & 3211.9 & 2511.7 & 12:34:14.21 & 08:09:56.10 & 24.48 & 0.01 & 24.424 & 0.026 & 0 & 0 & 1 & 1 & 0 & 0.3 & 0.4 & 272403.0 & -\\
  V76 & 3082.3 & 3172.1 & 12:34:13.63 & 08:10:28.50 & 25.666 & 0.031 & 23.556 & 0.012 & 0 & 1 & 0 & 1 & 0 & 0.5 & 0.3 & 274328.0 & candidate RSG\\
  V77 & 3130.7 & 3023.8 & 12:34:13.68 & 08:10:23.60 & 24.592 & 0.01 & 24.076 & 0.017 & 0 & 0 & 0 & 1 & 0 & 0.3 & 0.4 & 269743.0 & -\\
  V78 & 3399.3 & 2258.3 & 12:34:14.03 & 08:09:40.70 & 24.594 & 0.01 & 24.398 & 0.022 & 1 & 0 & 1 & 0 & 0 & 0.4 & 0.4 & 269688.0 & -\\
  V79 & 3113.2 & 3277.4 & 12:34:13.37 & 08:10:31.40 & 24.969 & 0.013 & 23.808 & 0.014 & 1 & 1 & 1 & 1 & 0 & 0.8 & 0.8 & 268433.0 & -\\
  V80 & 3276.8 & 2325.4 & 12:34:14.30 & 08:09:46.40 & 25.255 & 0.017 & 23.749 & 0.014 & 1 & 1 & 1 & 1 & 0 & 0.9 & 0.4 & 273070.0 & -\\
  V81 & 2354.8 & 2818.4 & 12:34:16.32 & 08:10:29.00 & 25.22 & 0.021 & 24.019 & 0.018 & 0 & 1 & 1 & 1 & 0 & 0.5 & 0.2 & 271654.0 & -\\
  V82 & 2959.2 & 3031.6 & 12:34:14.20 & 08:10:24.15 & 25.194 & 0.02 & 24.237 & 0.021 & 0 & 1 & 0 & 1 & 0 & 0.3 & 0.6 & 274605.0 & -\\
  V83 & 2586.2 & 2925.3 & 12:34:15.45 & 08:10:28.40 & 25.169 & 0.017 & 24.331 & 0.023 & 0 & 0 & 1 & 0 & 0 & 0.6 & 0.6 &  & -\\
  V84 & 2930.6 & 3102.2 & 12:34:14.16 & 08:10:28.00 & 25.59 & 0.025 & 24.189 & 0.019 & 0 & 0 & 1 & 0 & 0 & 0.9 & 0.4 &  & -\\
  V85 & 2460.6 & 3430.8 & 12:34:15.07 & 08:10:53.90 & 25.319 & 0.019 & 25.211 & 0.055 & 0 & 0 & 0 & 1 & 0 & 0.5 & 0.6 & 275292.0 & -\\
  V86 & 3377.4 & 2209.6 & 12:34:14.17 & 08:09:38.90 & 25.79 & 0.025 & 24.886 & 0.033 & 1 & 1 & 1 & 1 & 0 & 0.8 & 0.2 &  & -\\
  V87 & 3075.6 & 2887.8 & 12:34:14.03 & 08:10:16.90 & 25.698 & 0.025 & 25.072 & 0.039 & 0 & 0 & 0 & 1 & 0 & 0.4 & 0.8 &  & -\\
  V88 & 3446.0 & 2871.3 & 12:34:13.04 & 08:10:07.20 & 25.657 & 0.022 & 25.445 & 0.056 & 0 & 1 & 0 & 1 & 0 & 0.6 & 0.8 & 267078.0 & -\\
  V89 & 3132.4 & 2726.2 & 12:34:14.14 & 08:10:06.46 & 26.101 & 0.037 & 25.509 & 0.061 & 0 & 1 & 1 & 1 & 0 & 0.6 & 0.8 & 267829.0 & -\\
  V90 & 3156.2 & 3128.5 & 12:34:13.44 & 08:10:24.40 & 26.716 & 0.067 & 25.492 & 0.054 & 0 & 1 & 1 & 1 & 0 & 0.6 & 1.0 &  & -\\
\hline

\end{tabular}}
\begin{tablenotes}
\item \textbf{Note.} -- Units of right ascension are hours, minutes, and seconds, and units of declination are degrees, arcminutes, and arcseconds. 
\\Numbers 1 or 0 in columns MAD, IQR and inverse von Neumann ratio show whether the star was characterized as a variable in the respective index.
\end{tablenotes}
\label{wf3vars}
\end{sidewaystable*}

\begin{sidewaystable*} [ht!]
%
%
%
%
\caption{Properties of newly discovered variable candidates in the WF4 chip ordered by average magnitudes.}
\scalebox{0.8}{
\begin{tabular}{l c c c c c c c c c c c c c c c c c c}
\hline
{ID} & {X}& {Y} &{RA} & {Dec} &  {V}& {$\rm \sigma$ $_{V}$} &{I}&{$\rm \sigma$ $_{I}$}&  {MAD$_{V}$} & {MAD$_{I}$} & {IQR$_{V}$} & {IQR $_{I}$} & {1/$\eta$ $_{V,I}$} & {$\Delta$V} & {$\Delta$I}&{HSC MatchID}&{Notes}  \\ 
%
%

& & & (J2000) & (J2000) & (mag) & (mag) & (mag)& (mag) & & &  &  & & (mag) & (mag) & \\
\hline

   V91 & 678.62 & 2393.6 & 12:34:21.97 & 08:10:47.80 & 19.9 & 0.001 & 19.721 & 0.002 & 0 & 0 & 0 & 0 & 1 & 0.1 & 0.12 & 274121.0 & candidate LBV\\
  V92 & 802.5 & 2744.7 & 12:34:21.30 & 08:10:57.90 & 21.399 & 0.002 & 20.68 & 0.003 & 0 & 0 & 0 & 0 & 1 & 0.24 & 0.12 & 273179.0 & candidate YSG/YHG\\
  V93 & 666.8 & 2375.6 & 12:34:22.04 & 08:10:47.30 & 22.575 & 0.005 & 21.854 & 0.005 & 1 & 0 & 0 & 1 & 0 & 0.3 & 0.1 & 277881.0 & -\\
  V94 & 685.0 & 2638.6 & 12:34:21.60 & 08:10:57.80 & 22.535 & 0.003 & 22.167 & 0.006 & 0 & 0 & 1 & 1 & 0 & 0.08 & 0.1 & 277200.0 & candidate YSG/YHG\\
  V95 & 1306.7 & 3128.3 & 12:34:18.98 & 08:11:06.60 & 22.641 & 0.004 & 22.231 & 0.006 & 0 & 0 & 0 & 0 & 1 & 0.13 & 0.12 & 274046.0 & candidate YSG/YHG\\
  V96 & 1305.7 & 2883.6 & 12:34:19.35 & 08:10:55.40 & 22.95 & 0.005 & 22.376 & 0.007 & 0 & 0 & 0 & 0 & 1 & 0.2 & 0.1 &  & -\\
  V97 & 1587.5 & 3283.2 & 12:34:17.47 & 08:11:12.20 & 23.101 & 0.006 & 23.105 & 0.012 & 1 & 1 & 0 & 0 & 0 & 0.2 & 0.2 & 277140.0 & -\\
  V98 & 731.7 & 2192.9 & 12:34:22.11 & 08:10:37.80 & 23.401 & 0.005 & 23.052 & 0.011 & 1 & 1 & 0 & 0 & 0 & 0.3 & 0.1 & 271930.0 & -\\
  V99 & 1676.5 & 2339.7 & 12:34:19.14 & 08:10:26.00 & 23.401 & 0.005 & 23.104 & 0.01 & 1 & 1 & 0 & 0 & 0 & 0.1 & 0.3 & 270248.0 & -\\
  V100 & 1714.8 & 2163.6 & 12:34:19.22 & 08:10:14.00 & 23.737 & 0.007 & 22.803 & 0.012 & 0 & 0 & 1 & 1 & 0 & 0.1 & 0.1 &  & -\\
  V101 & 738.3 & 2240.3 & 12:34:22.02 & 08:10:39.60 & 23.632 & 0.006 & 23.687 & 0.018 & 0 & 0 & 1 & 1 & 0 & 0.2 & 0.2 & 272910.0 & -\\
  V102 & 1430.1 & 3233.8 & 12:34:18.21 & 08:11:00.50 & 23.84 & 0.007 & 23.851 & 0.018 & 0 & 0 & 1 & 1 & 0 & 0.3 & 0.2 & 271722.0 & -\\
  V103 & 2014.1 & 2431.4 & 12:34:17.92 & 08:10:19.60 & 25.915 & 0.039 & 22.93 & 0.009 & 0 & 0 & 0 & 1 & 1 & 0.5 & 0.1 & 277155.0 & -\\
  V104 & 844.3 & 2760.5 & 12:34:20.91 & 08:11:00.30 & 23.934 & 0.009 & 23.782 & 0.019 & 1 & 0 & 1 & 0 & 0 & 0.4 & 0.2 & 428271.0 & -\\
  V105 & 979.2 & 2976.9 & 12:34:20.19 & 08:11:07.00 & 24.129 & 0.01 & 23.479 & 0.015 & 1 & 1 & 1 & 1 & 0 & 0.5 & 0.4 & 428369.0 & -\\
  V106 & 645.9 & 3406.8 & 12:34:20.58 & 08:11:32.90 & 24.644 & 0.02 & 23.88 & 0.024 & 1 & 1 & 1 & 1 & 0 & 0.6 & 0.4 &  & -\\
  V107 & 684.9 & 2672.3 & 12:34:21.61 & 08:10:55.80 & 25.761 & 0.046 & 23.546 & 0.017 & 0 & 1 & 0 & 1 & 0 & 0.2 & 0.3 &  & candidate RSG\\
  V108 & 1258.3 & 2947.4 & 12:34:19.40 & 08:10:58.40 & 24.672 & 0.014 & 24.2 & 0.031 & 0 & 1 & 0 & 0 & 0 & 0.4 & 0.4 & 274124.0 & -\\
  V109 & 1166.8 & 3620.4 & 12:34:18.65 & 08:11:31.30 & 24.662 & 0.016 & 24.129 & 0.032 & 0 & 1 & 0 & 1 & 0 & 0.4 & 0.9 &  & -\\
  V110 & 716.4 & 2539.8 & 12:34:21.64 & 08:10:53.40 & 24.616 & 0.015 & 24.468 & 0.035 & 1 & 1 & 0 & 1 & 0 & 0.4 & 0.4 & 428368.0 & -\\
  V111 & 1909.5 & 3327.8 & 12:34:16.87 & 08:11:00.80 & 24.815 & 0.014 & 23.907 & 0.018 & 0 & 0 & 1 & 1 & 0 & 0.25 & 0.8 & 274398.0 & -\\
  V112 & 679.5 & 2161.0 & 12:34:22.32 & 08:10:37.50 & 24.686 & 0.015 & 24.732 & 0.044 & 1 & 1 & 0 & 0 & 0 & 0.3 & 0.25 &  & -\\
  V113 & 776.5 & 3459.8 & 12:34:20.06 & 08:11:32.80 & 24.979 & 0.024 & 24.568 & 0.044 & 0 & 0 & 0 & 1 & 0 & 0.2 & 0.9 &  & -\\
  V114 & 1396.5 & 2788.1 & 12:34:19.22 & 08:10:49.30 & 25.238 & 0.02 & 24.355 & 0.027 & 0 & 0 & 0 & 1 & 0 & 0.8 & 0.75 & 2720288.0 & -\\
  V115 & 936.1 & 2764.9 & 12:34:20.63 & 08:10:58.40 & 25.028 & 0.022 & 24.699 & 0.045 & 0 & 1 & 1 & 1 & 0 & 0.5 & 0.8 &  & -\\
  V116 & 1284.8 & 2517.7 & 12:34:20.56 & 08:10:36.70 & 25.368 & 0.031 & 24.577 & 0.037 & 1 & 1 & 1 & 0 & 0 & 1.2 & 0.6 & 277159.0 & -\\
  V117 & 1049.3 & 2199.9 & 12:34:21.17 & 08:10:30.10 & 25.335 & 0.023 & 24.701 & 0.036 & 1 & 1 & 0 & 1 & 0 & 0.4 & 0.8 &  & -\\
  V118 & 1675.0 & 2201.2 & 12:34:19.28 & 08:10:16.80 & 25.616 & 0.03 & 24.489 & 0.031 & 0 & 1 & 1 & 1 & 0 & 0.5 & 1.2 &  & -\\
  V119 & 860.8 & 2549.6 & 12:34:21.18 & 08:10:50.60 & 25.584 & 0.031 & 24.962 & 0.056 & 0 & 1 & 0 & 1 & 0 & 0.4 & 0.8 &  & -\\
  V120 & 1285.1 & 2760.5 & 12:34:19.53 & 08:10:51.00 & 25.996 & 0.05 & 24.738 & 0.037 & 0 & 1 & 1 & 1 & 0 & 1.0 & 0.5 &  & -\\

\hline

\end{tabular}}
\begin{tablenotes}
\item \textbf{Note.} -- Units of right ascension are hours, minutes, and seconds, and units of declination are degrees, arcminutes, and arcseconds. 
\\Numbers 1 or 0 in columns MAD, IQR and inverse von Neumann ratio show whether the star was characterized as a variable in the respective index.
\end{tablenotes}
\label{wf4svars}
\end{sidewaystable*}

\

\begin{figure*}[h!tb]
\centering
\begin{tabular}{  l   c }
\includegraphics[scale=0.8]{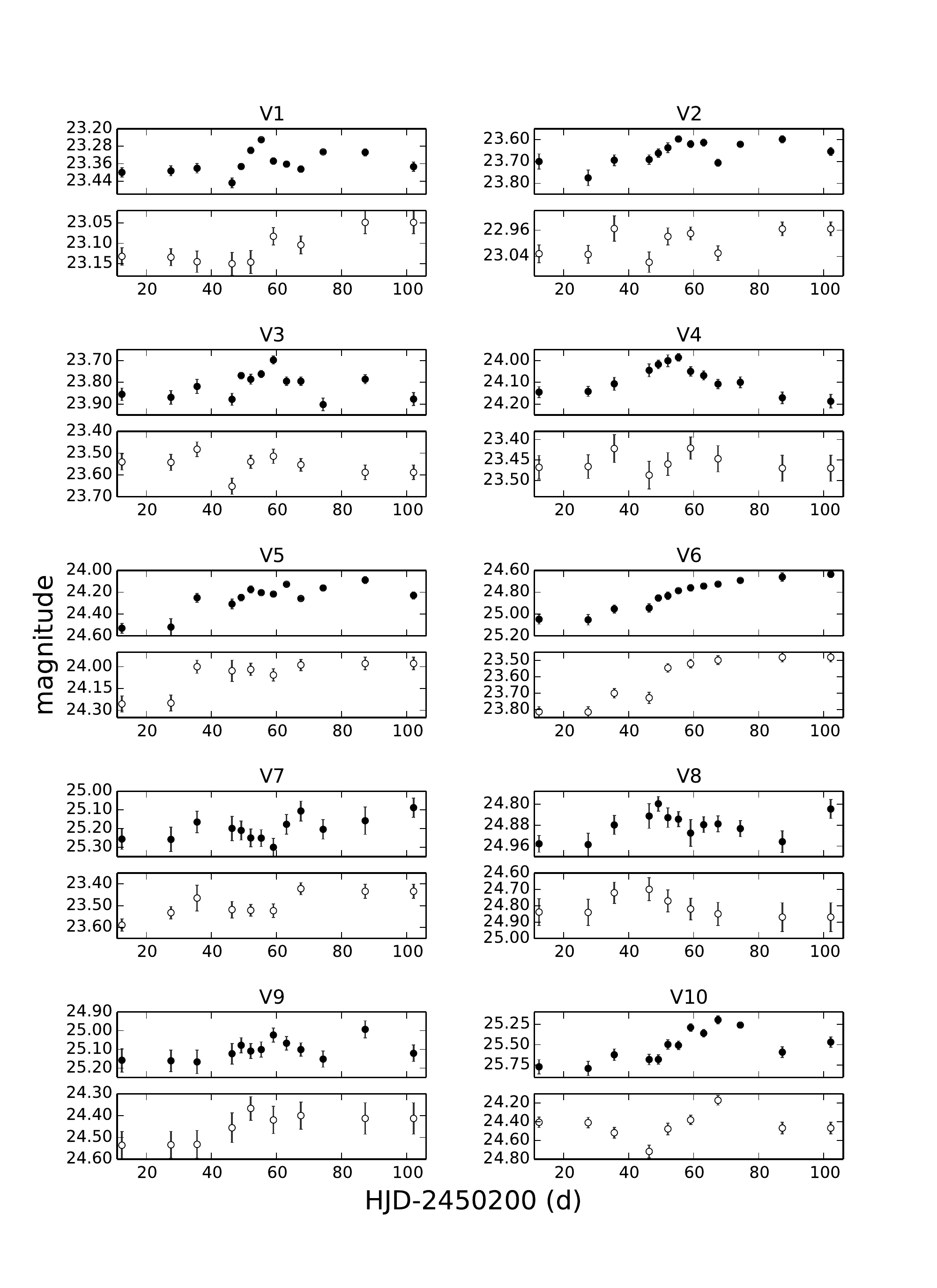}&
\end{tabular}
\caption{Light curves of the candidate variables in the PC chip in the V filter (filled circles) and I filter (open circles).}
\label{lcpc1}
\end{figure*}

 \begin{figure*}[h!tb]
 \setcounter{figure}{9}   
\centering
\begin{tabular}{  l   c }
\includegraphics[scale=0.8]{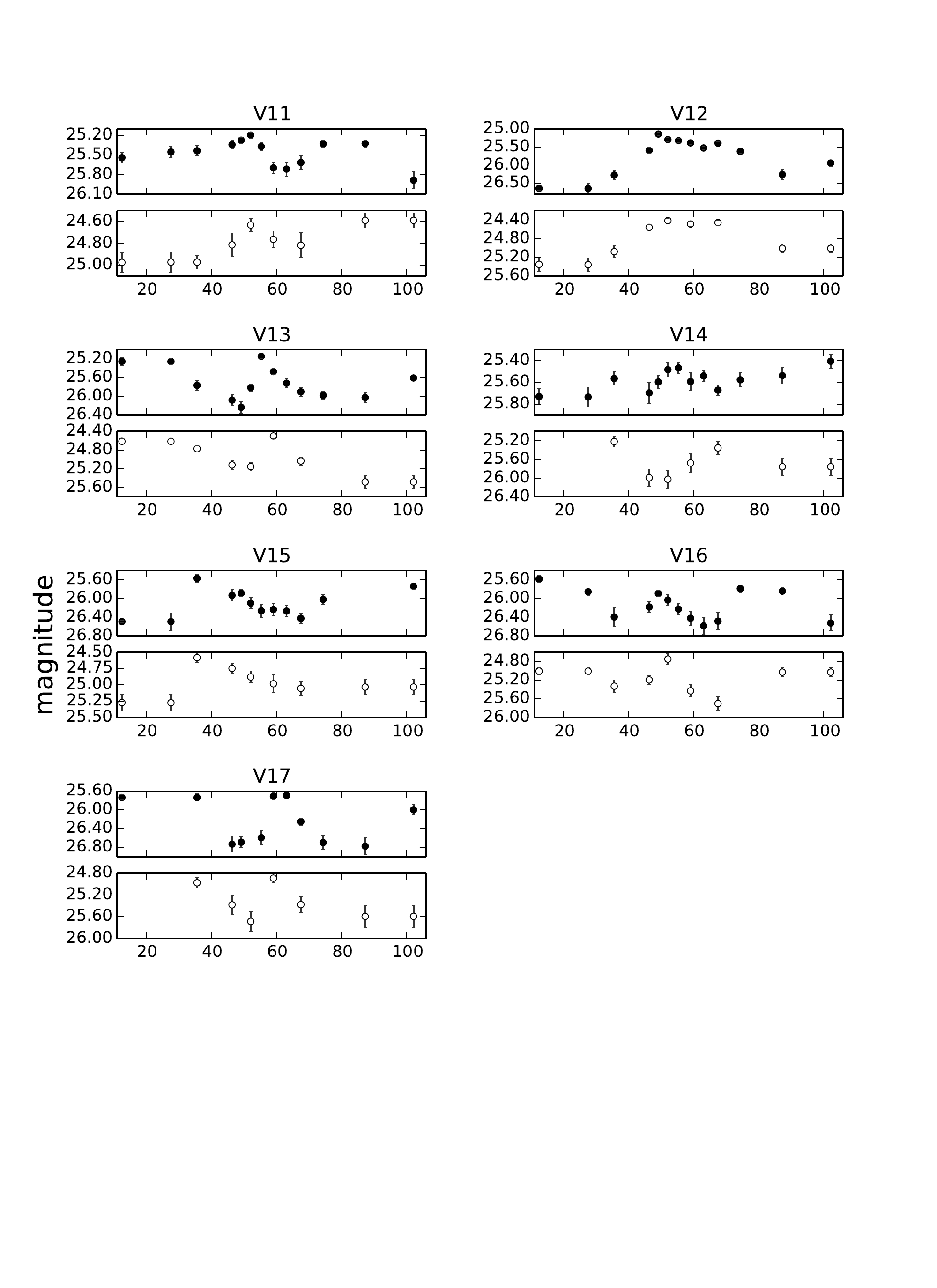}&
\end{tabular}
\caption{continued}
\label{lcpc2}
\end{figure*}
 
\clearpage
 
\begin{figure*}[h!tb]
\setcounter{figure}{10}  

\centering

\begin{tabular}{ l c c c c c}
\includegraphics[scale=0.8]{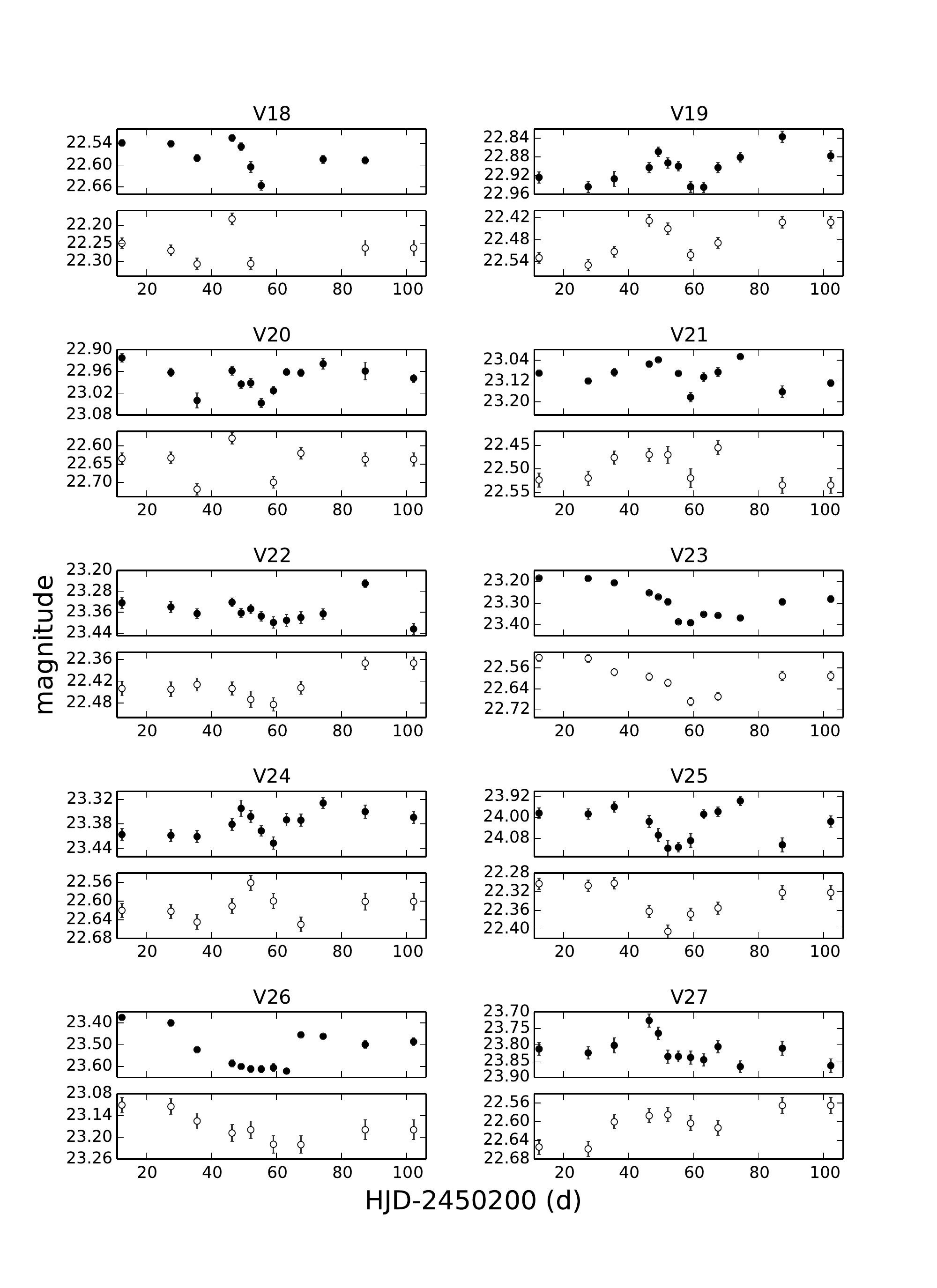}\\
\end{tabular}

\caption{Same as Figure~\ref{lcpc1} but for the candidate variables in the WF2 chip.}
\label{lcwf21}
\end{figure*}

\begin{figure*}[h!tb]
\setcounter{figure}{10}  

\centering

\begin{tabular}{ l c c c c c}
\includegraphics[scale=0.8]{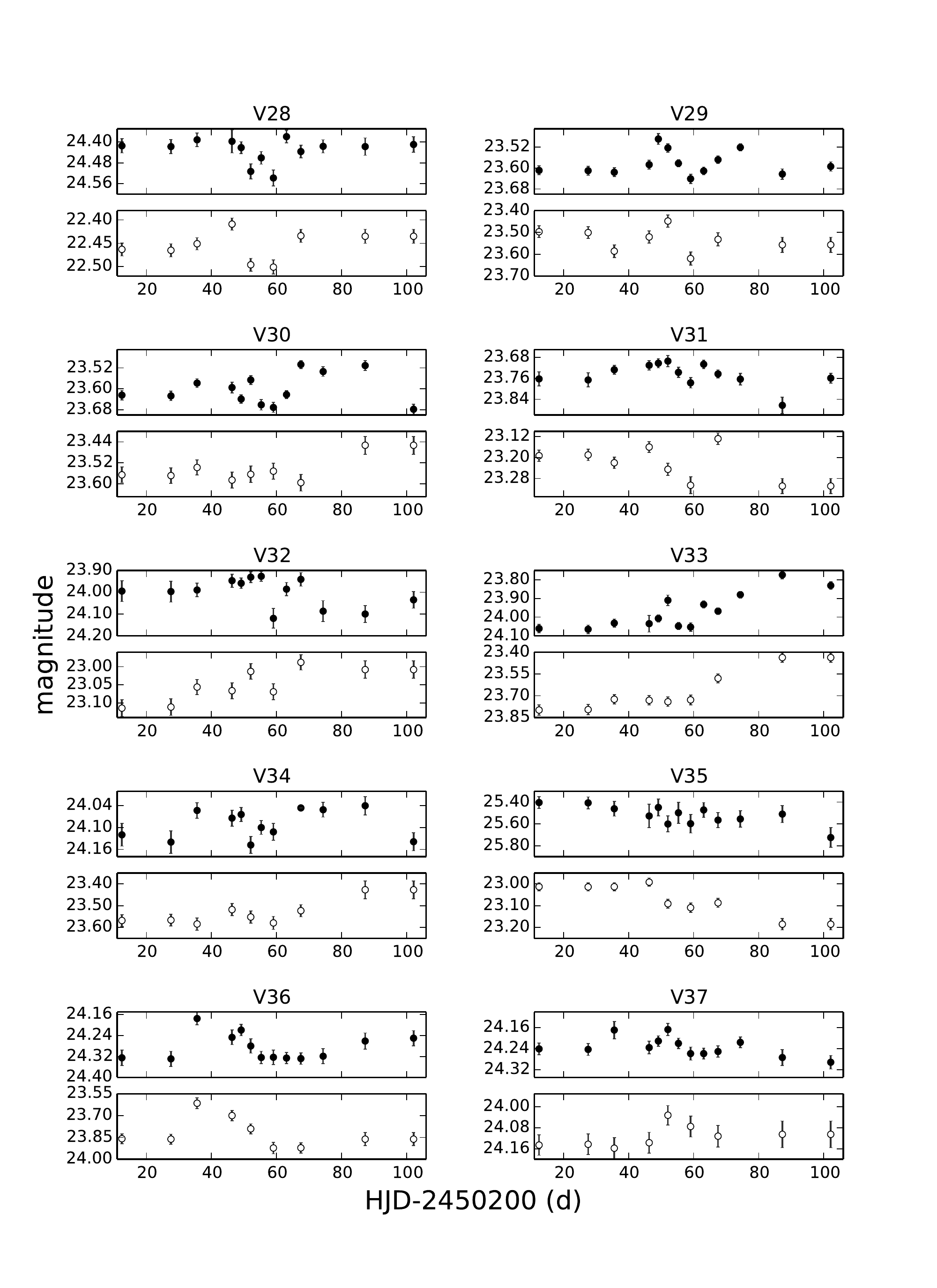} \\
\end{tabular}

\caption{continued}
\label{lcwf22}
\end{figure*}

\begin{figure*}[h!tb]
\setcounter{figure}{10}  

\centering

\begin{tabular}{ l c c c c c}
\includegraphics[scale=0.8]{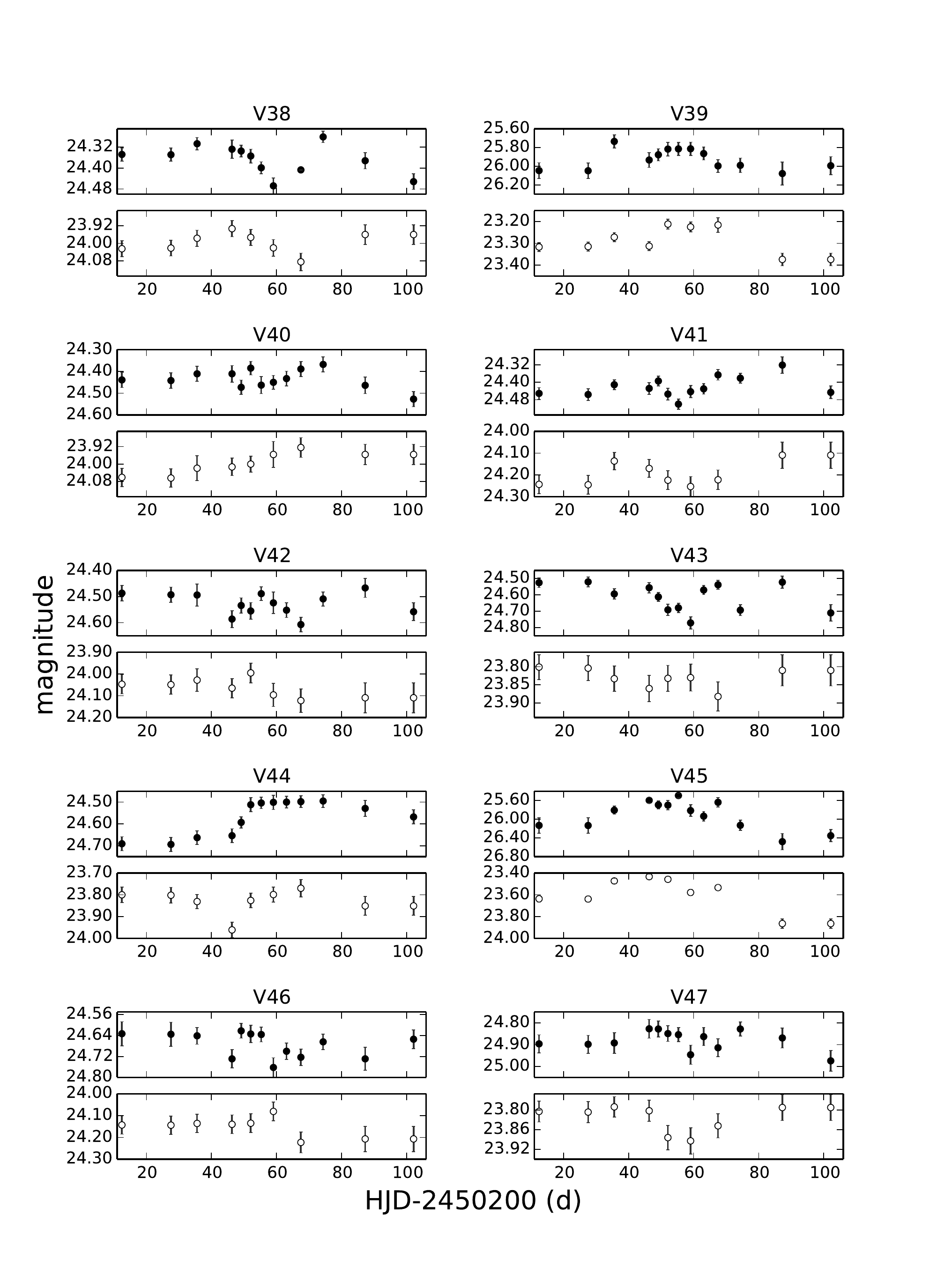}\\
\end{tabular}

\caption{continued}
\label{lcwf23}
\end{figure*}

\begin{figure*}[h!tb]
\setcounter{figure}{10}  

\centering

\begin{tabular}{ l c c c c c}
\includegraphics[scale=0.8]{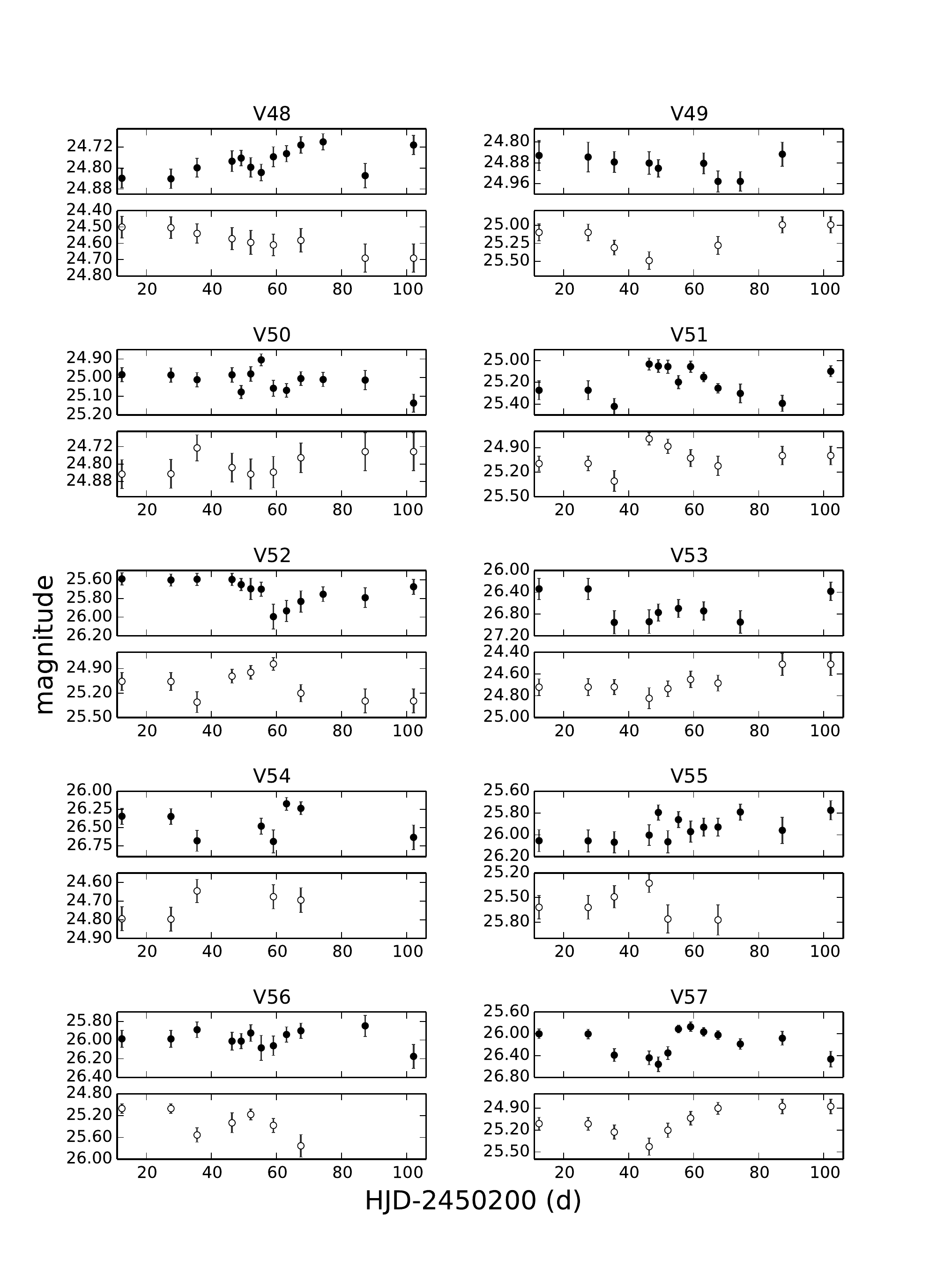}\\
\end{tabular}

\caption{continued}
\label{lcwf24}
\end{figure*}

\begin{figure*}[h!tb]
\setcounter{figure}{10}  

\centering
\begin{tabular}{l c}
\includegraphics[scale=0.8]{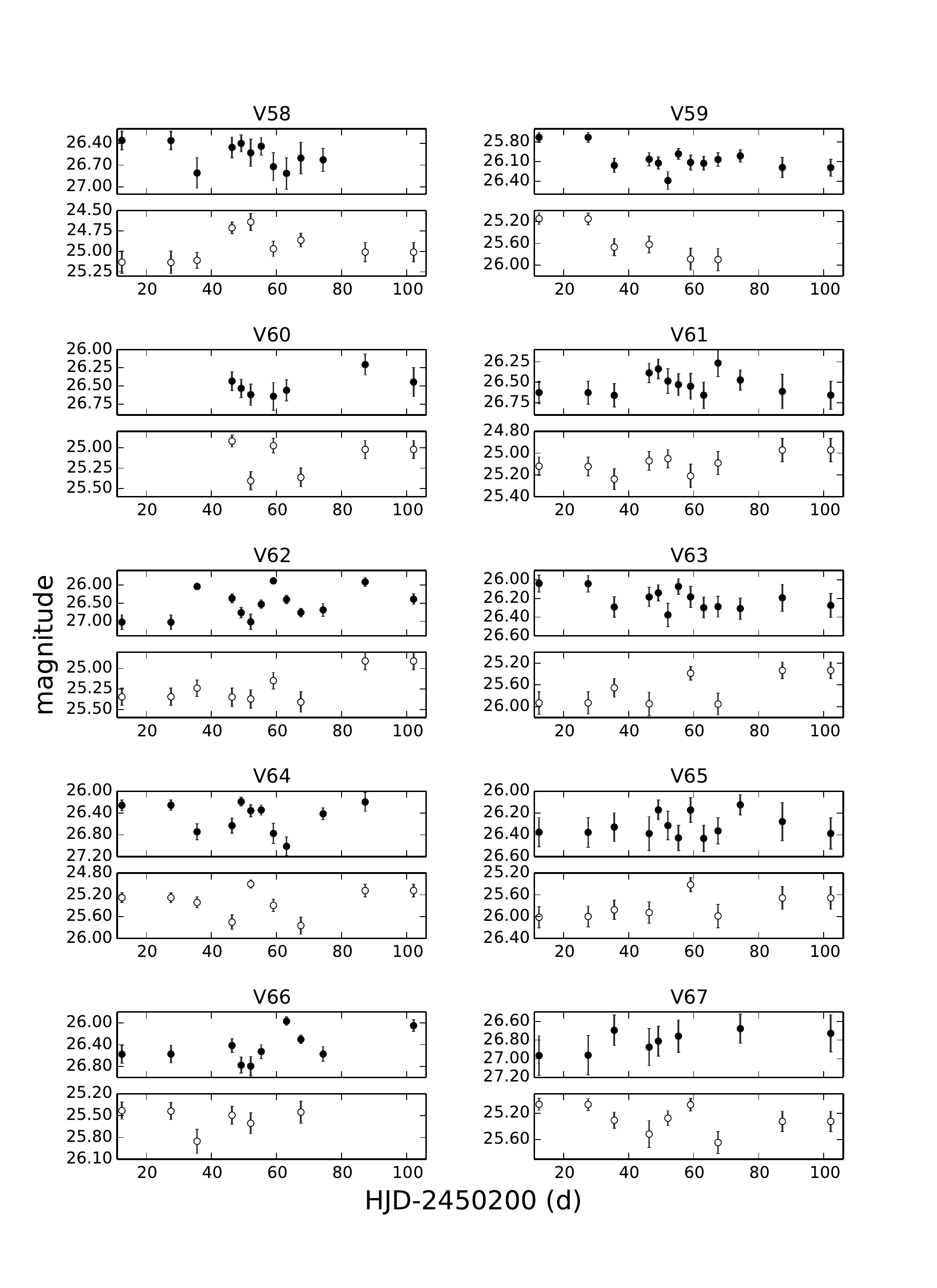}&
\\

\end{tabular}

\caption{continued}
\label{lcwf25}
\end{figure*}

\begin{figure*}[h!tb]
\setcounter{figure}{11} 

\centering
\begin{tabular}{ l   c  c c }
\includegraphics[scale=0.8]{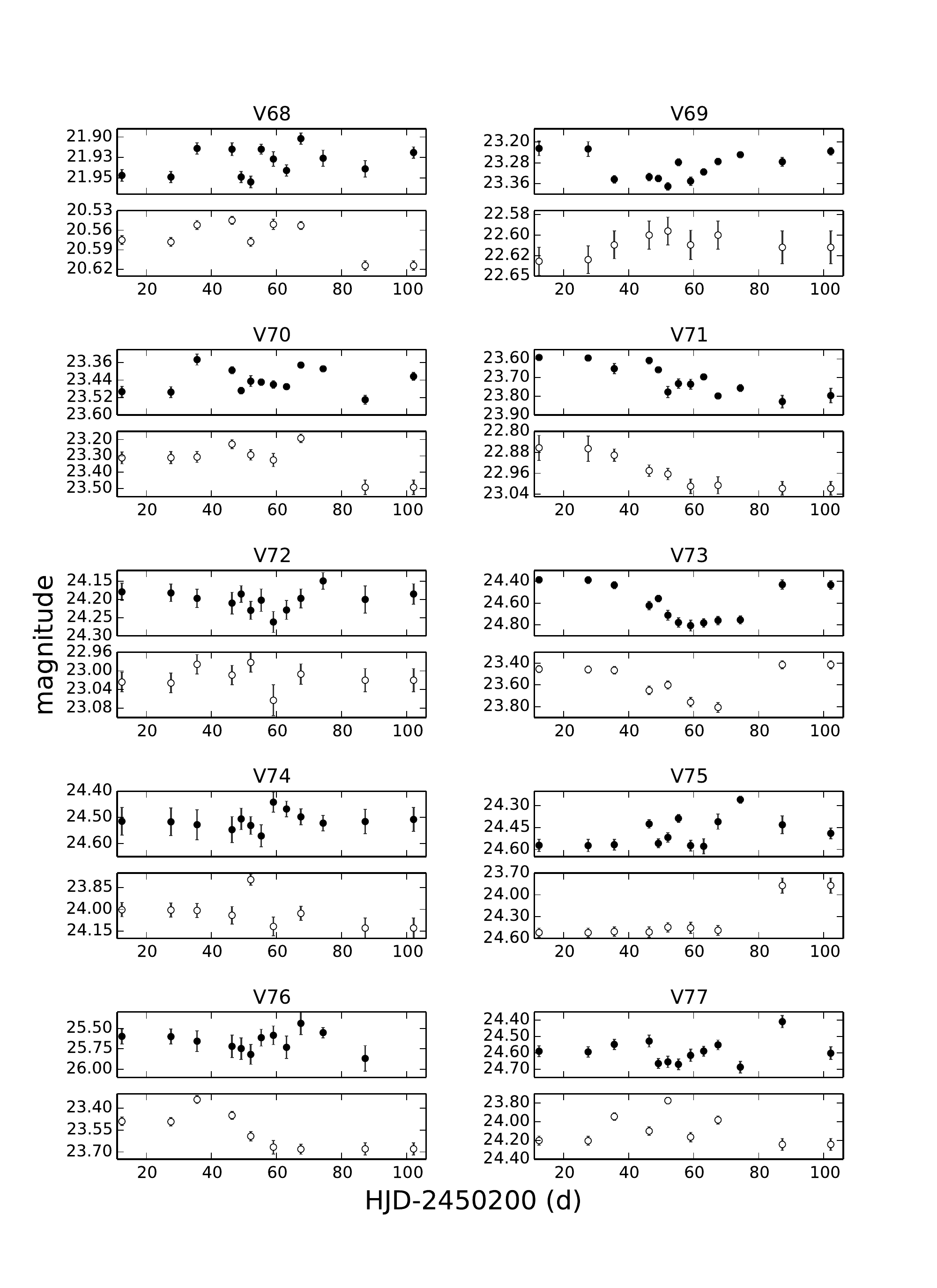}
\end{tabular}
\caption{Same as Figure~\ref{lcpc1} but for the candidates variables in the WF3 chip.}
\label{lcwf31}
\end{figure*}

\begin{figure*}[h!tb]
\setcounter{figure}{11} 

\centering
\begin{tabular}{ l   c  c c }
\includegraphics[scale=0.8]{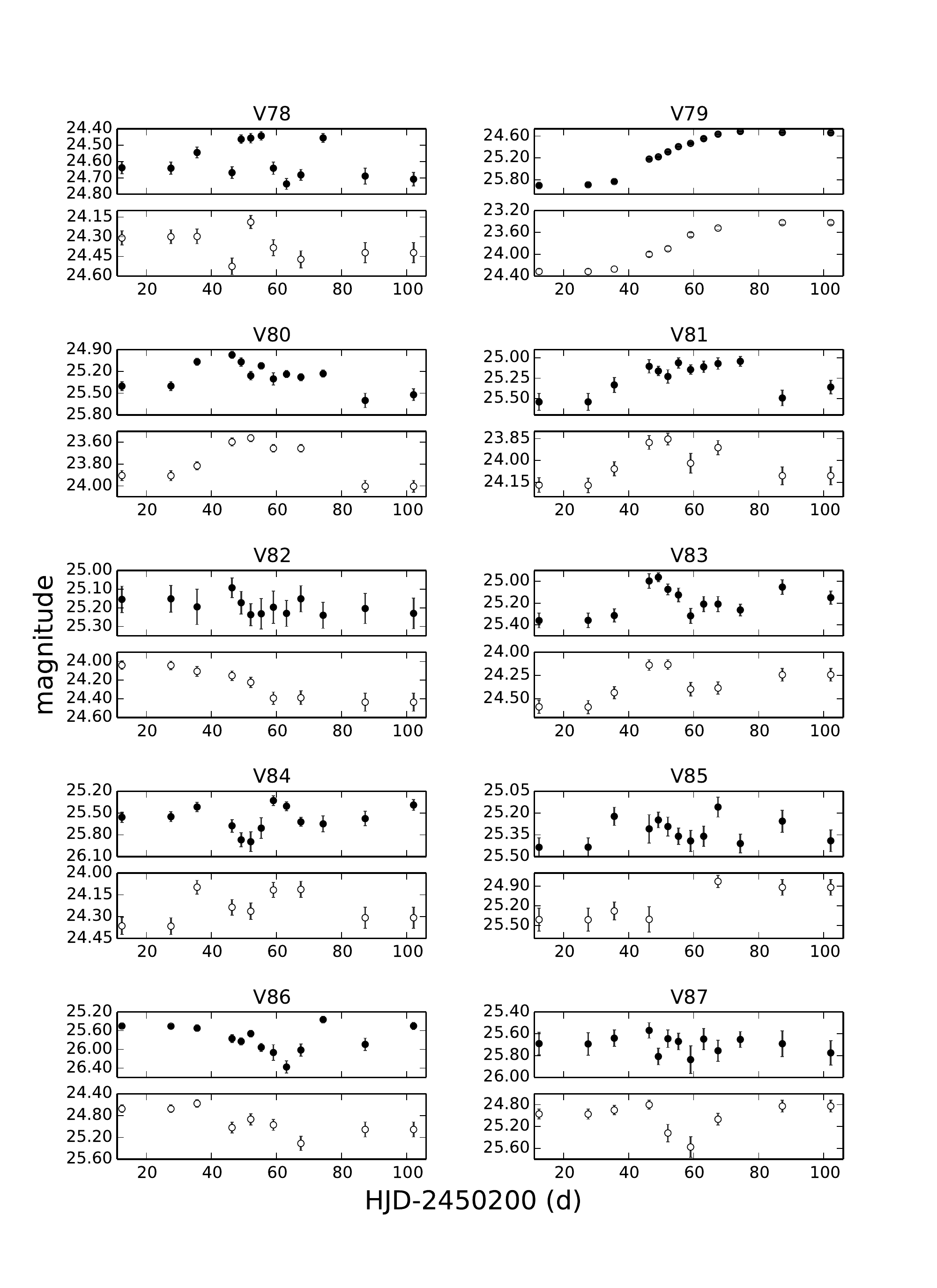} \\

\end{tabular}
\caption{continued}
\label{lcwf32}
\end{figure*}

\begin{figure*}[h!tb]
\setcounter{figure}{11} 

\centering
\begin{tabular}{ l   c  c c }
\includegraphics[scale=0.5]{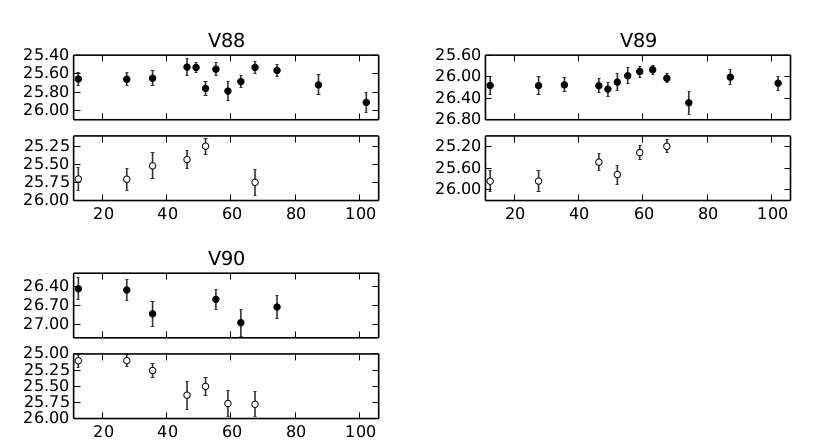} \\
\end{tabular}
\caption{continued}
\label{lcwf33}
\end{figure*}

\begin{figure*}[h!tb]
\setcounter{figure}{12} 

\centering
\begin{tabular}{  l   c  c c c }
\includegraphics[scale=0.8]{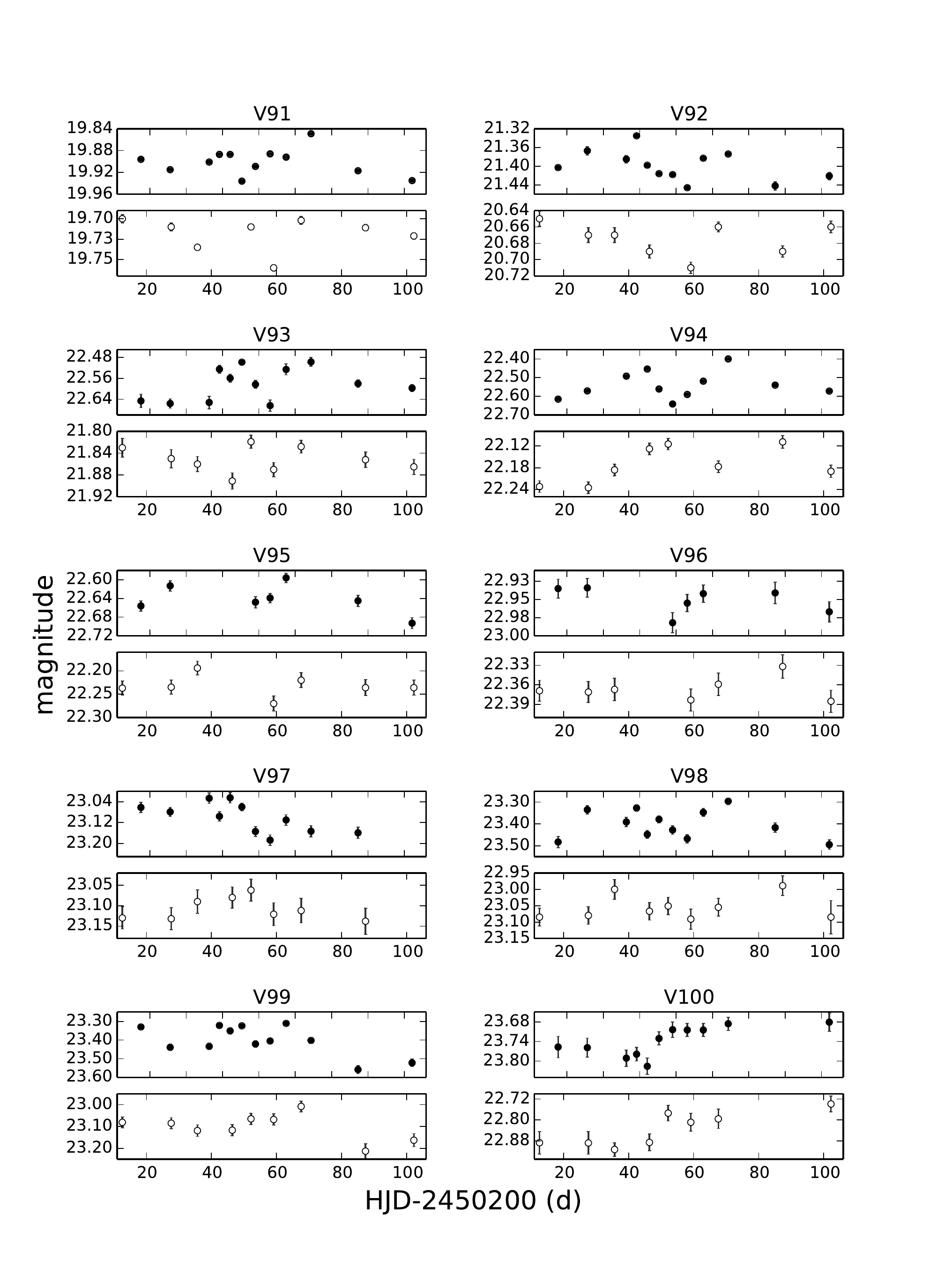} &

\end{tabular}
\caption{Same as Figure~\ref{lcpc1} but for the candidates variables in the WF4 chip.}
\label{lcwf41}
\end{figure*}

\begin{figure*}[h!tb]
\setcounter{figure}{12}

\centering
\begin{tabular}{  l   c  c c c }
\includegraphics[scale=0.8]{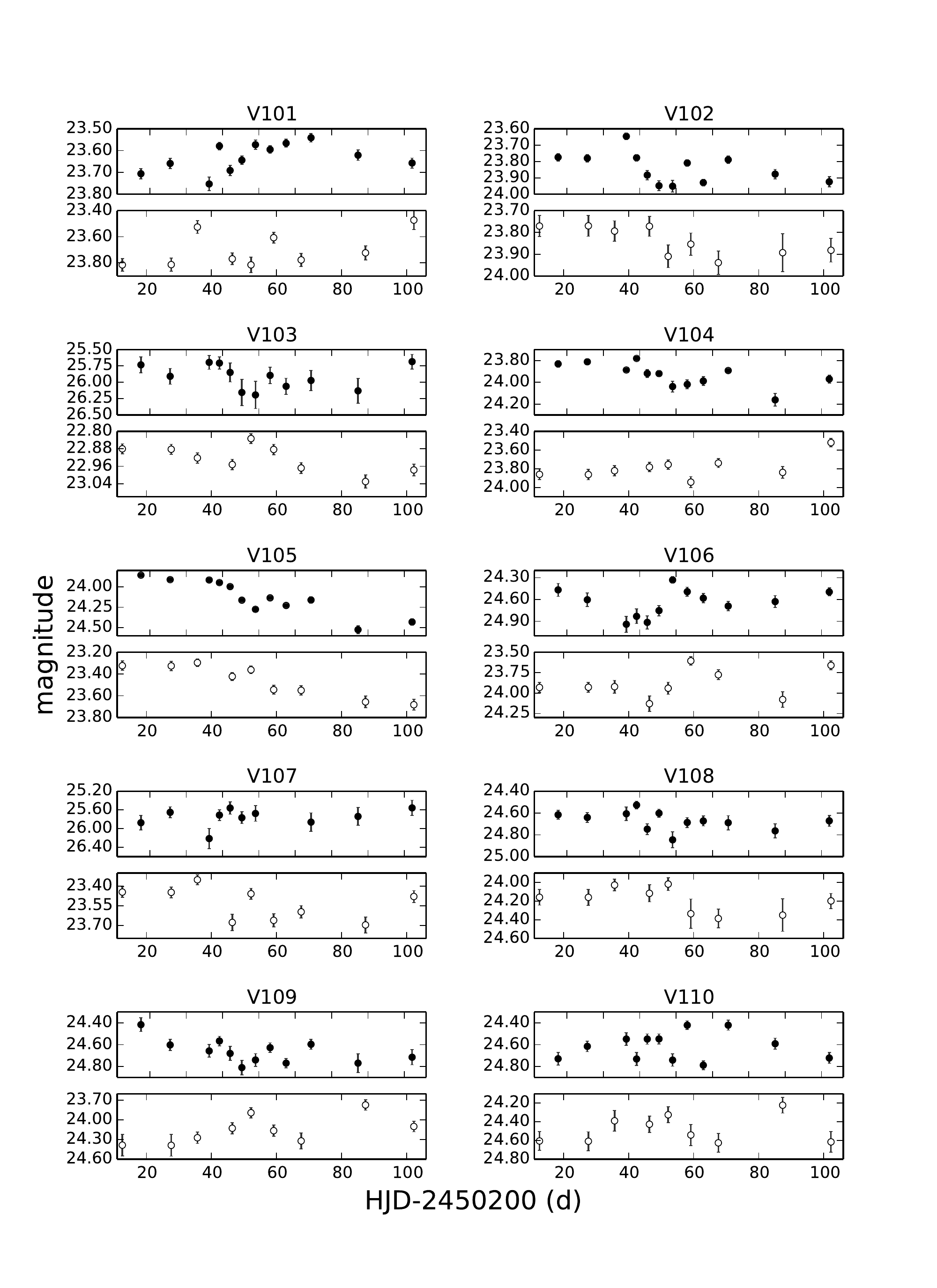} \\
\end{tabular}
\caption{continued}
\label{lcwf42}
\end{figure*}

\begin{figure*}[h!tb]
\setcounter{figure}{12} 

\centering
\begin{tabular}{  l c  c c c }

\includegraphics[scale=0.8]{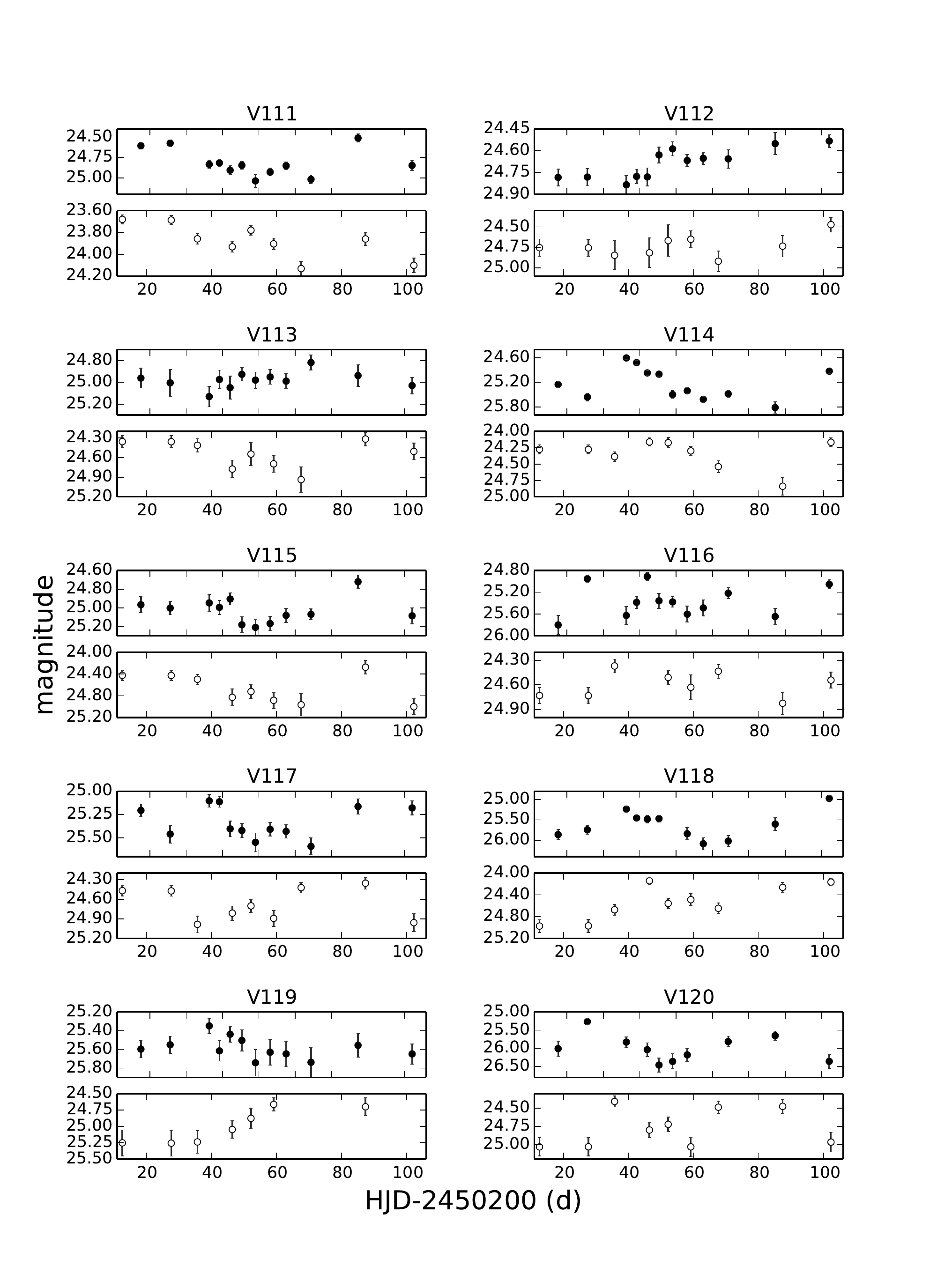} &
\setcounter{figure}{12} 
\end{tabular}
\caption{continued}
\label{lcwf43}
\end{figure*}

\clearpage
\begin{appendices}
\section{Appendix}

\begin{figure}[h!tb]
\vspace{2cm} 

\begin{tabular}{l c c }
\hspace{-10cm}
\includegraphics[trim=0cm 0cm 0.cm 0cm, width=0.5\textwidth]{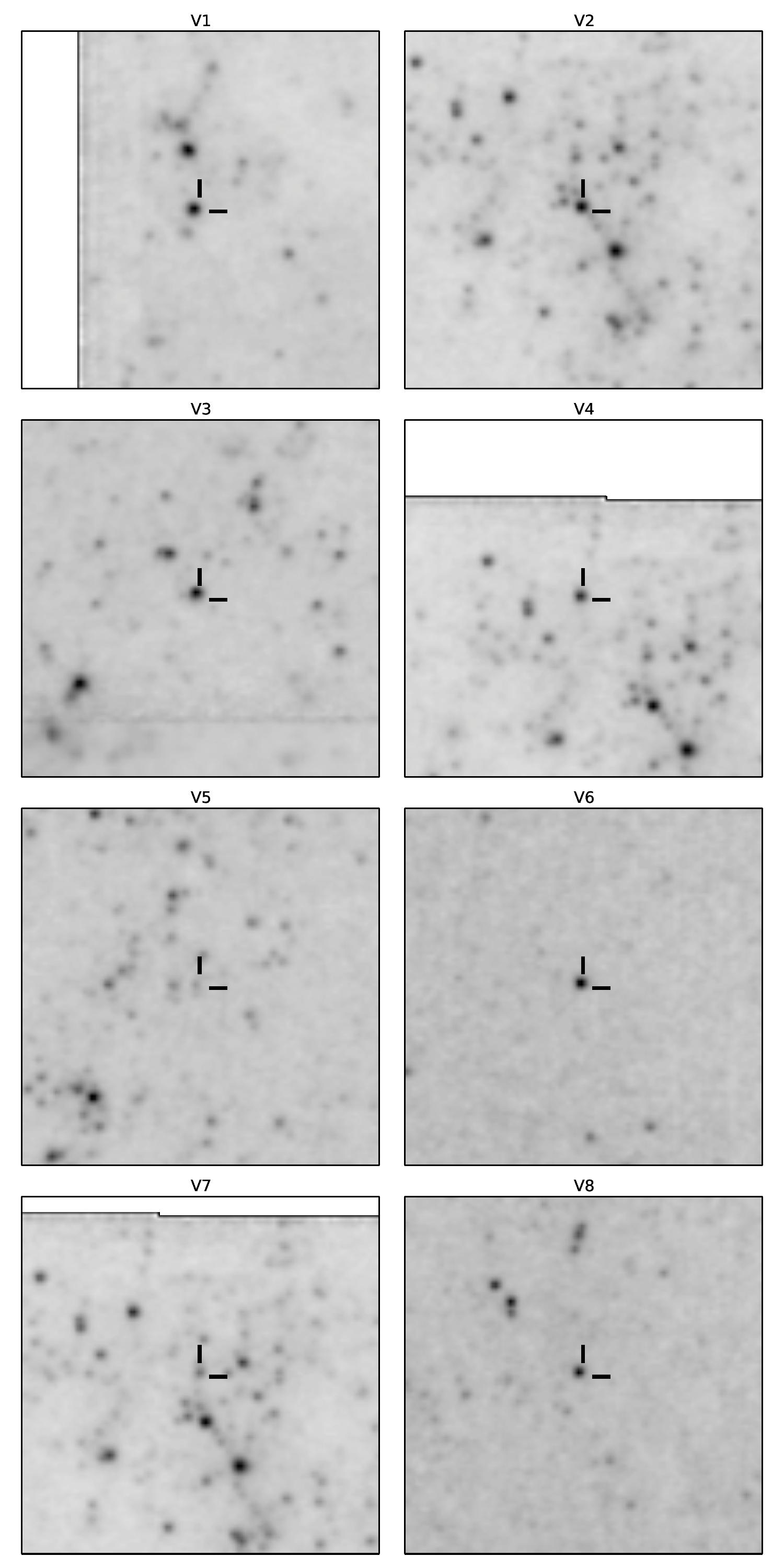} &
\includegraphics[trim=0cm 0cm 0.cm 0cm, width=0.5\textwidth]{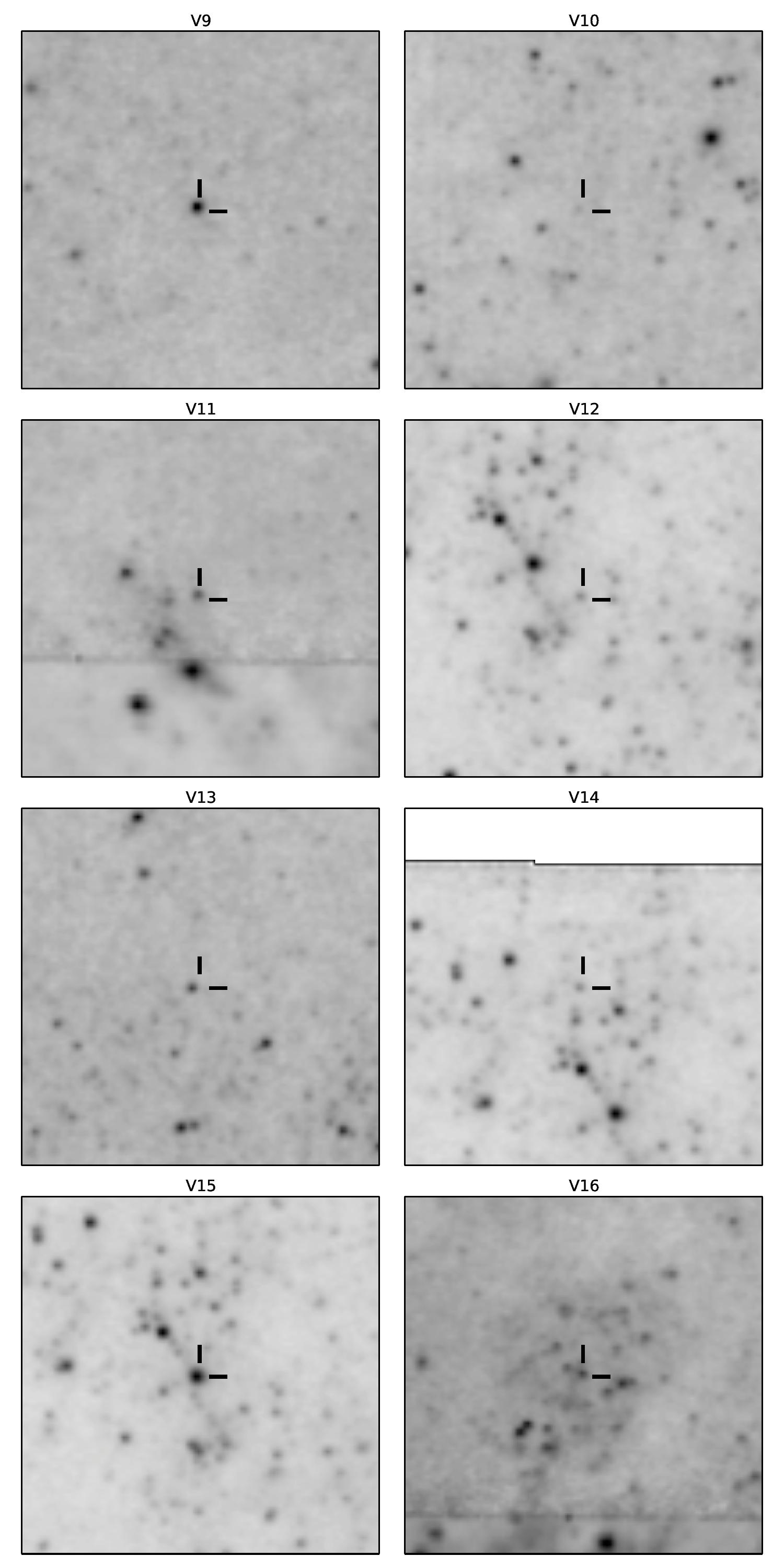}& \\
\hspace{-10cm}
\includegraphics[trim=0cm 0cm 0.cm 0cm, width=0.5\textwidth]{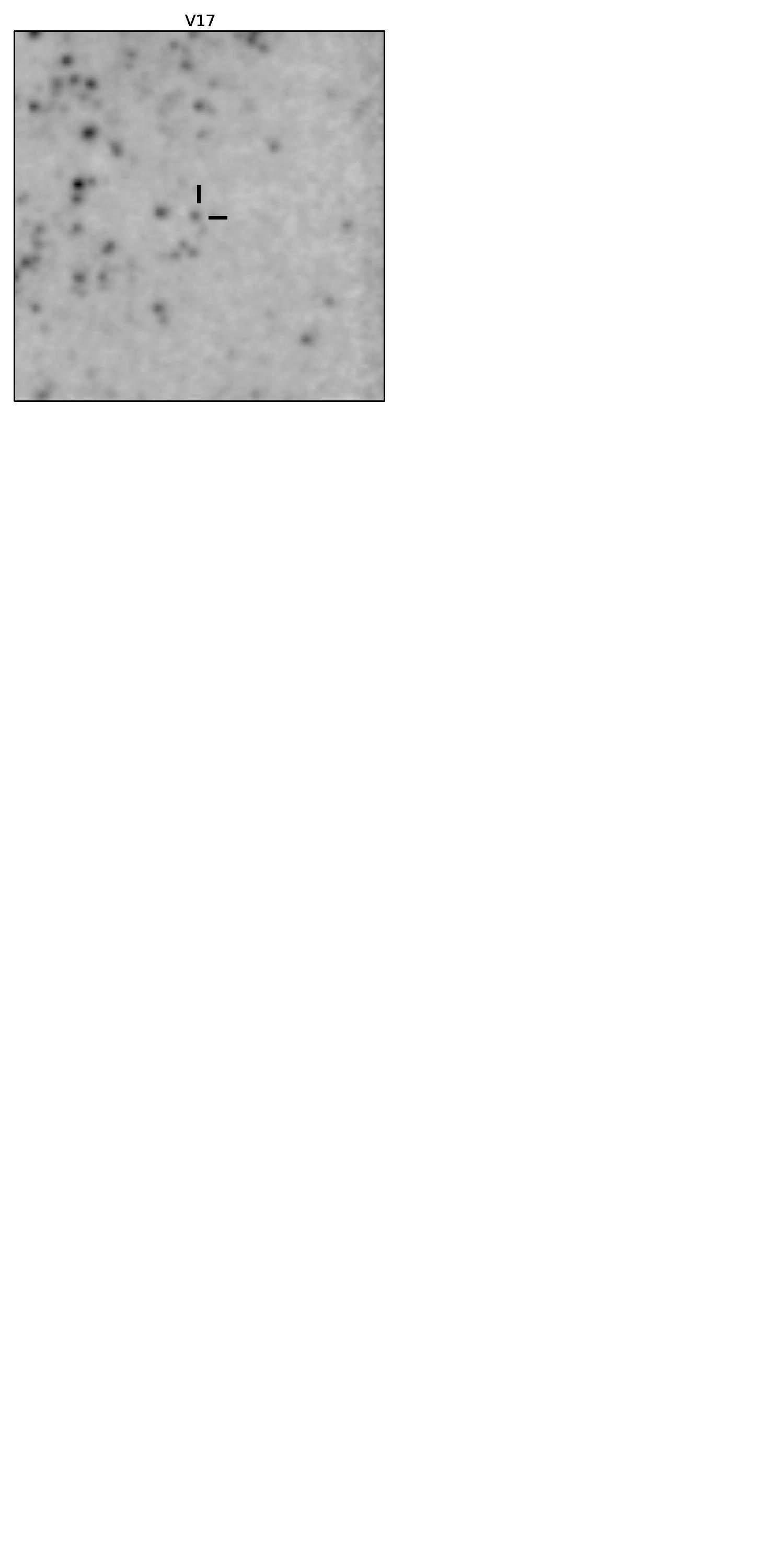}
\setcounter{figure}{13}
\end{tabular}
\caption{Finding charts for each variable star in the PC chip. The charts encompass a 50 $\times$ 50 arcseconds area around each variable.}
\end{figure}

\begin{figure*}[h!tb]
\centering

\begin{tabular}{l c c c c c c c}
\includegraphics[trim=0cm 0cm 0.cm 0cm, width=0.5 \textwidth]{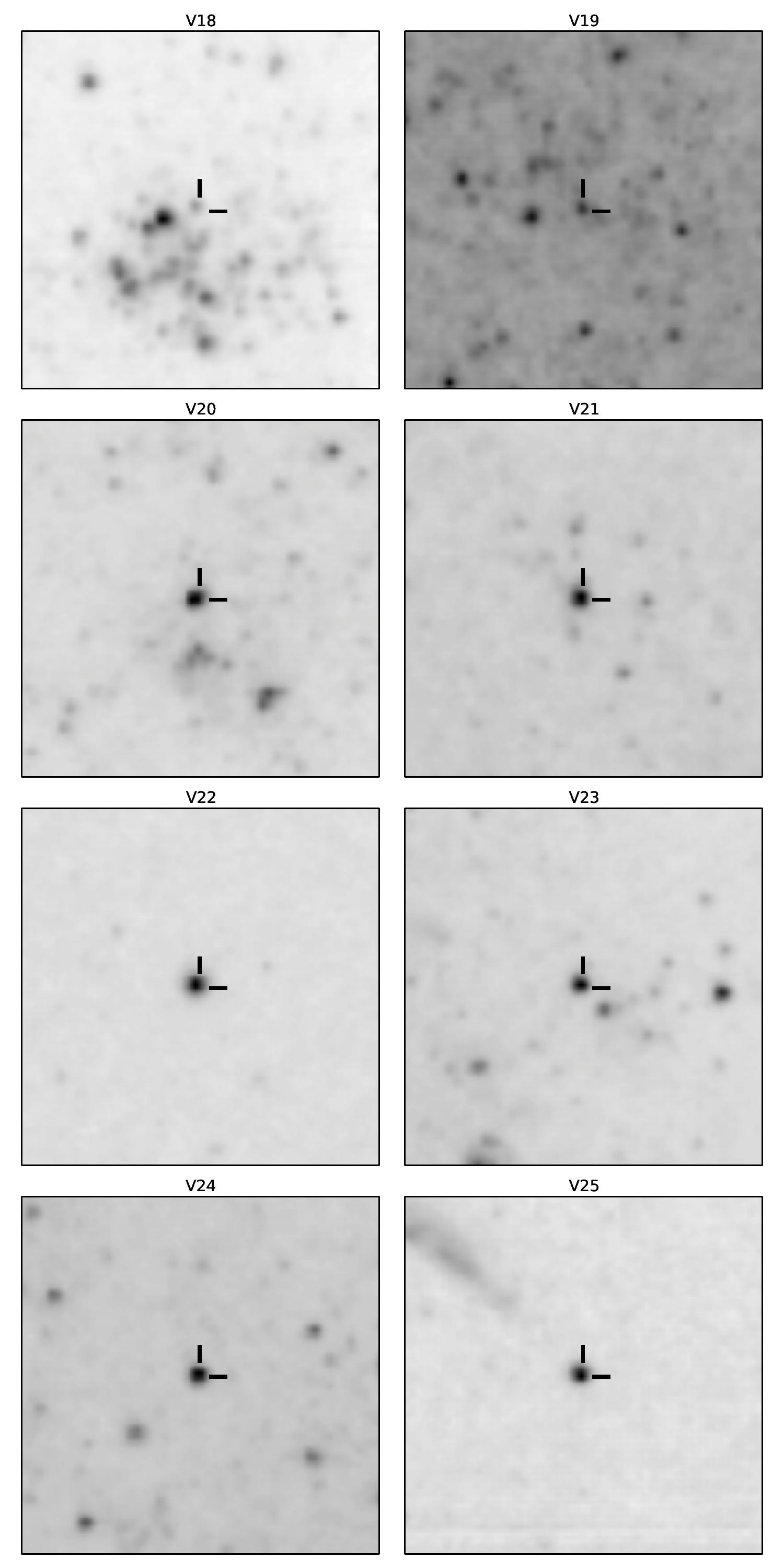} &
\includegraphics[trim=0cm 0cm 0.cm 0cm, width=0.5 \textwidth]{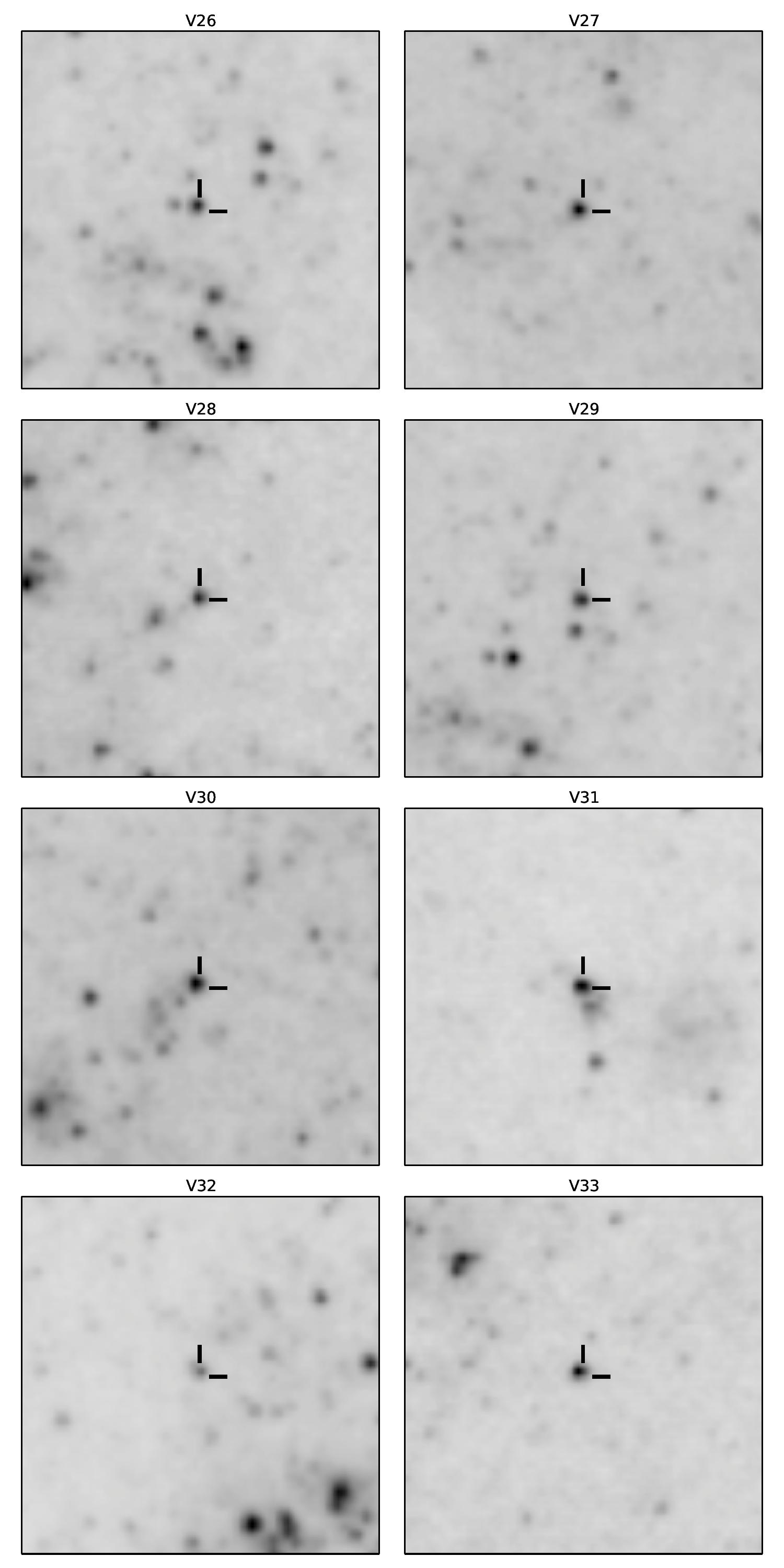}& \\
\end{tabular}
\caption{Finding charts for each candidate variable star in the PC and WF2 chips. The charts encompass a 50 $\times$ 50 arcseconds area around each variable.}
\label{chartspc}
\end{figure*}

\begin{figure*}[h!tb]
\begin{tabular}{l c c c c c c c}
\includegraphics[trim=0cm 0cm 0.cm 0cm, width=0.50\textwidth]{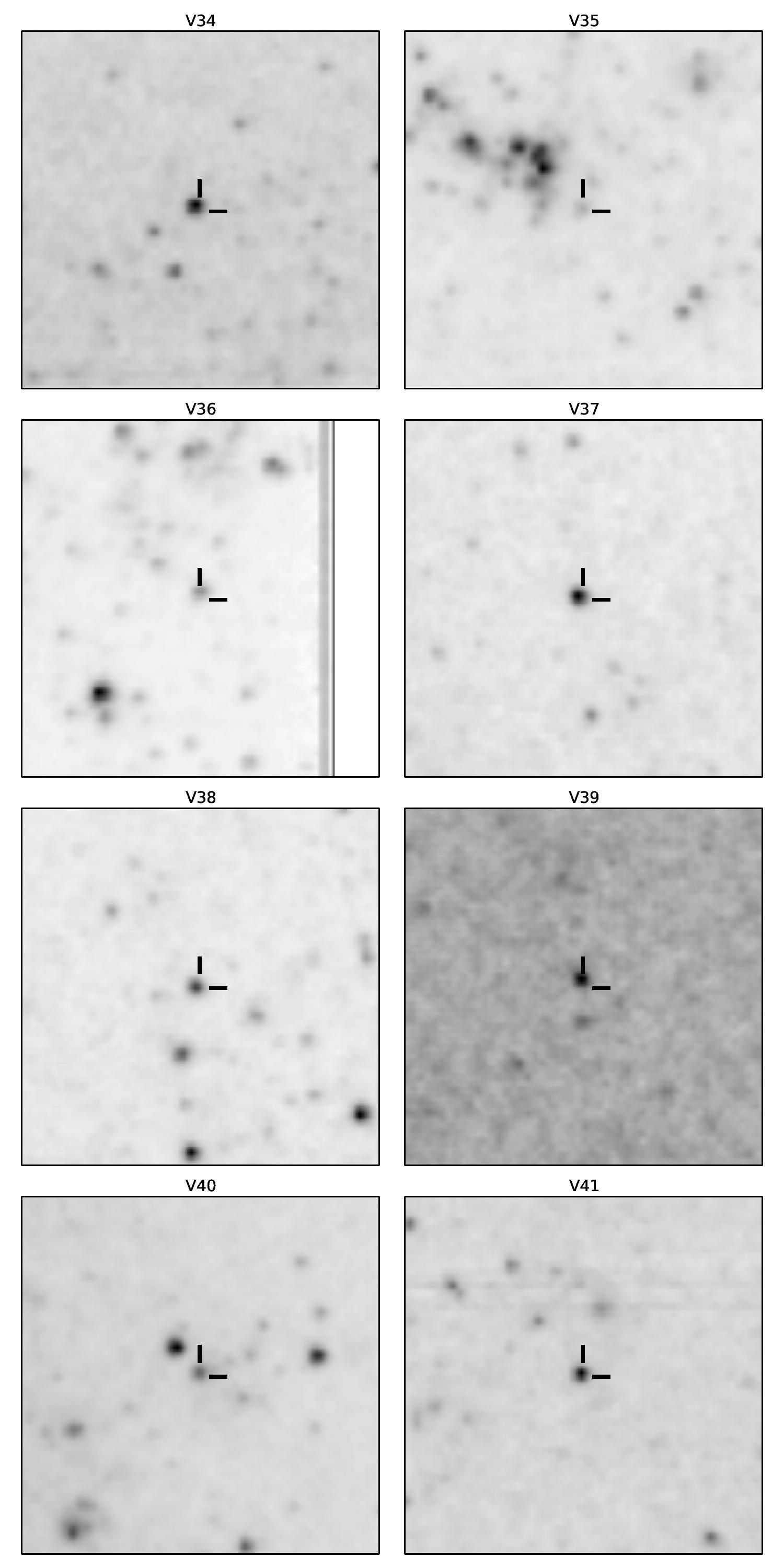} &
\includegraphics[trim=0cm 0cm 0.cm 0cm, width=0.50\textwidth]{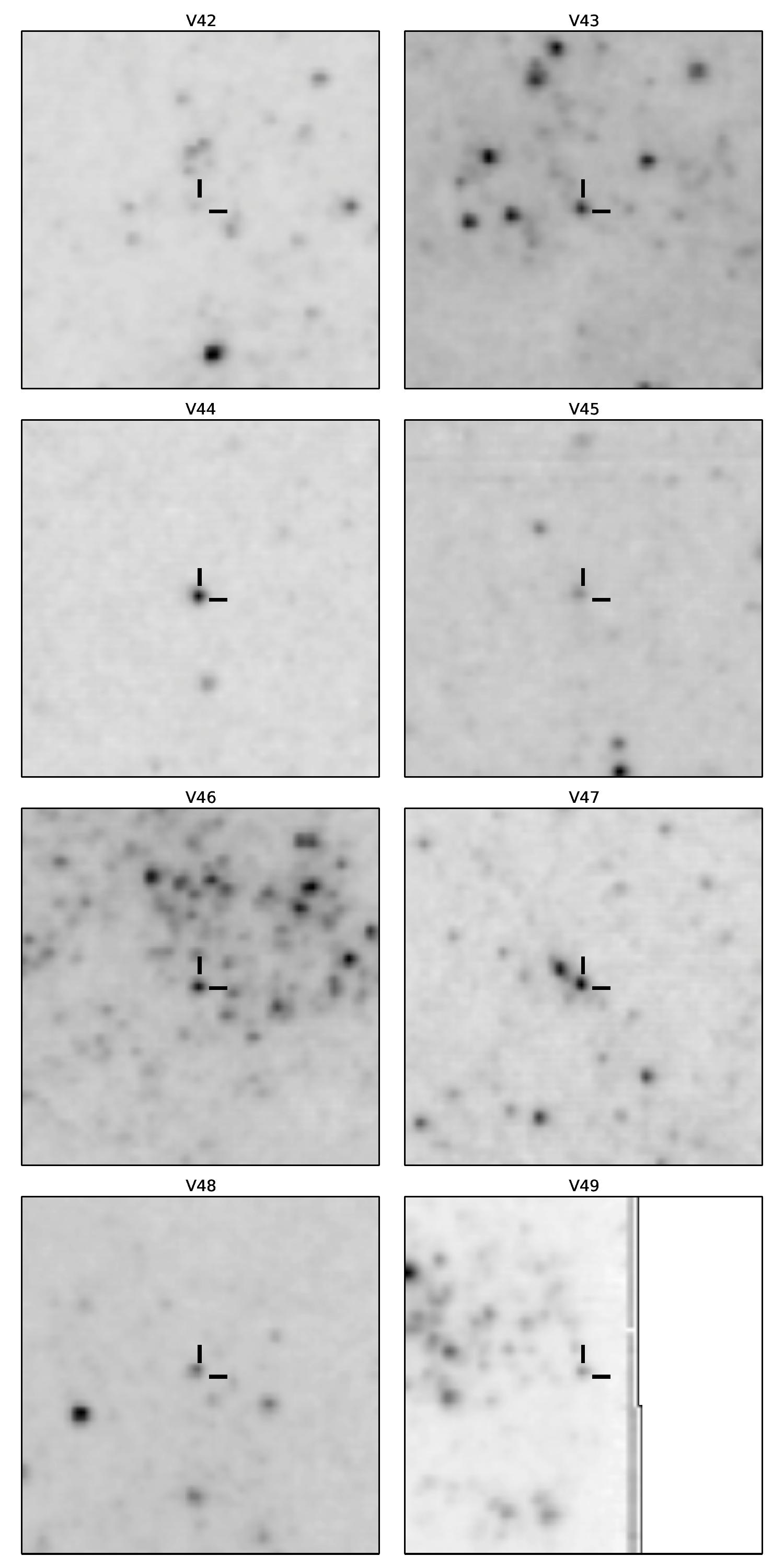}& \\
\setcounter{figure}{14}

\end{tabular}

\caption{continued}
\end{figure*}

\begin{figure*}[h!tb]
\begin{tabular}{l c c c c c c c}
\includegraphics[trim=0cm 0cm 0.cm 0cm, width=0.50\textwidth]{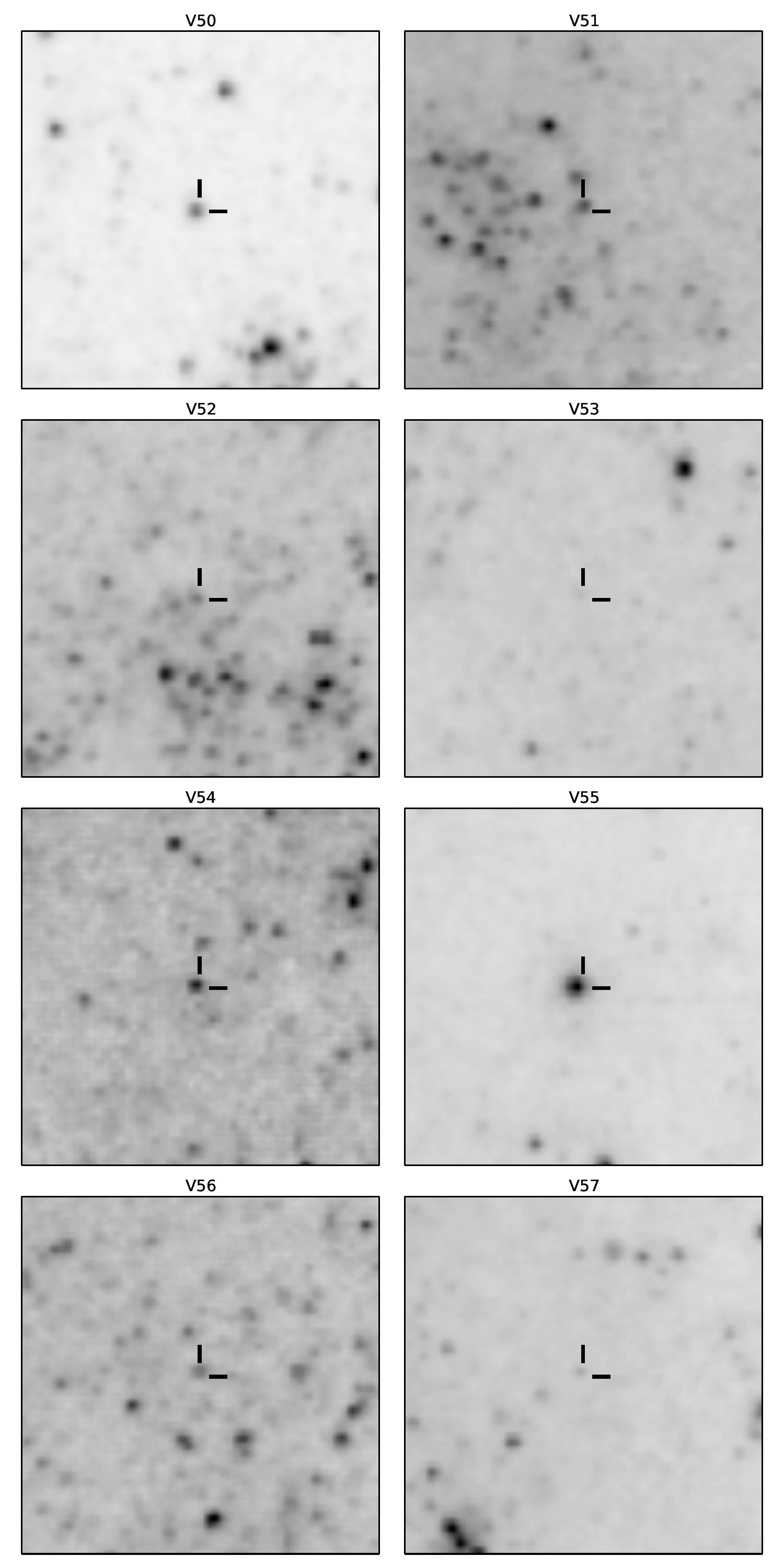} &
\includegraphics[trim=0cm 0cm 0.cm 0cm, width=0.50\textwidth]{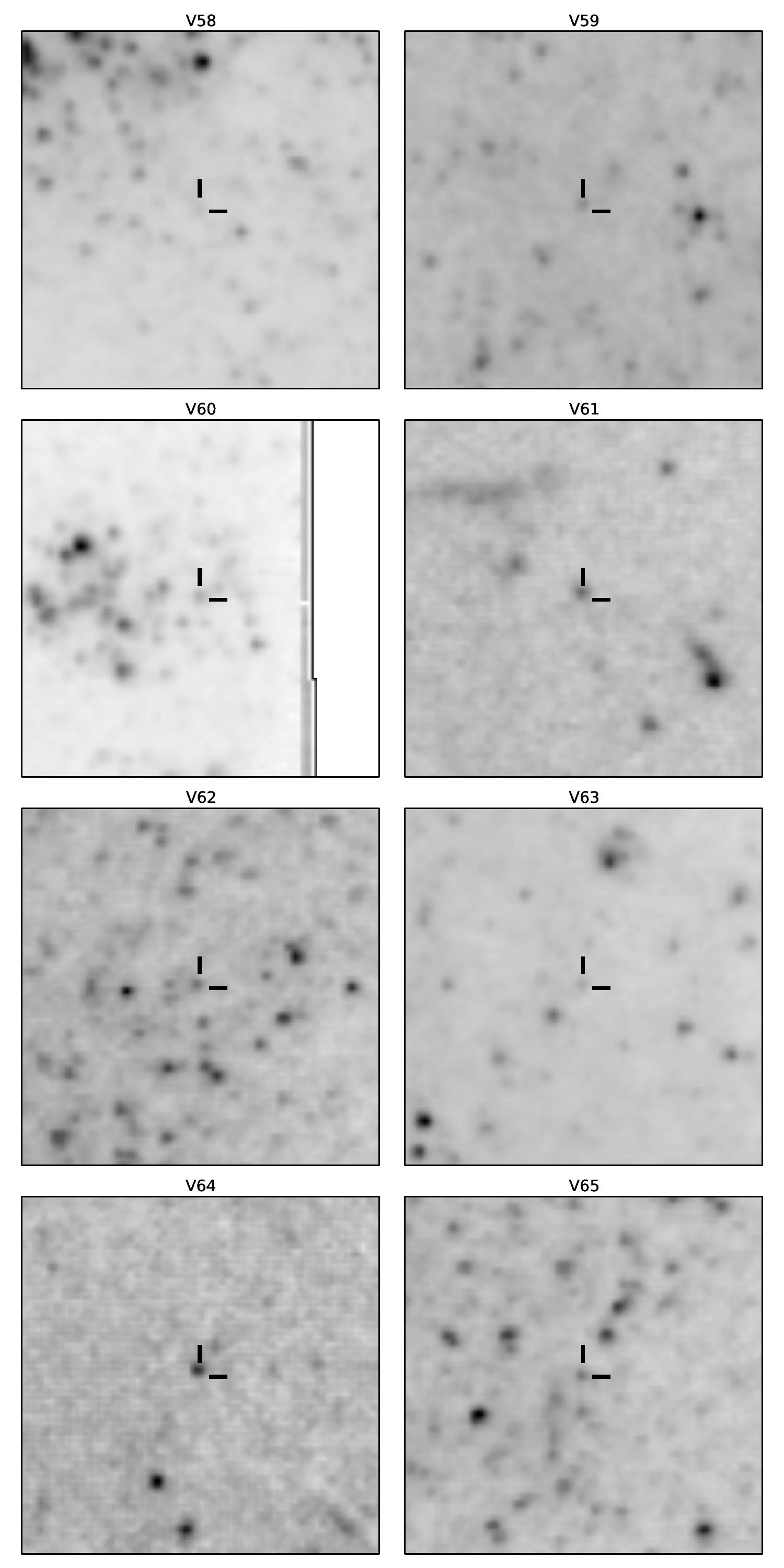}& \\
\includegraphics[trim=0cm 0cm 0.cm 0cm, width=0.50\textwidth]{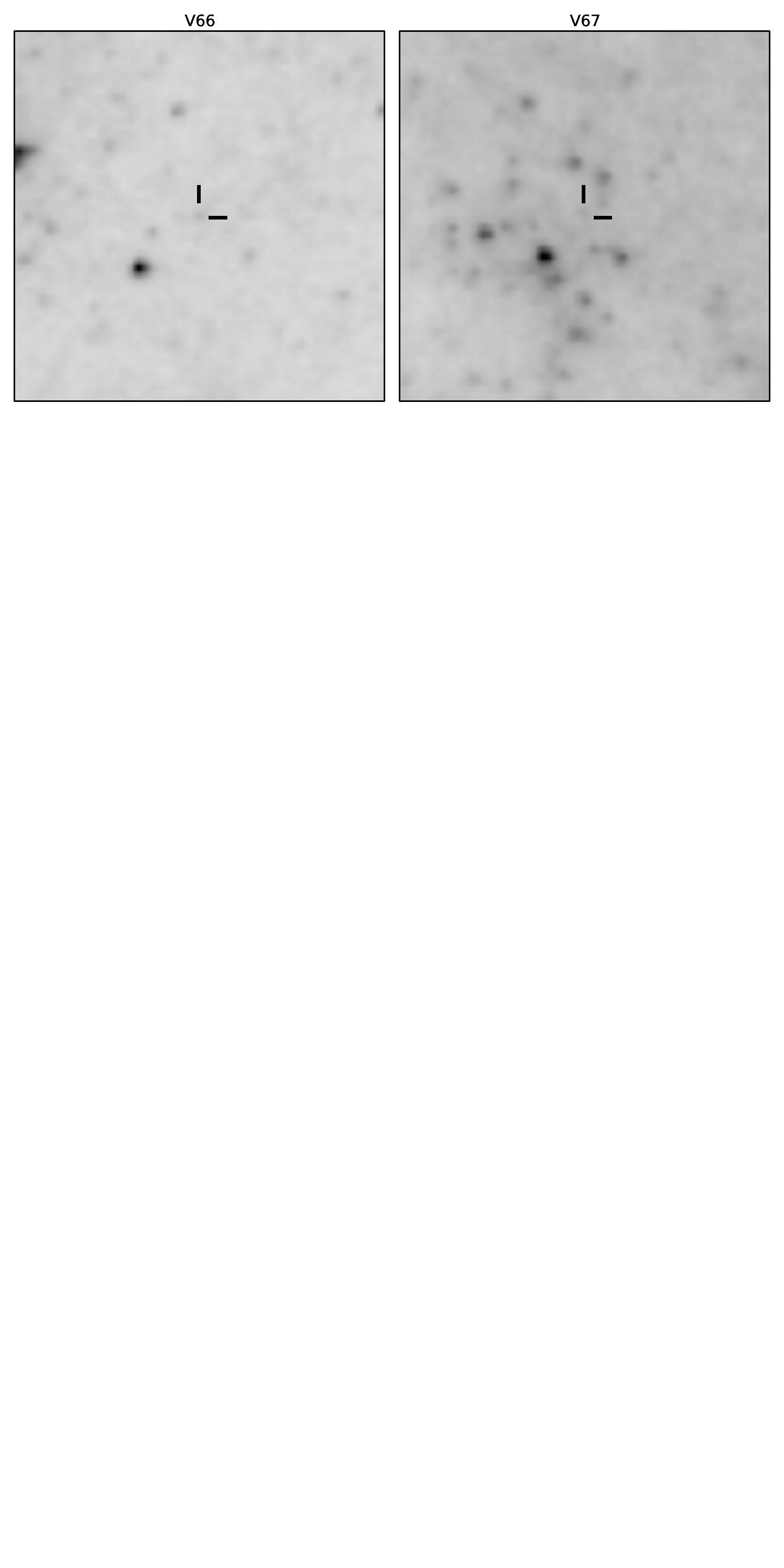} &
\setcounter{figure}{14}

\end{tabular}
\vspace{4cm} 

\caption{continued}
\end{figure*}

\begin{figure*}[h!tb]
\centering
\begin{tabular}{l c c c c c c c}
\includegraphics[trim=0cm 0cm 0.cm 0cm, width=0.5 \textwidth]{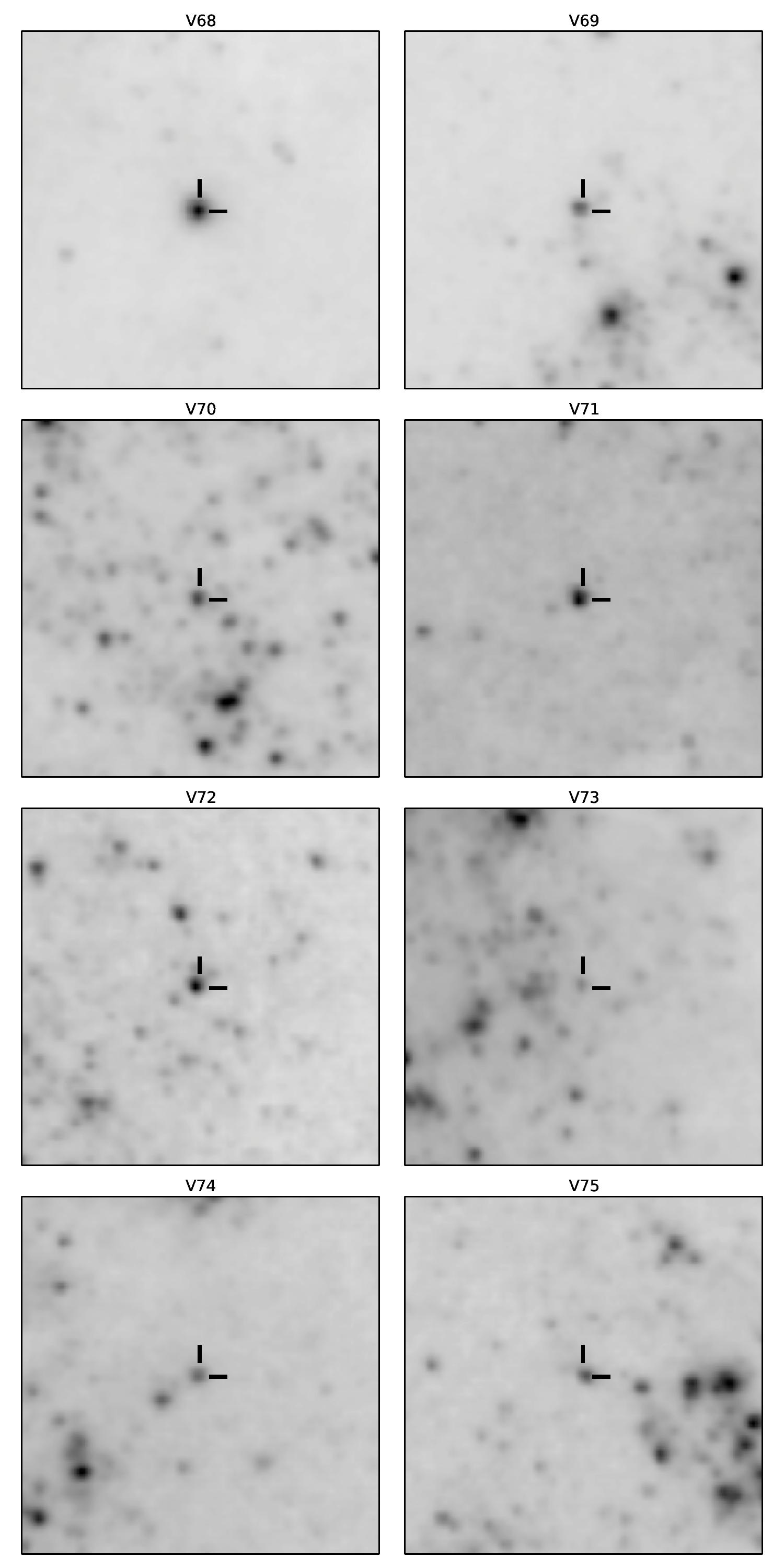} &
\includegraphics[trim=0cm 0cm 0.cm 0cm, width=0.5 \textwidth]{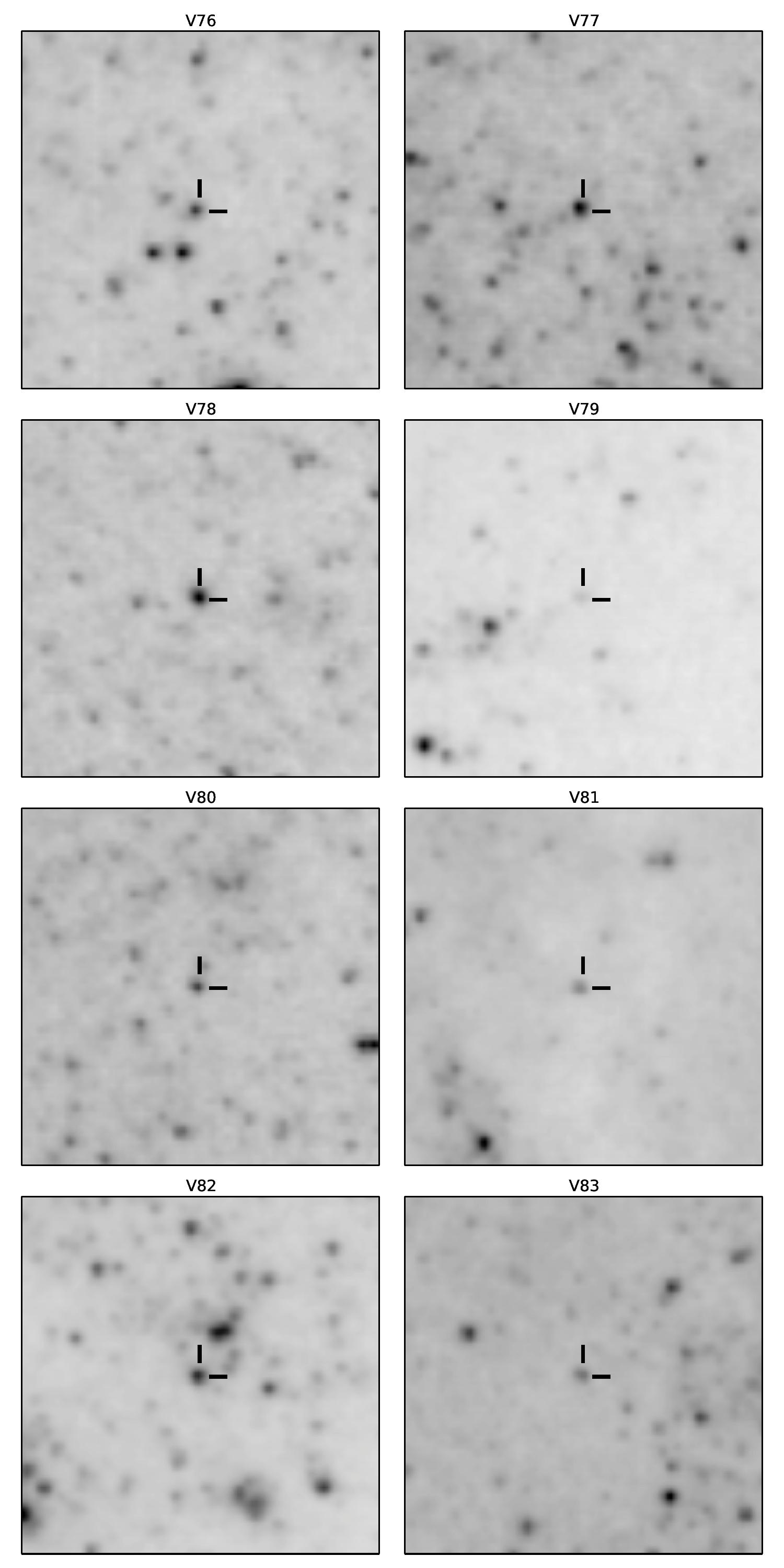}& \\
\end{tabular}

\caption{Same as Figure~\ref{chartspc} but for the candidate variables in the WF3 chip.}
\end{figure*}

\begin{figure*}[h!tb]
\centering
\begin{tabular}{l c c c c c c c}
\includegraphics[trim=0cm 0cm 0.cm 0cm, width=0.5 \textwidth]{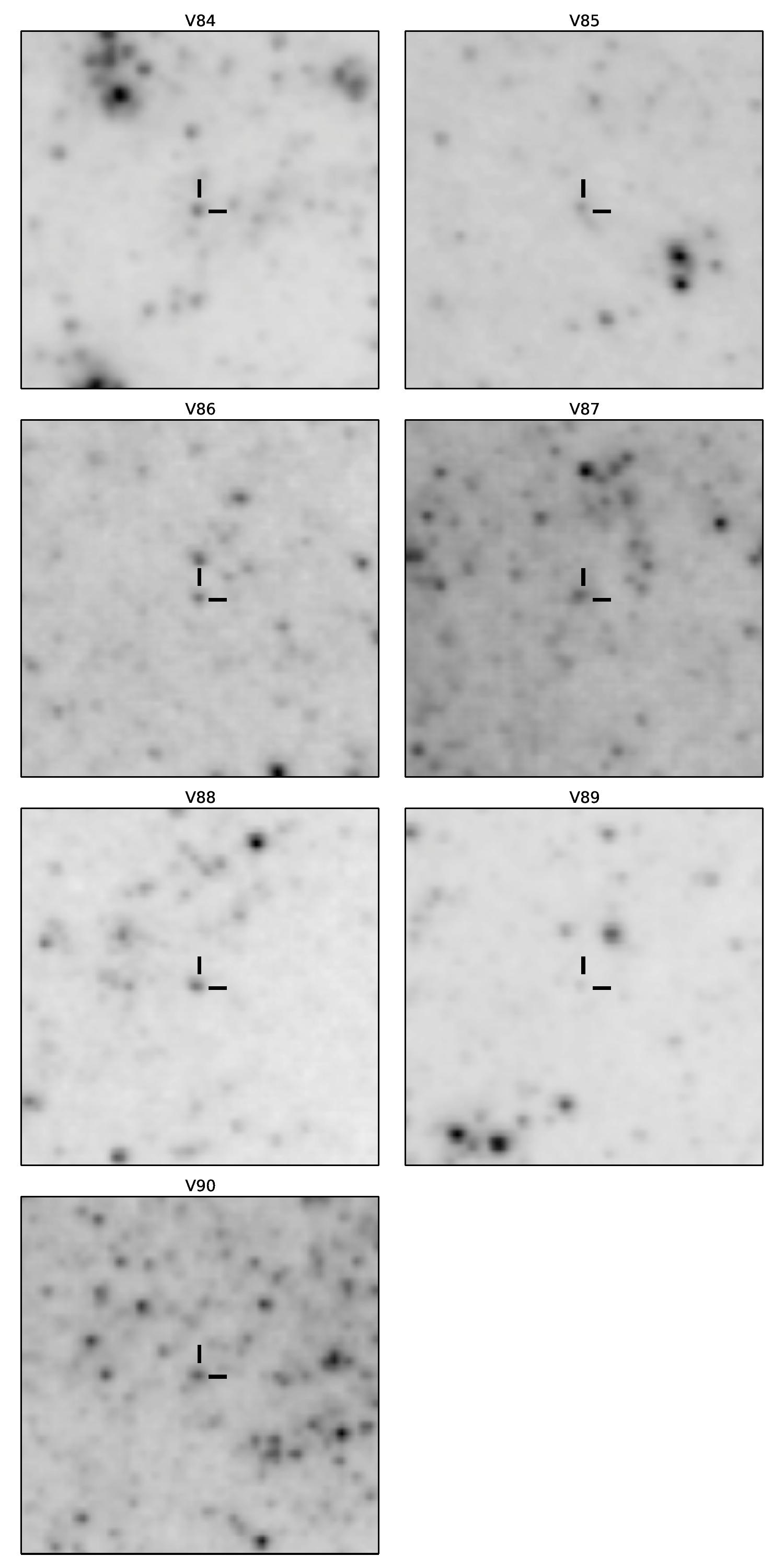}&
\includegraphics[trim=0cm 0cm 0.cm 0cm, width=0.5 \textwidth]{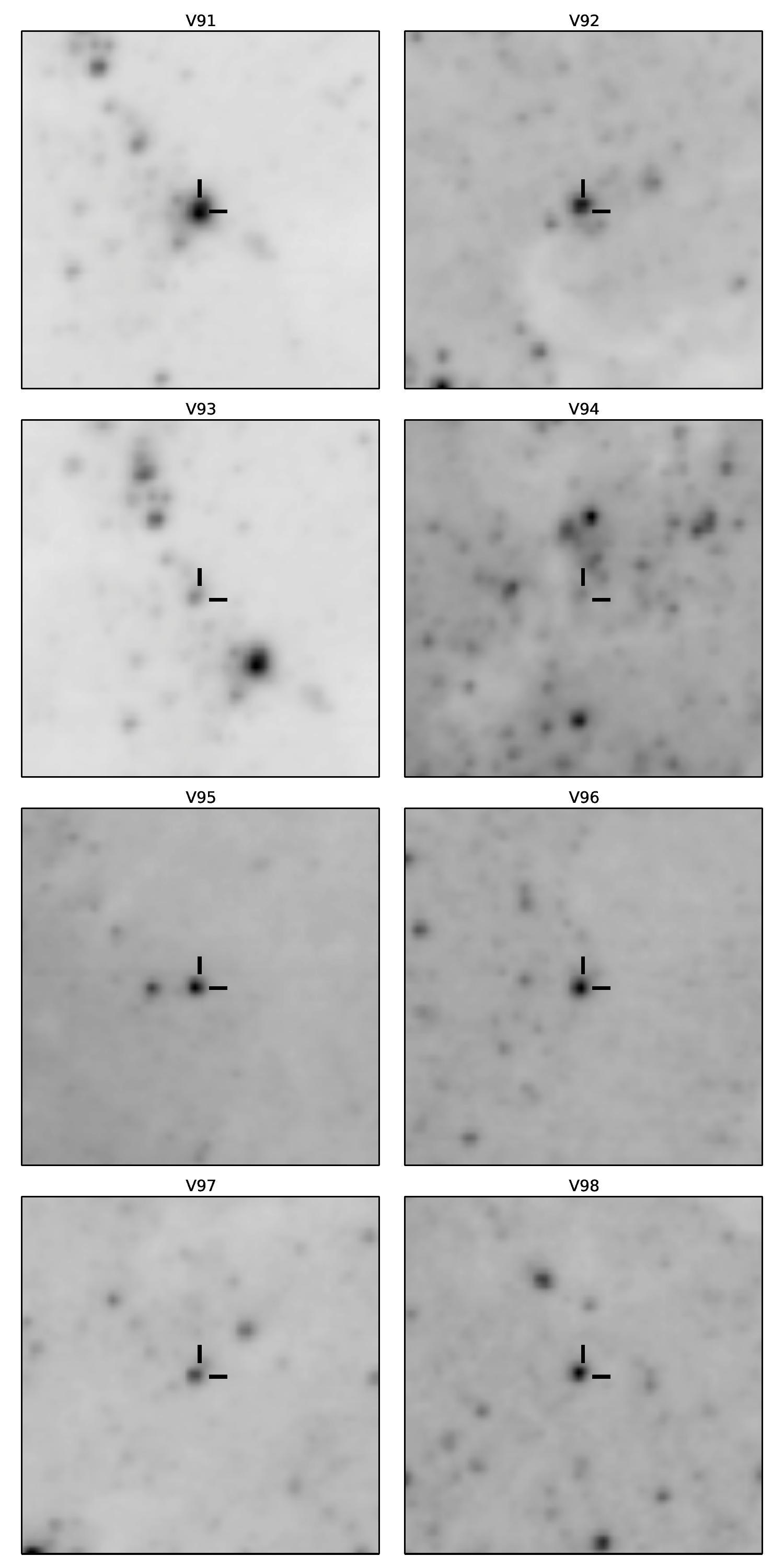} &\\

\end{tabular}

\caption{Same as Figure~\ref{chartspc} but for the candidate variables in the WF4 chip.}
\end{figure*}

\begin{figure*}[h!tb]
\centering
\begin{tabular}{l c c c c c c c}
\includegraphics[trim=0cm 0cm 0.cm 0cm, width=0.5 \textwidth]{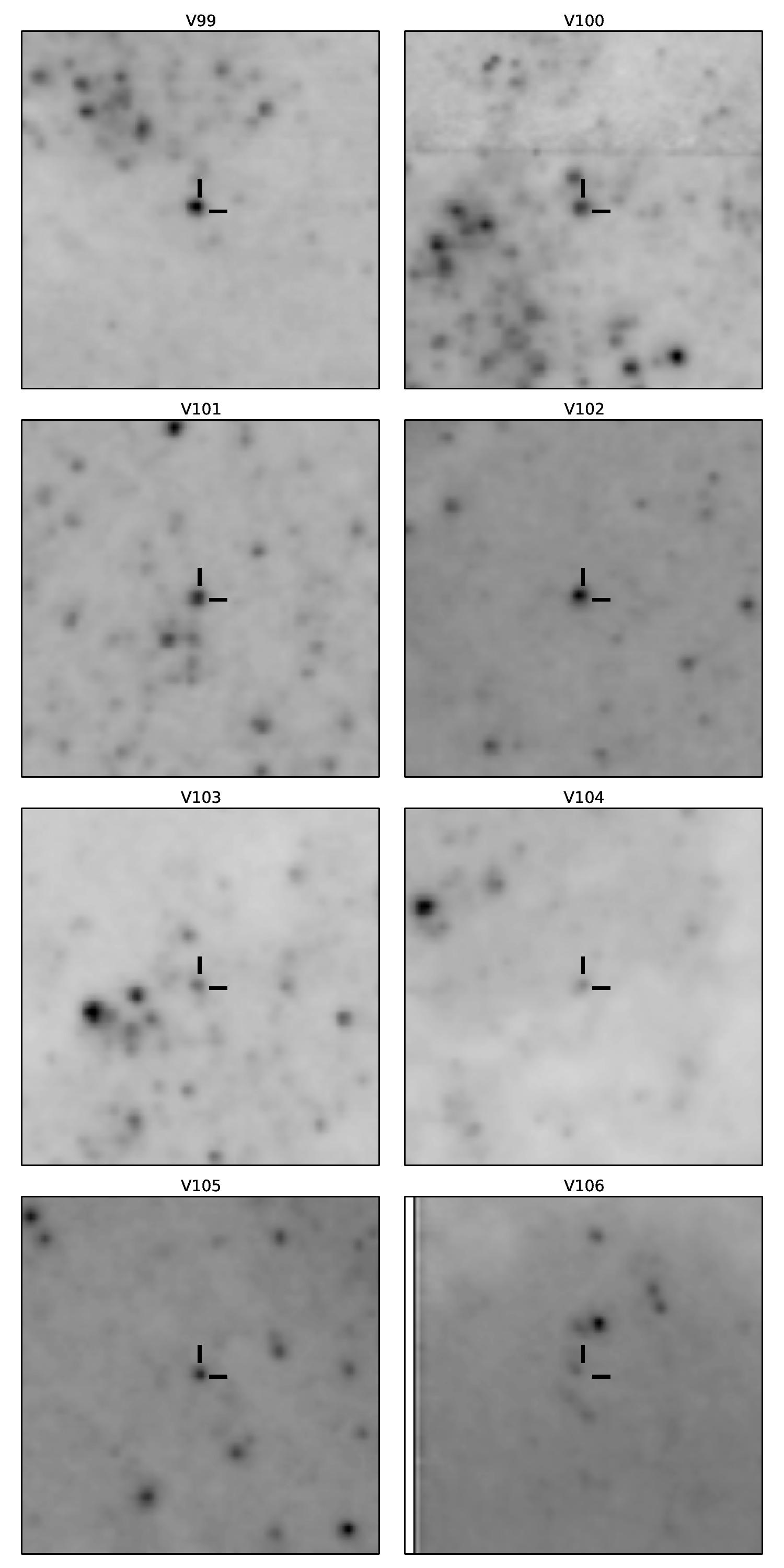}& 
\includegraphics[trim=0cm 0cm 0.cm 0cm, width=0.5 \textwidth]{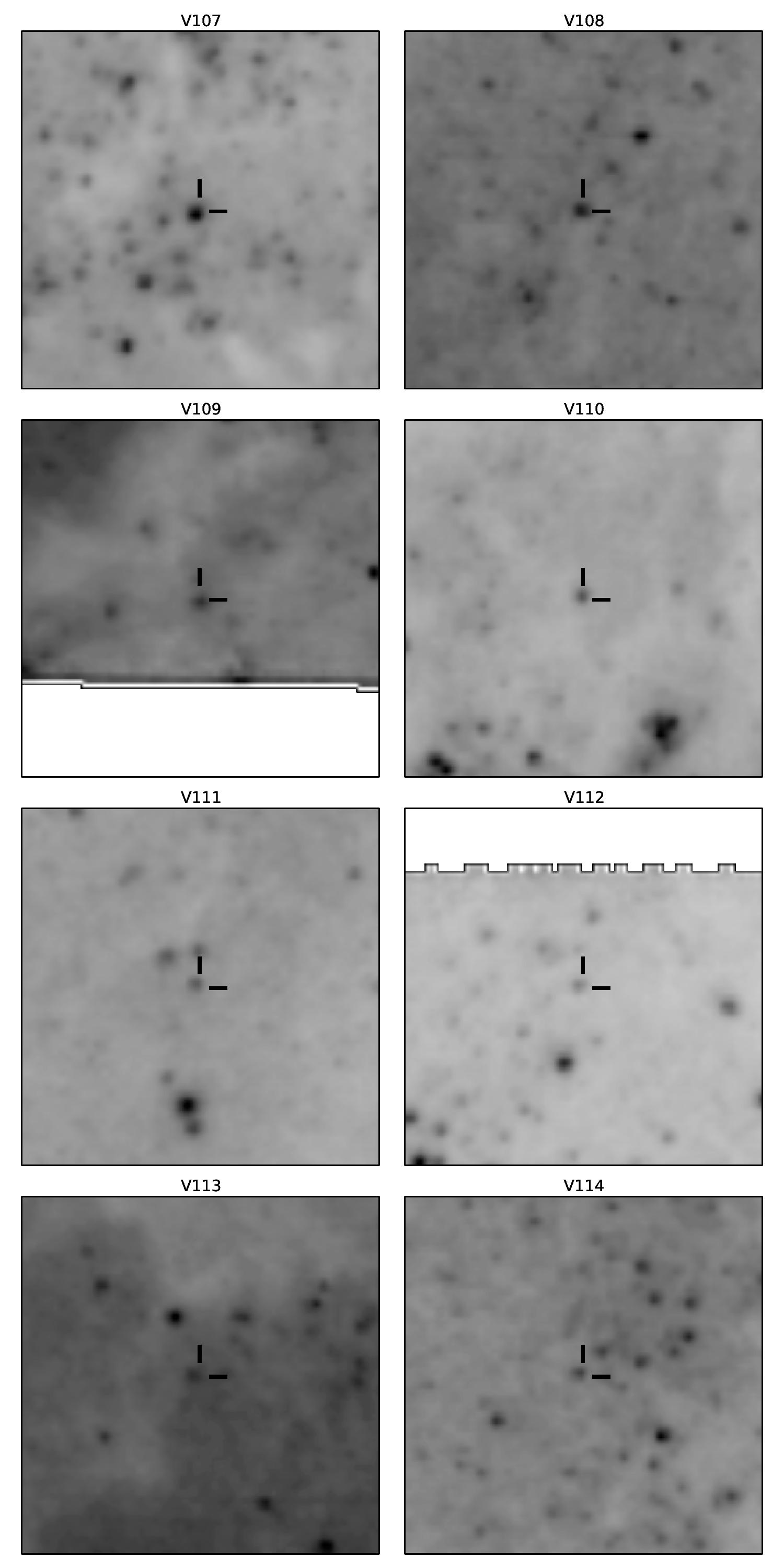} & \\
\setcounter{figure}{16}

\end{tabular}
\caption{continued}

\end{figure*}

\begin{figure*}[h!tb]
\centering
\begin{tabular}{l c c c c c c c}
\includegraphics[trim=0cm 0cm 0.cm 0cm, width=0.5 \textwidth]{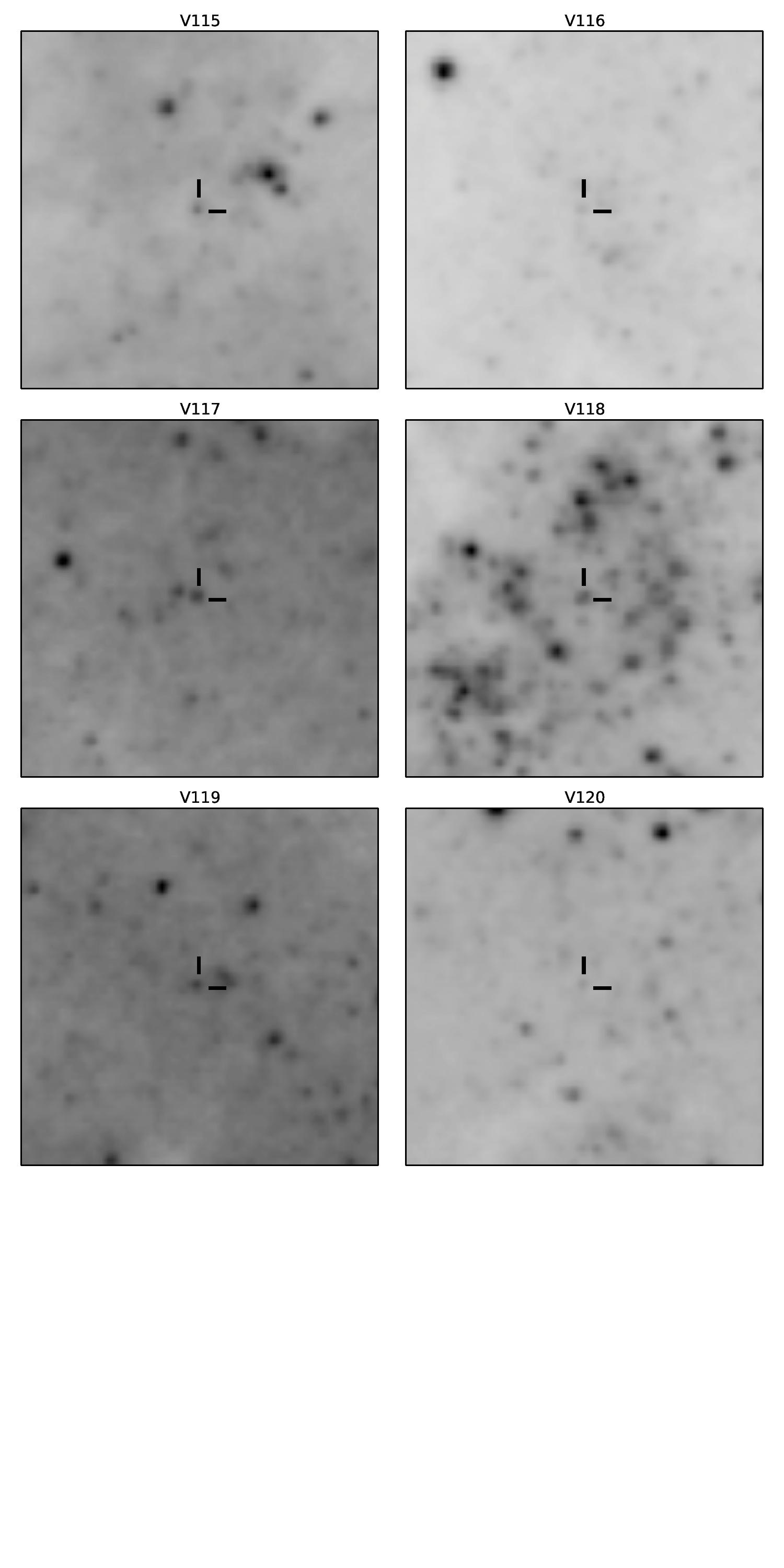}& \\
\setcounter{figure}{16}
\end{tabular}

\caption{continued}
\end{figure*}

\begin{figure*}[h!tb]
\centering
\begin{tabular}{l c c c c c c c}
\includegraphics[trim=0cm 0cm 0.cm 0cm, width=0.5 \textwidth]{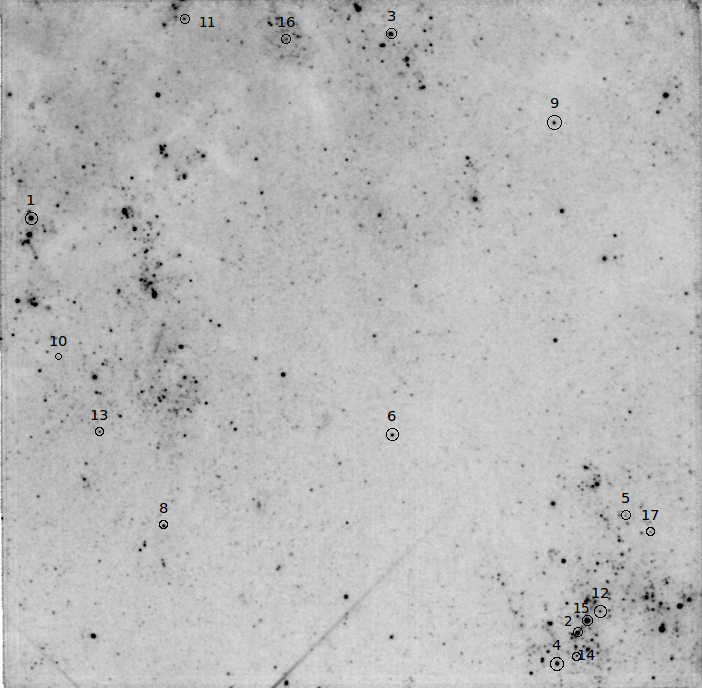}& 
\includegraphics[trim=0cm 0cm 0.cm 0cm, width=0.5 \textwidth]{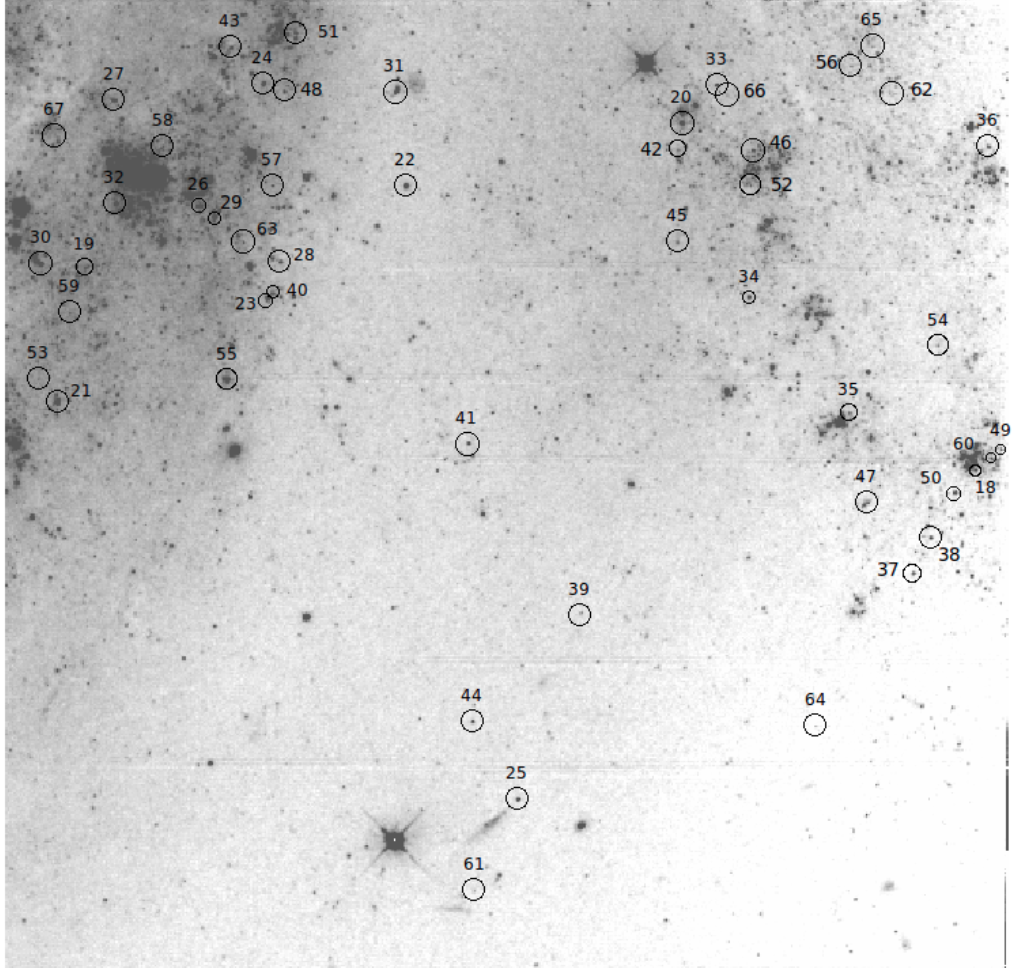} & \\
\includegraphics[trim=0cm 0cm 0.cm 0cm, width=0.5 \textwidth]{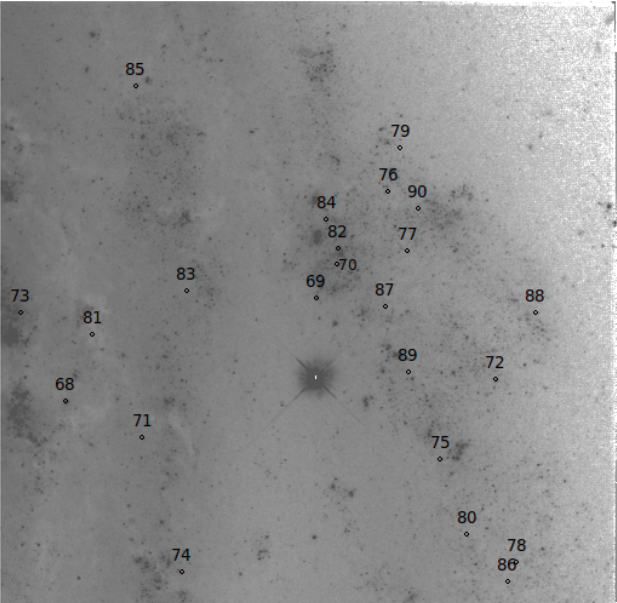} & 
\includegraphics[trim=0cm 0cm 0.cm 0cm, width=0.5 \textwidth]{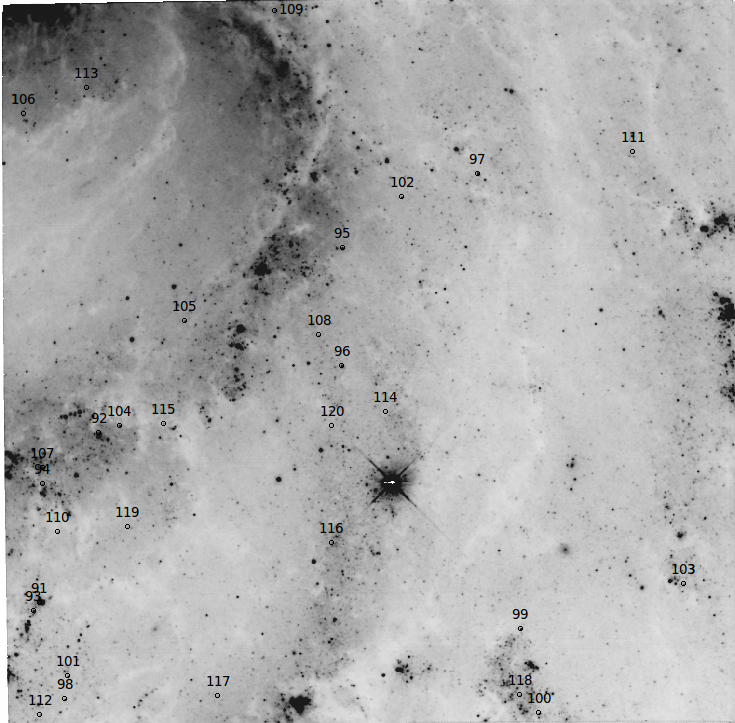} & \\
\end{tabular}
\caption{Field charts of the four WFPC2 chips. The circles indicate the position of each of the Cepheids, labeled as in Tables~\ref{pcvars} - \ref{wf4svars} }
\end{figure*}

\end{appendices}

\end{document}